%% file: main.tex
\documentclass{jpp}

\input{preamble}

\title{Electron root optimisation for stellarator reactor designs}

\author{E. Lascas Neto\aff{1}
  \corresp{\email{eduardo.neto@tecnico.ulisboa.pt}},
  R. Jorge\aff{2},
  C. D. Beidler\aff{3}
 \and J. Lion\aff{4}}

\affiliation{\aff{1}Instituto de Plasmas e Fusão Nuclear, Instituto Superior Técnico, Universidade de Lisboa, 1049-001 Lisboa, Portugal
\aff{2}Department of Physics, University of Wisconsin-Madison, Madison, Wisconsin 53706, USA
\aff{3}Max-Planck-Institut für Plasmaphysik, D-17491 Greifswald, Germany
\aff{4}Proxima Fusion GmbH, 81671 Munich, Germany}

\begin{document}
\maketitle
    
\begin{abstract}
In this work, we propose a method of optimising stellarator devices to favour the presence of an electron root solution of the radial electric field. Such a solution can help avoid heavy impurity accumulation, improve neoclassical thermal ion confinement and helium ash exhaust and possibly reduce turbulence. This study shows that an optimisation for such a root is possible in quasi-isodynamic stellarators. Examples are shown for both vacuum and finite plasma pressure configurations.
\end{abstract}

\section{Introduction \textbf}
Stellarators are a strong candidate for a fusion power plant reactor as they are not prone to current-driven disruptions, have lower recirculating power fraction, and have an inherent flexibility of their plasma shape \citep{Helander2014}. Such flexibility allows for optimisation of the different plasma and magnetic configurations to achieve the desired reactor properties.
This flexibility also comes with the caveat of the complexity of the optimisation problem and the three-dimensional geometry physics. In particular, coil complexity, increased neoclassical transport and turbulence have been seen as obstacles to the design of a reactor-size device of this type. However, recent work has shown that these difficulties can be surpassed. Improved coil optimisation techniques have shown that simple coil systems can be achieved \citep{Wechsung2022, Jorge2023SingleStage, Kappel2023}. Experimental results in W7-X have shown reduced levels of neoclassical transport \citep{Beidler2021}. New optimisation techniques have made precise quasi-symmetric (QS) \citep{Landreman2022} and quasi-isodynamic (QI) \citep{Goodman2023} stellarator configurations numerically achievable. Advances in turbulence optimisation have also been made \citep{Jorge2023, Kim2023,Clark2023}. With this work, we intend to show that stellarators can also be optimised for properties that entail less heavy impurity accumulation, improved helium ash exhaust, good energy confinement with less restrictive optimisation criteria, and possibly, reduced turbulence.

In a fusion reactor, the ion temperature will need to be large in order to achieve sufficient deuterium-tritium reactivity for the fusion reactions to be sustainable. In most reactor-relevant designs, the large heat load reaching the wall is redirected to the divertor. A carbon divertor in a stellarator has been studied in the W7-X experiment \citep{Pedersen2019}, and partial tungsten coating of the divertor tiles has only been initiated recently in LHD \citep{TOKITANI2019} and in W7-X \citep{Naujoks2023}. Tungsten is the first wall material of choice for high-temperature reactor-like designs as it allows for good power exhaust while allowing for lower tritium retention. Experimentally, tungsten divertors have been studied in large tokamak experiments such as ASDEX-U and JET \citep{Putterich2013}. In these experiments, it is seen that the sputtering of the heavy tungsten ions followed by their accumulation in the plasma core, can degrade and even terminate operation due to radiative collapse \citep{Putterich2013}. Heavy impurity transport is thus bound to be an issue in the path to a future stellarator reactor design.  

The importance of the radial electric field and the impurity density asymmetries as the main drive mechanisms of a strong heavy impurity core accumulation has been known and considered extensively 
\citep{Maassberg1993,Helander2017PRL,Calvo2018,buller2018}. Optimisation of the heavy impurity density asymmetries through control of the parallel variation of the electrostatic potential has been considered in quasi-isodynamic (QI) stellarators \citep{Buller2021}. However, such an optimisation technique requires a considerable number of Fourier components of the electrostatic potential \citep{Buller2021}. Moreover, the available degrees of freedom to change the electrostatic potential are limited in QI designs, as strong flows are likely not present \citep{Helander2008}. Thus, ICRH and fast particle asymmetries are the few options to control the heavy impurity density asymmetries \citep{Buller2021}. 
 
Another approach to optimise for low impurity confinement is to control the ambipolarity constraint so that the plasma will operate with a core positive radial electric field, often called the electron root. In general, for heavy impurities, due to their high charge, the main drive of neoclassical radial transport is the electric field. Thus, the electron root can help maintain low core heavy impurity accumulation for most situations, as the neoclassical pinch of impurities can point radially outwards \citep{Tamura2016}, thus helping flush heavy tungsten impurities out of the plasma core.
Such an electric field solution may also improve helium ash exhaust, which, if not performed effectively may lead to intolerable fuel dilution \citep{Beidler2024}. 
Additionally, the transition from a positive to negative electric field often observed when an electron root is present in the plasma core, may generate large enough sheared flows with the potential to suppress turbulence. The effect of an electron root on ITG turbulence was studied in Ref. \citet{Fu2021}. Electron root plasma operation has been considered experimentally in Refs. \cite{Pablant2018,Fujiwara_2001,Yokoyama_2007} and was seen to have beneficial effects on the electron heat confinement \citep{Yokoyama_2007}. However, such electron root scenarios were mostly obtained at low densities or with high electron temperatures provided by auxiliary heating methods such as NBI and ECRH. Optimising for an electron root operation in new optimised stellarator reactor designs without the need for such auxiliary mechanisms is thus an attractive solution to tackle the heavy impurity accumulation problem that is foreseen in future devices.

Finding the electric field solution to the ambipolarity constraint in three-dimensional magnetic configurations is a complex task. This is because, at the values of temperature and density relevant to the reactor plasma core, the ions and electrons can be at different collisionality regimes, leading to a nonlinear equation in the electric field which can have positive (electron root) and negative (ion root) solutions, or both, throughout the radial extent of the device.  This equation depends on the radial transport coefficients of the electrons and ions, which in turn depend on the magnetic configuration and plasma parameters. Theoretical work to unravel the relation between the solutions of this equation with the magnetic configuration and plasma parameters is scarce. Efforts have been made to obtain the ion and electron monoenergetic transport coefficients to help identify the space of solutions of the ambipolarity in various stellarator devices \citep{Beidler2011} and the study of helium ash transport has been recently done for different W7-X-like configurations in Ref. \citet{Beidler2024}. In this work, we propose a new method to optimise for such radial electric field solutions based on maximising the plasma parameter space of electron root solutions. We start by reviewing the physics behind the cost function and the numerical methods used. Then, we describe the optimisation method used as well as the method used to obtain the electric field solution in the optimised configurations. Finally, we discuss our results by analysing different optimised configurations, comparing the respective electric field solutions and discussing how they affect several properties of interest for a fusion reactor design. 

\section{Theoretical Background \textbf}
In the standard approach, stellarator optimisation is performed by finding a magnetic configuration that conforms to different aspects of interest defined by a given cost function. Such magnetic configurations are usually bound to obey the MHD equilibrium equation \citep{Freidberg2014}
\begin{equation} \label{MHD}
\bm{J}\times \bm{B}=\bm{\nabla} P,
\end{equation}
where $\bm{J}=\nabla \times \bm{B}/\mu_0$ is the plasma current density, $\bm{B}$ is the magnetic field and $P$ is the plasma pressure.
Equation \ref{MHD} governs the plasma equilibrium of fusion reactors in the absence of strong flows or strong pressure anisotropies. If the equilibrium is assumed to be described by nested closed surfaces of constant toroidal magnetic flux $\psi$, the plasma pressure $P=P(\psi)$ is a flux function. The plasma pressure is the sum of both electron and ion contributions $P=n_eT_e+n_iT_i$, where $n_a(\psi)$ and $T_a(\psi)$ are the density and temperature of species $a$, respectively. In the set of Boozer flux coordinates $(\psi,\theta,\phi)$ \citep{Boozer1981} the magnetic field $\bm{B}$ can be written in its contravariant and covariant forms respectively as 
\begin{equation} \label{B_field}
\bm{B}=\nabla \psi\times \nabla \alpha =B_{\psi}(\psi,\theta,\phi)\nabla \psi +I(\psi)\nabla \theta +G(\psi) \nabla \phi \numberthis
\end{equation}
where $\alpha=\theta-\iota(\psi) \phi$ is the field line label, $\iota$ is the rotational transform, $G$ is $\mu_0/(2\pi)$ times the poloidal current outside the surface, $I$ is $\mu_0/(2\pi)$  times the toroidal current inside the surface, and $B_{\psi}$ is associated with the Pfirsch-Schl\"{u}tter current.

In fusion reactors, it is of interest to have a magnetic configuration in which the plasma is well confined in order to be able to sustain fusion reactions. In toroidal magnetic confinement devices such as stellarators, particles can drift radially, leading to unwanted transport. This transport is usually dictated by the particles trapped in the magnetic field wells (neoclassical effects) and the anomalous transport due to turbulence. State-of-the-art stellarator configurations are optimised to reduce neoclassical transport across flux surfaces \citep{Beidler2021,Landreman2022,Goodman2023}. Magnetic configurations for which the neoclassical transport across flux surfaces vanishes are called omnigenous. Such magnetic configurations are obtained by tailoring the magnetic field strength properties such that the bounce average of the radial drift of the trapped particles along their bounce orbit is zero. Two particular cases of interest are QS \citep{Landreman2022} and QI \citep{Jorge2022,Camacho2022,Goodman2023} configurations. In QS devices, the magnetic field strength obeys $B=B(M\theta-N\phi)$, with $M$ and $N$ unique integers. Such a symmetry allows for the drift across flux surfaces to vanish along the bounce orbit of the trapped particles as the magnetic well is symmetric about its minimum value, similar to tokamaks \citep{HelanderSigmar}. On the other hand, QI configurations are less restrictive and omnigenity is achieved by imposing 
\begin{align}\label{QI_omni1}
 &\left. \frac{\partial B_\text{min}(\psi,\alpha)}{\partial \alpha}\right|_{\psi} \overset{!}{=}\left.\frac{\partial B_\text{max}(\psi,\alpha)}{\partial \alpha}\right|_{\psi} \overset{!}{=}0 \hspace{2mm}\forall \hspace{2mm} \psi, \numberthis\\ 
 &\left.\frac{\partial l_b(\psi,B,\alpha)}{\partial \alpha}\right|_{\psi,B} \overset{!}{=}0   \hspace{2mm} \forall \hspace{2mm} \psi,B \numberthis,\label{QI_omni2}
\end{align}
in which $B_\text{min}(\psi,\alpha)$ and $B_\text{max}(\psi,\alpha)$ are respectively the minimum and the maximum of the magnetic field strength in a particular flux surface and magnetic field line, and $l_b(\psi,B,\alpha)$ is the distance along the particle orbit between two consecutive bounce points. These constraints impose certain properties in the magnetic wells \citep{CaryQI}, which are less restrictive than the symmetry imposed by QS. Quasi-isodynamic devices also impose an extra constraint that the contours of $B$ close poloidally to target the lowest possible bootstrap current in an omnigenous configuration \citep{Helander2014}. While perfect omnigenity was shown to be impossible to achieve \citep{CaryQI}, precise QS and QI magnetic fields can be achieved numerically \citep{Landreman2022,Goodman2023} with error fields as small as the geomagnetic field.

Ambipolarity in a fusion plasma is achieved when the net radial currents in the plasma vanish. Such a condition is linked to the temporal evolution of the radial electric field, which is governed by the electric field transport equation. Such an equation is obtained by flux averaging Ampère's law
\begin{align}\label{Faraday_radial}
\varepsilon\frac{\partial E_{\psi}}{\partial t} +J_{\psi}=0,
\end{align}
in which $E_{\psi}=\bm{E}\cdot \nabla \psi$ and $J_{\psi}=\bm{J}\cdot \nabla \psi$ are the components of the electric field $\bm{E}$ and the plasma current density across flux surfaces, and $\varepsilon$ is the plasma dielectric constant. In \cref{Faraday_radial}, the fact that the plasma obeys to \cref{MHD} is used. Thus, the currents generated by the magnetic field have no component across the flux surface, i.e.\, $(\nabla \times \bm{B})\cdot \nabla \psi=0$. Therefore, in \cref{Faraday_radial}, $J_{\psi}$ is only generated by the plasma (neglecting external sources) and can be obtained from the distribution function $f^a$ of species $a$. Expanding the distribution function in $\Delta_{\psi}/L$, with $\Delta_{\psi}$ the orbit width and $L$ the typical macroscopic length of the plasma gradients, we can write $f^a=f_M^a+f_1^a+f_h^a$, with $f_M^a$ a zeroth order Maxwellian. The radial current density due to the plasma can then be written as 
\begin{align}\label{current_f}
&J_{\psi}=\sum_a \int Z_a e \bm{v_d}\cdot \nabla \psi f^a d^3 v\nonumber \\
&=\sum_a \int Z_a e \bm{v_d}\cdot \nabla \psi f^a_1 d^3 v+\sum_a \int Z_a e \bm{v_d}\cdot \nabla \psi f^a_h d^3 v \nonumber \\ 
&= \sum_a e Z_a \Gamma_a + \varepsilon \left(\frac{dV}{d\psi}\right)^{-1}\frac{\partial}{\partial \psi}\left[\frac{dV}{d\psi}\sum_a (e Z_aD_{E_{\psi}}^a) \frac{\partial E_{\psi}}{\partial\psi}\right], \numberthis
\end{align}
where $f^a_1$ is the first order part of $f_a$ and $f_h^a$ contains higher order terms \citep{Hastings1984,Shaing1984}. In \cref{current_f}, $\Gamma_a$ is the usual particle-flux-density across flux surfaces obtained at lowest order in $\Delta_{\psi}/L$, and $D_{E_{\psi}}^a$ is the electric field diffusivity associated with species $a$ which comes from considering higher order orbit width effects \citep{Hastings1984,Shaing1984}, $e$ is the electron charge, $Z_a$ is the charge number of species $a$, $\bm{v_d}$ is the drift velocity of a particle and $V(\psi)$ is the volume enclosed by the flux surface $\psi$. If the higher order orbit width effects from $f_h^a$ are neglected, the steady-state radial electric field is obtained from the ambipolarity constraint
\begin{align}\label{ambi}
 \sum_a e Z_a \Gamma_a =0,
\end{align}
which can have multiple solutions due to the different collisionality regimes of the electrons and ions. The choice between these multiple solutions depends on thermodynamical considerations \citep{Shaing1984,Hastings1985Entropy} and large transitions of the electric field sign between neighbouring flux surfaces can occur. In these regions, the gradient of the electric field gets so large that higher order orbit width effects from $f_h^a$ become important. When considering this correction, one obtains the electric field diffusion term in \cref{current_f} and information about the electric field gradient is retained \citep{Hastings1984}. This extra term is related to the minimisation of the heat transfer rate \citep{Shaing1984}. Taking into account this extra term, the electric field transport equation can be written as
\begin{equation}\label{Er_transp}
\varepsilon\frac{\partial E_{\psi}}{\partial t} - \varepsilon\left(\frac{dV}{d\psi}\right)^{-1}\frac{\partial}{\partial \psi}\left[\frac{dV}{d\psi}\sum_a \left(e Z_aD_{E_{\psi}}^a\right) \frac{\partial E_{\psi}}{\partial \psi}\right] = \sum_a e Z_a \Gamma_a, 
\end{equation}
yielding a single solution for the radial electric field. However, the electric field affects the power balance in a fusion reactor and vice-versa. Therefore, to obtain a consistent solution of the electric field, \cref{Er_transp} has to be solved together with a set of energy transport equations. A common form of the energy transport equation for species $a$ is \citep{Turkin2011}
\begin{equation}\label{energy}
\frac{\partial n_{a}T_{a}}{\partial t}+\left(\frac{dV}{d\psi}\right)^{-1}\frac{\partial}{\partial \psi}\left[\frac{dV}{d\psi}\left(Q_a+\Gamma_aT_a-\chi_a \frac{\partial T_a}{\partial \psi}\right)\right]=\mathcal{P}_a+e Z_a \Gamma_a E_{\psi}, 
\end{equation}
where $Q_a$ is the energy-flux-density of species $a$ across flux surfaces and $\chi_a$ is the anomalous heat diffusivity coefficient of species $a$. On the right-hand side of \cref{energy}, the first term, $\mathcal{P}_a$, is the total power balance of external energy sources and sinks related to species $a$ and the second term is the work done by the electric field in the fluid of species $a$. 
If the density of the different species is allowed to vary, then a set of continuity equations also needs to be solved. A common form of the continuity equation of species $a$ is \citep{Turkin2011}
\begin{align}\label{continuity}
\frac{\partial n_{a}}{\partial t} +\left(\frac{dV}{d\psi}\right)^{-1}\frac{\partial}{\partial \psi}\left[\frac{dV}{d\psi}\left(\Gamma_a-\mathcal{D}_a \frac{\partial n_a}{\partial \psi}\right)\right]=S_a,
\end{align}
in which $\mathcal{D}_a$ is the particle diffusivity of species $a$ associated with the anomalous transport and $S_a$ is the source term of particles of species $a$. To close the system of equations \crefrange{Er_transp}{continuity} we use the quasi-neutrality constraint $\sum_a Z_a n_a=0$. The particle and energy-flux-densities can be prescribed using a reduced model or obtained from solving the drift kinetic equation, while the anomalous diffusivities are obtained either from a reduced model or from solving the gyrokinetic equation.
We note that in the previous discussion, we have assumed that the fluxes contributing to the ambipolarity term of the radial current density in \cref{current_f} are the neoclassical fluxes. Such an assumption comes from the fact that in stellarators, at the usual gyrokinetic orderings, the anomalous transport is intrinsically ambipolar \citep{Helander2014,Sugama1996} and will
thus have no influence on solutions of \cref{ambi}. Thus, unless a perfect omnigenous magnetic configuration is achieved to reduce neoclassical fluxes to extremely low levels, the neoclassical fluxes are the most important players in the ambipolarity constraint. 
Therefore, we now show how the neoclassical theory can help obtain the distribution function $f_a$, and find $\Gamma_a$ and $Q_a$.

The usual local neoclassical theory allows to obtain the steady-state solution of the first order distribution function $f_1^a$ which is dictated by the drift kinetic equation \citep{SFINCS}
\begin{align}\label{DK}
&\dot{\psi}\frac{\partial f_1^a}{\partial \psi}+\dot{\theta}\frac{\partial f_1^a}{\partial \theta}+\dot{\phi}\frac{\partial f_1^a}{\partial \phi}+\dot{x}\frac{\partial f_1^a}{\partial x}+\dot{\xi}\frac{\partial f_1^a}{\partial \xi}-C(f^a _1) \nonumber\\
&=\frac{T_ac}{ZeB^3\sqrt{g}}x^2(1+\xi^2)\left(G\frac{\partial B}{\partial \theta}-I\frac{\partial B}{\partial \phi}\right)f_M^a (A_1^a+x^2A_2^a)-\frac{v_{\parallel}}{B_0} A_3^a f_M, \numberthis
\end{align}
where the velocity variables are the normalised velocity $x=v/v_{th}^a$ and the pitch angle $\xi=v_{\parallel}/v$, with $v_{th}^a=\sqrt{2T_a/m_a}$ the thermal velocity of species $a$, $m_a$ the mass of species $a$ and $c$ the speed of light. The Jacobian determinant $\sqrt{g}$ in \cref{DK} is written as $\sqrt{g}=(G+\iota I)/B^2$ in Boozer coordinates.
In \cref{DK}, the thermodynamical forces are given by
\begin{align*} 
&A_1^a=\frac{1}{n_a}\frac{d n_a}{d \psi}-\frac{Z_a e E_{\psi}}{T}-\frac{3}{2}\frac{1}{T_a}\frac{d T_a}{d \psi},\label{A1} \numberthis  \\  
&A_2^a=\frac{1}{T_a}\frac{d T_a}{d \psi}, \label{A2} \numberthis\\   
&A_3^a=-\frac{Z_a e B_0}{T_a} \frac{\langle E_{\parallel}B\rangle}{\langle B^2\rangle}, \label{A3} \numberthis
\end{align*}
with the flux surface average of a quantity $X$ defined as $\langle X \rangle = \int_0^{2\pi} \int_0^{2\pi} d\theta d\phi X \sqrt{g} /(dV/ d\psi)$.
In the local approximation, $\dot{\psi}$ is usually neglected, and at a large aspect ratio we can write the poloidal velocity of the particles as
\begin{equation} \label{theta_dot}
\dot{\theta}=\frac{xv_{th}^a\xi}{B\sqrt{g}}\iota -\frac{c}{B^2 \sqrt{g}}E_{\psi}G=\left(x\xi-E^*_{\psi}\right)\frac{\iota v_{th}^a}{B\sqrt{g}},
\end{equation} 
and the toroidal velocity as
\begin{equation} \label{phi_dot}
\dot{\phi}=\frac{xv_{th}^a\xi}{B\sqrt{g}} +\frac{c}{B^2 \sqrt{g}}E_{\psi}=\left(x\xi+E^*_{\psi}\frac{I \iota}{G}\right)\frac{ v_{th}^a}{B\sqrt{g}},
\end{equation} 
where the magnetic drift tangent to the flux surfaces has been neglected, and thus the poloidal and toroidal velocity of the particles combine the motion along field lines and the $E\times B$ drift. In \cref{theta_dot,phi_dot}, $E^*_{\psi}=cE_{\psi}G/(\iota v_{th}^a B)$ is a measure of the strength of the $E\times B$ drift (drift frequency) relative to the parallel orbit motion of the particle (transit frequency). The parallel acceleration of the particles is given by
\begin{align*}  \label{xi_dot}
\dot{\xi}=-\frac{x(1-\xi^2)v_{th}^a}{2B^2\sqrt{g}}\left(\iota\frac{\partial B}{\partial \theta}+\frac{\partial B}{\partial \phi}\right)-\frac{\iota v_{th}^a}{2 \sqrt{g}B^2}\xi(1-\xi^2)\left(\frac{\partial B}{\partial \theta}-\frac{I}{G}\frac{\partial B}{\partial \phi}\right)E^*_{\psi}, \numberthis
\end{align*}
and the kinetic energy temporal variation by
\begin{equation} \label{x_dot}
\dot{x}=-\frac{\iota v_{th}^a}{2 \sqrt{g}B^2}x(1+\xi^2)\left(\frac{\partial B}{\partial \theta}-\frac{I}{G}\frac{\partial B}{\partial \phi}\right)E^*_{\psi}.
\end{equation}
These two equations show that trapped/passing domains are influenced by the parallel mirror force as well as by the effect of a strong radial electric field. The set of \crefrange{theta_dot}{x_dot} for the orbits of the particles in the phase space are usually called "full trajectories" \citep{SFINCS}. If we neglect the electric field corrections in \cref{xi_dot,x_dot}, we obtain the so-called "partial trajectories" \citep{SFINCS}. 

It is useful to linearise the drift kinetic equation, \cref{DK}, to obtain an approximate, but more expeditiously and less numerically expensive result. Such a task is accomplished by looking at the right-hand side term of \cref{DK}, which is associated with the thermodynamical forces in \crefrange{A1}{A3}. In \cref{DK}, it can be observed that each thermodynamical force term is associated with a particular dependency on the velocity space variables $(x,\xi)$. The first-order correction to the distribution function can thus be linearised as 
\begin{equation} \label{f1_linear}
f_1^a=\frac{n_a}{\pi^{\frac{3}{2}}{v^a_{th}}^{3}}\sum_{p=1}^3 f_{1p}^a A_p^a 
\end{equation}
which leads to the set of equations 
\begin{align*} \label{nonmono}
&(\xi x-E^*_{\psi})\iota \frac{\partial f_{1p}^a}{\partial \theta}+\left(\xi x+\frac{\iota I}{G}E^*_{\psi}\right)\frac{\partial f_{1p}^a}{\partial \phi}-\frac{x(1-\xi^2)}{2}\left(\frac{\iota}{B}\frac{\partial B}{\partial \theta}+\frac{1}{B}\frac{\partial B}{\partial \phi}\right)\frac{\partial f_{1p}^a}{\partial \xi}\\ 
&-\frac{(1-\xi^2)}{2}\left(\frac{\iota}{B}\frac{\partial B}{\partial \theta}-\frac{\iota I}{G B}\frac{\partial B}{\partial \phi}\right)\left[\xi\frac{f_{1p}^a}{\partial \xi}+x\frac{f_{1p}^a}{\partial x}\right]E_{\psi}(\psi)=\frac{B_0}{B}\nu'C(f_{1p}^a)+\mathcal{S}^a_p \numberthis ,
\end{align*} 
where
\begin{align} \label{nonmono1}
&\mathcal{S}_1^a=\frac{T_a cG}{Z_aeB_0}x^2(1+\xi^2)e^{-x^2}\frac{B_0}{B}\left(\frac{1}{B}\frac{\partial B}{\partial \theta}-\frac{ I}{G B}\frac{\partial B}{\partial \phi}\right),\numberthis \\ \label{nonmono2} 
& \mathcal{S}_2^a=\frac{T_a cG}{Z_aeB_0}x^4(1+\xi^2)e^{-x^2}\frac{B_0}{B}\left(\frac{1}{B}\frac{\partial B}{\partial \theta}-\frac{ I}{G B}\frac{\partial B}{\partial \phi}\right), \numberthis \\ \label{nonmono3}
&\mathcal{S}_3^a=-x\xi e^{-x^2}\frac{(G+\iota I)}{B_0}.
\end{align}
In \cref{nonmono}, $\nu^{\prime}=\nu_{aa}(G+\iota I)/v_{th}^a$ and $\nu_{aa}=n_a(Z_a e)^4\ln \Lambda/(4\pi^4\epsilon_0^2m_a^2{v_{{th}}^a}^3)$ is the typical self-collisions frequency of species $a$ \citep{HelanderSigmar}, $\ln \Lambda$ is the Coulomb logarithm and $B_0$ is the magnetic field strength on-axis. The set of \crefrange{nonmono}{nonmono3} describes the linear approximation of the 4D drift kinetic equation, which, throughout this work, we will refer to as the "4D approach" method to solve the drift-kinetic equation.
The radial particle-flux-density of species $a$ can then be calculated as
\begin{equation} \label{Gammaa}
\Gamma_a=-n_a\left[L_{11}^a A_1^a+L_{12}^a A_2^a+L_{13}^a A_3^a\right],
\end{equation}
the radial energy-flux-density as
\begin{equation} \label{qa}
Q_a=-n_a T_a\left[L_{21}^a A_1^a+L_{22}^a A_2^a+L_{23}^a A_3^a\right],
\end{equation}
and the parallel flow as 
\begin{equation} \label{vpara}
\frac{\langle U_{a_{\parallel}}B\rangle}{B_0}=-\left[L_{31}^a A_1^a+L_{32}^a A_2^a+L_{33}^a A_3^a\right].
\end{equation}
The thermal transport coefficients are thus defined as
\begin{align*} \label{L1i}
&L_{1p}^a=-\left\langle\int dx^3 \pi^{-\frac{3}{2}}f_{1p}^a \bm{v_d}\cdot \nabla \psi \right\rangle, \numberthis \\ \label{L2i}
&L_{2p}^a=-\left\langle\int dx^3 \pi^{-\frac{3}{2}} f_{1p}^a x^2 \bm{v_d}\cdot \nabla \psi \right\rangle, \numberthis \\ \label{L3i}
&L_{3p}^a=-\left\langle\int dx^3 \pi^{-\frac{3}{2}} \frac{B}{B_0} v_{th}^a f_{1p}^a \xi x \right\rangle. \numberthis
\end{align*}
These coefficients contain the pieces of the distribution function related to the neoclassical radial particle-flux-density, radial energy-flux-density and parallel flow, respectively. We note that these coefficients depend on the flux coordinate $\psi$, the magnetic configuration, the inverse mean free path $\nu_{aa}/v_{th}^a$ and the normalised $E\times B$ velocity, $E_{\psi}/(v_{th}^aB)$, or equivalent parameters like $\nu'$ or collisionality $\nu^*=\nu_{aa} R_0/(\iota v_{th}^a)$, and $E^*_{\psi}$, where $R_0$ is the plasma major radius. 
The monoenergetic approximation further simplifies \crefrange{nonmono}{nonmono3} by assuming the particle velocity to be a parameter. This assumption implies the energy $x$ to be a parameter rather than a variable. With this assumption, the set of equations \crefrange{nonmono}{nonmono3} only needs to be solved for the $p=1$ and $p=3$ components, because the components $p=1$ and $p=2$ carry the same information as they have identical dependencies in $\xi$. Additionally, because this monoenergetic description assumes $\dot{x}=0$, the electric field terms in \cref{x_dot,xi_dot} have to be dropped for consistency, and thus only the "partial trajectories" can be considered in such a description. The monoenergetic transport coefficients can then be defined as
\begin{align*} 
&D_{1p}=-\frac{1}{2}\left\langle\int^{1}_{-1} d\xi f_{1p}^a  e^{x^2} \bm{v_d}\cdot \nabla \psi \right\rangle, \\
&D_{3p}=-\frac{1}{2}\left\langle \int^{1}_{-1} d\xi v_{th}^a\frac{B}{B_0}  f_{1p}^a e^{x^2} \xi x  \right\rangle,
 \hspace{10mm} p=1,3, \label{D3i} \numberthis
\end{align*}
with $D_{22}=D_{12}=D_{21}=D_{11}$, $D_{23}=D_{13}$ and $D_{32}=D_{31}$.
The thermal coefficients can then be evaluated by energy convolution of the monoenergetic coefficients
\begin{equation} \label{Lij}
L_{ij}^a=\frac{2}{\sqrt{\pi}}\int_0^{\infty} dx^2 D_{ij}(x) xe^{-x^2}h_ih_j,
\end{equation}
with $h_1=h_3=1$ and $h_2=x^2$.

The DKES code solves the drift kinetic equation by solving the monoenergetic linearised set of equations using a variational principle \citep{DKES}. Therefore, the set of the two monoenergetic equations in such a code has to be cast in a conservative form. To achieve this, the DKES code makes use of the "partial trajectories" with the extra assumption that the $E\times B$ drift is incompressible such that 
\begin{equation} \label{vE_DKES}
v_{E}\approx\frac{\bm{B}\times \bm{E}}{\langle B^2 \rangle}.
\end{equation}
With this extra approximation, one has the so-called "DKES trajectories". The SFINCS code \citep{SFINCS} uses a different numerical scheme and is able to solve \cref{DK}, or both the 4D and monoenergetic approaches to the set of \crefrange{nonmono}{nonmono3}, using the various options which SFINCS provides for describing the trajectories. 

A moment should be taken here to note the difference between the thermal and monoenergetic transport coefficients, which will become important when understanding choices made in the optimisation method.
Notice that the thermal coefficients are written with a superscript associated with the species $a$ while the monoenergetic ones are not. This is due to the thermal velocity dependence of the $\nu_{aa}/v_{th}^a$ and $E_{\psi}/(B v_{th}^a)$
variables. We thus need to specify the species temperature and density to obtain the thermal transport coefficients. In the monoenergetic case, the variables of interest become $\nu_{ii}(v=v_{th}^i)/v$ (in which the ion-ion collision frequency of ions is chosen as the reference value) and $E_{\psi}/(B v)$, where the $v$ is a parameter and appears only for normalisation purposes. Therefore, the monoenergetic coefficients are species-independent. This makes the 4D approach quite useful when we seek to compute values of the transport coefficients at specific collisionalities and electric field values for different species easily. Yet, the monoenergetic approach is more suitable in cases where we intend to create a full species-independent scan of the transport coefficients as a basis to analyse a specific magnetic configuration. 

The different terms in the drift kinetic equation will lead to different collisionality regimes in which the transport coefficients scale differently with the collisionality. While these regimes have been extensively analysed \citep{Shaing2015}, we point here some physical aspects of the main regimes. At high collisionality, the most important mechanism for radial transport is collisions and the radial transport scales linearly with collisionality. When collisionality is close to unity, a plateau regime may form due to the resonance between the orbit movement along the magnetic field and collisions. In this case, radial transport is independent of collisionality and proportional to the transit frequency $\omega_T=\iota v_{th}^a/R_0$. At lower collisionalities, several regimes may exist. If the $E\times B$ drift is less important than collisions, the latter will limit the radial drift of particles. The stronger the collisions the more the radial excursions of trapped particles are reduced and thus the transport coefficients scale with the inverse of collisionality. This is the $1/\nu^*$ regime that is problematic in stellarators as the heat fluxes increase strongly with temperature in this regime \citep{Beidler2011,Beidler2024}. When the electric field magnitude is strong enough for the $E\times B$ poloidal drift to be more important than collisions, the radial excursion of the particles will be limited by this poloidal drift. Two asymptotic scalings can be identified in such case, a scaling with $\sqrt{\nu^*}$ and a linear scaling with $\nu^*$ depending on which de-trapping mechanism is more relevant, collisions or drift-associated collisionless trapping de-trapping processes. In both of these regimes, the radial excursions of trapped particles are also reduced with increasing $E\times B$ drift frequency. These two asymptotic regimes may not be well separated and in such cases, other intermediate scalings may coexist. The fact that different species may be at different regimes in low collisionalities is the reason why the ambipolarity constraint can have multiple solutions.

Finally, we discuss the collision operator $C(f_1)$. There are three widely used forms of the linearised collision operator considered in drift kinetic calculations. The first is the full Fokker-Plank collision operator \citep{SFINCS}, which considers both energy and momentum collisional transfer between the first-order distribution and the zeroth-order Maxwellian distribution, as well as the respective correction terms that ensure energy and momentum conservation. If the transfer of energy is neglected, one has the second form which considers only a Lorentz scattering operator and a momentum correction term. The third form drops the momentum correction term and considers only a Lorentz scattering operator. In this work, we will consider the third and simplest form of the collision operator, which can be written as:
\begin{equation} \label{Lorentz}
C(f_1)=\frac{\nu_D(x)}{2\nu_{aa}}\frac{\partial}{\partial \xi}\left[(1-\xi^2)\frac{\partial}{\partial \xi}\right]
\end{equation} 
where $\nu_D(x)=\nu_{aa}\left(\text{erf}(x)-\frac{\text{erf}(x)}{2x^2}+\frac{2}{2x}(d [\text{erf}(x)]/dx)\right)$.

We can now determine the ambipolarity constraint for a simple deuterium (D) and tritium (T) plasma. In this case, quasi-neutrality implies that $n_e=n_i=n_D+n_T$. In general, the tritium concentration will be a fraction of deuterium, $n_T=\gamma n_D$, with a usual choice of equal concentrations $\gamma=1$. Assuming ion species of equal temperature such  that $T_D=T_T=T_i$ we thus can write the ambipolarity constraint as $\Gamma_e=\Gamma_D+\Gamma_T$ which, as shown in Ref. \citet{Beidler2024}, has the solution
\begin{align} \label{Er_ambi}
\frac{e E_{\psi}}{T_i}=\frac{1}{\mathcal{A}}\left[\left(1-\frac{L_{11}^e}{L_{11}^i}\right)\frac{1}{n_i}\frac{d n_i}{d \psi} +\delta_{12}^i\frac{1}{T_i}\frac{d T_i}{d \psi}-\frac{L_{11}^e}{L_{11}^i}\delta_{12}^e \vphantom{\frac{1}{T_e}\frac{d T_e}{d \psi}}\right], \numberthis
\end{align} 
with
\begin{equation} \label{ambi_den}
\mathcal{A}=1+\frac{L_{11}^e}{L_{11}^i}\frac{T_i}{T_e},
\end{equation}
\begin{equation} \label{L11av}
L_{11}^i=\frac{L_{11}^D+\gamma L_{11}^T}{1+\gamma},
\end{equation}
\begin{equation} \label{delta12e}
\delta_{12}^a=\frac{L_{12}^a}{L_{11}^a},
\end{equation}
and
\begin{equation} \label{delta12I}
\delta_{12}^i=\frac{L_{12}^D\delta_{12}^D +\gamma L_{11}^T\delta_{12}^T }{L_{11}^D +\gamma L_{11}^T}.
\end{equation}
As stated in Ref. \citet{Beidler2024}, for relevant fusion plasmas, it is usually verified that $\delta_{12}^a > 0$, $dn_a/d\psi n_a^{-1}<0$ and $dT_a/d\psi T_a^{-1}<0$. Thus, a condition that maximises the parameter space of positive electric field solutions (electron root) is to maximise the ratio $L_{11}^e/L_{11}^i$. In the following, we consider the ambipolar electric field to be set by \cref{Er_ambi} and, therefore seek stellarator solutions that maximise this ratio. 

\section{Optimisation Method \textbf}
Our main goal is to obtain a magnetic field equilibrium that allows for an electron root solution at a large space of plasma parameters. For this purpose, we use the MHD VMEC equilibrium code \citep{Hirshman1983}, together with the optimisation framework SIMSOPT \citep{Landreman2021b}. The VMEC code uses the set of flux coordinates $(s,\theta_V,\phi_V)$ where $s=\psi/\psi_b$ is the flux label with $\psi_b$ the toroidal flux at the plasma boundary, $\phi_V$ is the toroidal angle in cylindrical coordinates and $\theta_V$ is a poloidal angle. VMEC discretises the flux surface label $s$ with a uniform radial grid and treats the $(\theta_V,\phi_V)$ in the Fourier space. A flux surface $S$ is then parameterised using the usual cylindrical coordinate system as $S=[R(\theta_V,\phi_V)\cos(\phi_V),R(\theta_V,\phi_V)\sin(\phi_V),Z(\theta_V,\phi_V)]$ with 
\begin{align*}
&R(\theta_V,\phi_V)=\sum_{m=0}^{M_\text{pol}}\sum_{n=-N_{\text{tor}}}^{N_{\text{tor}}}\text{RBC}_{m,n}\cos(m\theta_V-n\phi_V),  \label{Rmn} \numberthis \\ 
&Z(\theta_V,\phi_V)=\sum_{m=0}^{M_\text{pol}}\sum_{n=-N_{\text{tor}}}^{N_{\text{tor}}}\text{ZBS}_{m,n}\sin(m\theta_V-n\phi_V). \label{Zmn} \numberthis
\end{align*}
The integers $M_\text{pol}$ and $N_\text{tor}$ refer to the number of poloidal and toroidal mode numbers used to describe the Fourier decomposition, respectively. The decomposition in \cref{Rmn,Zmn} set the $\sin$ terms of $R$ and the $\cos$ terms of $Z$ to zero to ensure stellarator-symmetry \citep{Dewar1998}. Such a symmetry simplifies the problem by reducing the degrees of freedom, while still achieving good optimised solutions. In this work, we use the fixed boundary mode of VMEC where, given the boundary of the plasma, the plasma pressure $P(\psi)$ and net toroidal current $I(\psi)$ profiles, VMEC will converge to a solution in which at each flux label $s$ the MHD force balance equation is satisfied. In stellarator optimisation, we are interested in finding solutions with net zero toroidal current and so we set $I(\psi)=0$. Optimisations are often done for the vacuum solution of the MHD force balance in which $P=0$. Solutions which include the correction effects of the plasma pressure are called finite-$\beta$ solutions with $\beta=P/(2\mu_0 B^2)$ the ratio between the plasma and magnetic energies. The magnetic configuration in VMEC coordinates is transformed to Boozer coordinates using the BOOZ\_XFORM code \citep{Sanchez2000a}. 

To target a precise QI magnetic geometry, we use the cost function 
\begin{align} \label{costQI}
J_\text{QI}=\frac{n_\text{fp}}{4\pi^2 (B_\text{max}-B_\text{min})^2}\int^{2\pi}_0 d\alpha \int^{2\pi n_\text{fp}}_0 d\phi\left[B( s,\alpha,\phi)-B_\text{QI}( s,\alpha,\phi)\right]^2 
\end{align}
proposed in Ref. \citet{Goodman2023}. In this function, $n_\text{fp}$ is the number of field periods and $B_\text{QI}$ is the target magnetic field constructed from $B(s,\alpha,\phi)$ in such a way that its magnetic wells satisfy \cref{QI_omni1,QI_omni2}. 
We follow Ref. \citet{Goodman2023} and avoid large stresses in the coils by constraining the overall mirror of the magnetic field
\begin{equation} \label{mirror}
\Delta_M=\frac{B_\text{max}-B_\text{min}}{B_\text{max}+B_\text{min}},
\end{equation}
to be below a certain threshold.
To avoid small minor radius solutions to be favoured during optimisation, we constrain the aspect ratio $A$ of the configuration to be close to the aspect ratio of the initial configuration.
We add the additional constraint to the mean of the rotational transform $\bar{\iota}$ in order to avoid low-order rational surfaces. Thus, the total cost function that directly targets the magnetic geometry properties can be written as 
\begin{align} \label{cost1}
    J_1 = w_\text{QI}J_\text{QI}+w_{\iota}(\bar{\iota}-\bar{\iota}_\text{target})^2
    &+w_{A}(A-A_\text{target})^2+w_{M}(\Delta_M-\Delta_{M_\text{target}})^2,
\end{align}
where $w_\text{QI}$, $w_{\iota}$, $w_{A}$ and $w_{M}$ are the weights that multiply the different cost functions and can be adjusted in order to give more focus to one or more of the targets during optimisation.

As seen in the previous section, maximising the plasma parameter space of the electron root solution can be achieved by selecting a magnetic configuration that maximises the ratio of the radial diffusion coefficients $L_{11}^e/L_{11}^i$. Thus, the choice here for a cost function to target magnetic configurations prone to have an electron root is
\begin{equation} \label{costEr}
J_{E_{\psi}} =  \left(\frac{L_{11}^i}{L_{11}^e} \right)^2,
\end{equation}
where we emphasise again the dependence of these coefficients in the magnetic configuration and the parameters $\nu'$ and $E_{\psi}^*$.
There are several codes which are suitable to calculate the transport coefficients $L_{11}^i$ and $L_{11}^e$. Two of these codes have been used consistently for stellarator transport calculations: DKES \citep{DKES} and SFINCS \citep{SFINCS}. We choose to use SFINCS for the calculation of our electron root cost function $J_{E_{\psi}}$. This is because DKES is only able to calculate the transport matrix in the monoenergetic approximation, while SFINCS is capable of solving the linearised drift kinetic equation in both monoenergetic and 4D approaches \citep{SFINCS}.
We preferably want to use the 4D approach to calculate the cost function $J_{E_{\psi}}$. To understand this choice, it is instructive to look again into the difference between the monoenergetic and thermal transport coefficients pointed out in the last section. Equation \ref{costEr} involves calculating at each optimisation loop the thermal radial transport coefficients at least for one ion species and electrons at one radial position, one value of collisionality and one value of the electric field. Using the 4D approach, to meet such requirements, one needs to undergo only two drift kinetic solver iterations per loop (one per species). However, in the monoenergetic approach, we need more iterations. As seen in the previous section, the monoenergetic coefficients are species-independent through the dependence on $\nu/v$ instead of $\nu/v_{th}^a$. Thus, to obtain the thermal transport coefficients $L_{ij}^a$ from the monoenergetic coefficients, one needs to perform an energy convolution of the monoenergetic coefficients for which we need to obtain $D_{ij}(v)$ for different values of $\nu/v$. We could simplify such a method by performing an interpolation for each species. Because in a reactor electrons are usually close to the $1/\nu^*$ transport regime, the behaviour of the $D_{11}$ coefficients for the electrons can be approximated by a linear interpolation and thus at least two values of $D_{11}$ are necessary to be computed at $E^*_{\psi}=0$. For the ion species, however, these can be in different collisionality regimes such as the $1/\nu^*$, $\nu^*$ or $\sqrt{\nu^*}$. Therefore, at least around five iterations of the solver would be needed to use interpolation methods. If we would choose the monoenergetic approximation to calculate the cost function, at each optimisation loop, a minimum of seven drift kinetic solver iterations would be needed, together with the interpolation and convolution steps. It is therefore more cost-effective to use the 4D approach in the optimisation loop. This leads to the choice of the SFINCS code to calculate $J_{E_{\psi}}$ during optimisation. The thermal transport coefficients $L^a_{{11}_S}$ outputted by SFINCS are defined to be adimensional and thus they are slightly different from the definitions already presented. In terms of these SFINCS thermal transport coefficients, we may write the cost function in \cref{costEr} as
\begin{align} \label{costEr2}
J_{E_{\psi}} = \left(\frac{L^i_{{11}_S}}{L^e_{{11}_S} T_e^{-2}v_{th}^e} \right)^2= \left(\frac{L^D_{{11}_S}T_D^{-2}v_{th}^D+\gamma L^T_{{11}_S}T_T^{-2}v_{th}^T }{(1+\gamma)L^e_{{11}_S}T_e^{-2}v_{th}^e} \right)^2 \numberthis.
\end{align}      
Furthermore, even if we are aiming for a reactor scenario we may consider only deuterium as the ion species in calculating the electron root cost function. Note that the tritium species has a slightly smaller thermal velocity and thus its normalised electric field is larger than that of deuterium, for the same electric field value. Therefore, the radial transport coefficient of the tritium species will always be equal to or slightly smaller than its deuterium counterpart. This makes the weighted average ion species radial transport coefficient, \cref{L11av}, equal or smaller when considering both species than when considering deuterium alone. Therefore, for minimising the ratio of ion to electron radial transport coefficients in a reactor, the worst case scenario occurs when the tritium species is absent and we may substitute the ion-weighted average radial thermal transport coefficient in \cref{L11av} by the radial thermal transport coefficient of deuterium in the cost function ($\gamma=0$), and skip one drift kinetic solver iteration. The total cost function for electron root optimisation can then be defined as 
\begin{equation} \label{cost_neo}
J_{2} =\sum_{j=1}^3 w_{{E_{\psi}}_j}\left[ \frac{L^i_{{11}_S} (s=s_j)}{L^e_{{11}_S} (s=s_j)}\frac{T_e^2 v_{th}^i}{T_i^2 v_{th}^e}\right]^2,
\end{equation}  
in which we consider deuterium $D$ as the only ion species $i$ and we choose to optimise for at least three flux surfaces, in order to better target the electron root over a wider extent of the plasma radius. More surfaces could of course be targeted, but three surfaces were found to be a reasonable choice as we will see in the next sections. 

Collisionality and electric field values are necessary as input for SFINCS to solve the drift kinetic equation. The collisionality values are defined by the temperature and density profiles of the deuterium and electron species. In fusion reactors, one would envisage to have an equithermal plasma with $T_e=T_i$. However, a clamp of the ion temperature is often observed in experiments \citep{Beurskens2021} so that $T_i<T_e$, but such behaviour is not expected in a reactor plasma. Regardless, during the optimisation, we consider the worst case scenario to achieve electron root, which occurs when $T_e=T_i$ \citep{Beidler2024} and thus we consider equal temperature profiles for ions and electrons. We also consider quasi-neutrality to be satisfied and thus $n_e=n_i$. To follow previous work on neoclassical transport optimisation \citep{LandremanBootstrap} we use the following profiles for temperature
\begin{align} \label{profilesT}
T=(T_0-T_b)(1-s)+T_b, 
\end{align}  
and density
\begin{align} \label{profilesN}
n=(n_0-n_b)(1-s^5)+n_b.
\end{align} 
Here, $T_0$ and $n_0$ are the temperature and density on-axis. $T_b$ and $n_b$  are the temperature and density at the boundary, which allow the collisionality parameter $\nu^*$ to remain finite at this position. We use an on-axis temperature $T_0=17.8\,\text{keV}$ and density $n_0=4.21\times 10^{20}\,\text{m}^{-3}$ yielding a volume averaged plasma $\beta$ of $4-5\%$ for a magnetic field configuration that does not deviate much from $B=8\,\text{T}$ and $a=1.2\,\text{m}$, which are the reactor-like targets used for scaling the initial configuration used for the optimisations in this work. At the boundary, we choose the temperature $T_b=0.7\,\text{keV}$ and the density $n_b=0.6\times 10^{20}\,\text{m}^{-3}$ which are consistent with the electron root solution in Ref. \citet{Beidler2024}.  
While different methods could have been chosen to obtain the boundary values, different choices of these values are not seen to have a significant impact on the electron root solution optimisation which mostly takes place in the core region. We only introduce these non-finite values so that at the testing step, we may use the same profiles as during the optimisation loop, as will be shown in the next section.  The electric field profile is unknown at the optimisation stage and thus we have to make a guess to this input. 
With regards to the choice of electric field values, we want to optimise for a finite electric field value that ideally would be as large as possible in order to obtain a large electron root. In this work, we use the value of electric field values for the three radial coordinates considered such that $eE_{r}/T=1$, which yielded favourable results for every optimisation considered.

Finally, we discuss the three remaining sets of inputs for SFINCS. The first is which trajectory set to choose when using the 4D approach. This approach allows the usage of any of the three types of trajectories discussed in the previous section. However, as we show in the next section, the testing of the optimisation is simplified by making use of the monoenergetic radial transport coefficients. The only set of trajectories compatible with the monoenergetic approach are the "DKES trajectories" as discussed in the previous section. Thus, to be consistent between the optimisation and testing steps, we choose the "DKES trajectories" during optimisation even if the 4D approach is used. The same argument is used to discuss the second set of inputs which concerns the type of collision operator to use. A Lorentz collision operator is used for consistency with the testing method. Nevertheless, the absence of the momentum correction term and energy terms in the collision operator are not expected to significantly change the results as these terms only become important at higher collisionalities and we expect our reactor plasma species to be at low collisionality values. The third set of inputs concerns the resolution used during the optimisation. The convergence of drift kinetic solvers such as SFINCS and DKES is extremely dependent on the magnetic configuration. During the optimisation, the magnetic configuration is always changing, making it difficult to set a fixed resolution for convergence.  In this work, we opted to use a resolution of $N_{\theta}=25$, $N_{\phi}=31$, $N_{\xi}=34$ and $N_x=4$. These are not enough to ensure complete convergence of SFINCS for every configuration. Yet, complete convergence of the radial coefficients is not strictly necessary for the electron root optimisation as we are interested in the minimisation of the ratio of the coefficients and we are not looking for an exact match of the value of radial particle flux of each species. Thus, the values of resolution are chosen as the minimum required resolution for a usual W7-X-like configuration in such a way that doubling the resolution for each coordinate will lead to less than a $6\%$ variation in the transport coefficients. As this is configuration-dependent, we leave here some general guidelines about the resolution parameters. At low collisionality, we expect $N_x=4$ and $N_{\theta}=25$ to be a generally good value for convergence in SFINCS. At the same time, the number of points in the $\phi$ direction should usually be equal or larger than in the $\theta$ direction and the resolution in the pitch angle variable $\xi$ should be equal or larger than the resolution in $\phi$.

The optimisations presented in this work are optimised using the total cost function $J=J_{1}+J_{2}$ subject to the Levenberg-Marquardt optimization algorithm. We use boundary Fourier coefficients RBC and ZBS as independent variables with $M_{pol}=N_{tor}=2$.  We use 24 flux surfaces for VMEC resolution during optimisation, unless stated otherwise. The QI cost function is targeted at $4$ flux surfaces, namely $s=1/24$, $3/24$, $9/24$ and $13/24$. For the calculation of this cost function, $141$ points are measured along each well, with $27$ wells and $51$ bounce points per well. For the Boozer coordinate transformation, $18$ poloidal and $18$ toroidal mode numbers are used. A maximum mirror ratio of $0.19$ is targeted and the target aspect ratio is $10$. To avoid low-order rational surfaces we target a mean iota of $\bar{\iota}=0.61$. The electron root optimisation is targeted at the radial positions $\rho=\sqrt{s}=r/a=0.2$, $0.29$ and $0.35$ with $a$ the minor radius as defined by VMEC. Unity weights are used for all cost functions except when targeting the mirror ratio in which a weight of $w_{M}=100$ is used. The optimisation method is carried out with SFINCS integrated into the SIMSOPT framework. All the optimisation scripts used in this work and the configurations obtained can be found in Ref. \citet{ErootGit}.

\section{Validation Method \textbf}
In order to measure the efficiency of the electron root cost function, it is necessary to obtain the electric field profile for the configurations obtained from the optimisation and check for the presence of an electron root. The 4D approach in SFINCS is able to reconstruct the ion and electron fluxes as a function of the electric field at different radial positions and assess the presence of the solutions of the ambipolarity constraint, i.e., the electric field values at which the ion and electron particle-flux-density curves cross. However, such a method is tedious to apply for several radial positions, is local, and does not provide any information about the transition zones between negative electric field values (ion root) and positive electric field values (electron root). Instead, we solve the electric field transport equation, \cref{Er_transp}, which holds information about ambipolarity and the "diffusion" of the electric field. This becomes important near root transitions. A code that solves this equation is NTSS \citep{Turkin2011}. 

The NTSS code is capable of solving the 1D radial transport equations, \crefrange{energy}{continuity}, for deuterium, tritium and electrons, and the electric field transport equation, \cref{Er_transp}. Helium ash can also be considered by solving the continuity equation, \cref{continuity}, for helium. The NTSS code requires as inputs the initial temperature and density of the species of interest, the magnetic configuration obtained from VMEC and converted in Boozer coordinates with BOOZ\_XFORM, and a database of monoenergetic radial transport coefficients at different radial positions. As a remark, we note that if the magnetic configuration is fixed, a database of the monoenergetic coefficients, which are species-independent (and therefore independent of the temperature and density profiles), is more relevant for the calculation of the particle-flux-densities and energy-flux-densities as these can be calculated solely by interpolating the monoenergetic coefficients database for different species, collisionality values, electric field values and radial positions. Such an interpolation is performed in NTSS in order to be able to solve the transport equations, in which the temperature and density can evolve over time. Such time evolution of the temperature and density would render the use of the radial thermal transport coefficients much less efficient than in the case of the optimisation loop. Historically, NTSS obtains such a database from DKES. Therefore, all the input files make use of DKES conventions. Since our optimisation is performed with the SFINCS code, to be consistent, we would preferably use SFINCS as a generator of the monoenergetic transport coefficients database. To replicate the DKES-like database with SFINCS, we use the typical range of values of the normalised $E\times B$ velocity $E_r/(vB)$ and inverse mean free path $\nu/v$ inputs used for generating a DKES-like NTSS database. Such values are chosen to be the ones used to obtain the DKES database for the magnetic configuration Hydra-Np04-20190108 in Ref. \citet{Beidler2024}, which contains an electron root. We convert these values to the corresponding values of the SFINCS inputs $\nu'$ and $E_{\psi}^*$ for the magnetic configuration of interest. We then perform a convergence study in order to find a reasonable resolution to describe the radial transport monoenergetic coefficients for the magnetic configuration of interest at the different electric field, collisionality and radial points. This is achieved by varying each resolution parameter by up to $150\%$ of the initial values and ensuring that the difference between the initial and final radial transport coefficient values varies $3\%$ or less. This convergence study is done for $E_{\psi}=0$ at the radial position which is nearest to the magnetic axis ($r/a=0.12247$), and for the lowest collisionality value of the database. This point in the database space is chosen because it is the point which is expected to need a larger resolution, and thus, it is the best point to ensure we are close to convergence at all points in the database. In practice, it is possible that this level of convergence will not be met at all points. However, doing such a convergence study for every database point is expensive and impractical. Therefore, after setting the resolution through this process and obtaining the database scan with that resolution, we check for any point that does not follow the expected trend in the scan and increase the resolution accordingly. Note that we are not interested in the convergence of $D_{31}$, $D_{13}$ and $D_{33}$ as they are not relevant for calculating the radial particle and energy-flux-densities of interest for this work if we neglect the collisional momentum correction which is expected to be small at the collisionalities of interest. As such, the resolution is based on the convergence of only the radial transport coefficient $D_{11}$ to reduce unnecessary computation time. After obtaining the monoenergetic radial coefficient database with SFINCS, we convert this database into a DKES-like database file which can be read as an input by NTSS. The typical DKES database file used in Ref. \citet{Beidler2024} uses $7$ radial positions at $r/a=[0.12247,0.25,0.375,0.5,0.625,0.75,0.875]$, and we follow the same pattern in this work. Moreover, such a database calculation is applied for the optimised configurations with a radial VMEC resolution of $201$ which is larger than the resolution of $24$ points used during optimisation.

\begin{figure}
     \includegraphics[width=.5\textwidth]{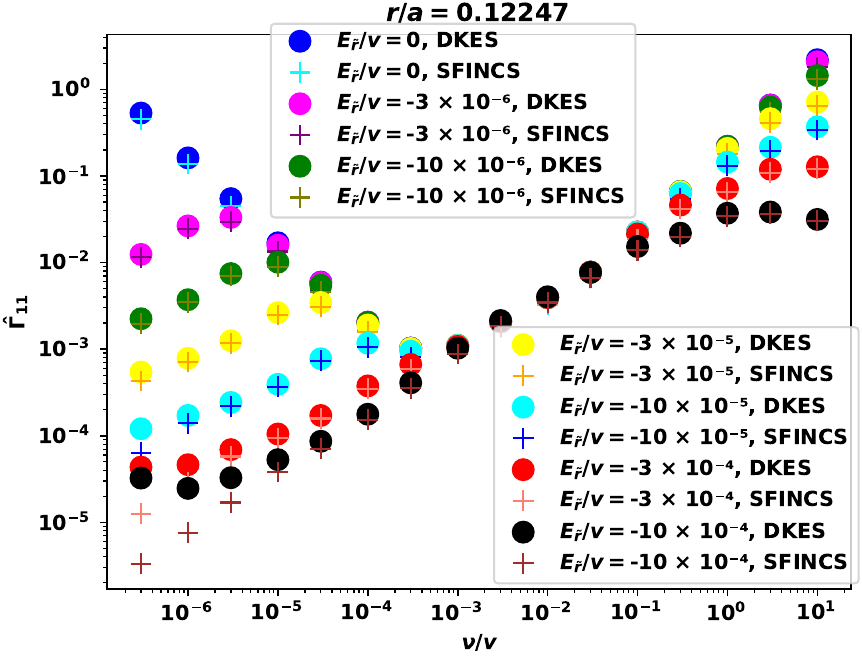}
    \includegraphics[width=.5\textwidth]{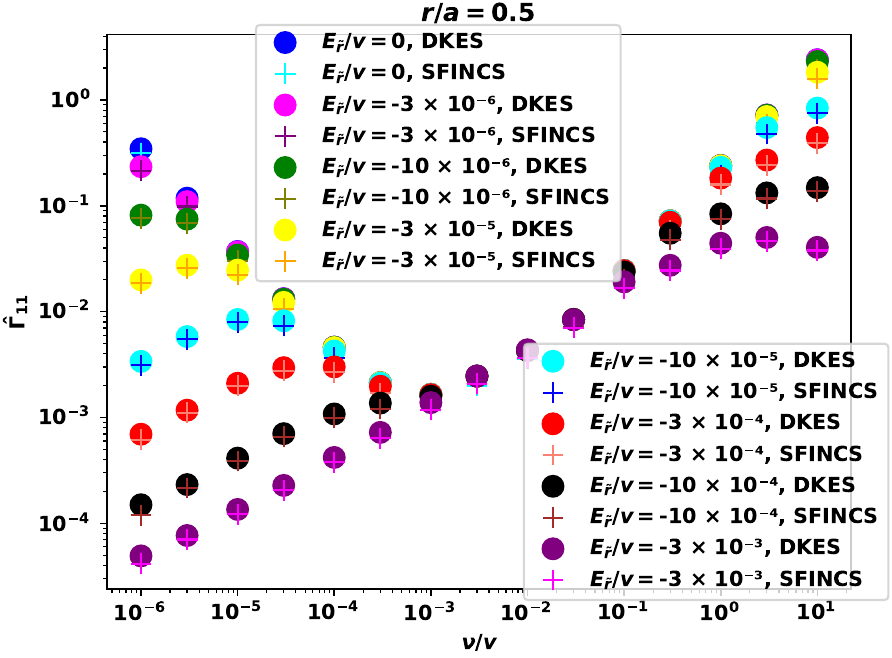}
   \caption{Comparison between SFINCS (crosses) and DKES (circles) monoenergetic radial transport coefficients (normalised to DKES conventions) at $r/a=0.12247$ and $r/a=0.5$ for the Hydra-Np04-20190108 configuration presented in Ref. \protect\citet{Beidler2024}. Values are shown for the electric fields away from resonant values. A SFINCS resolution of $N_{\theta}=39$, $N_{\phi}=59$ and $N_{\xi}=116$ is used.}
    \label{fig:bench1}
\end{figure}

\begin{figure}
     \includegraphics[width=.5\textwidth]{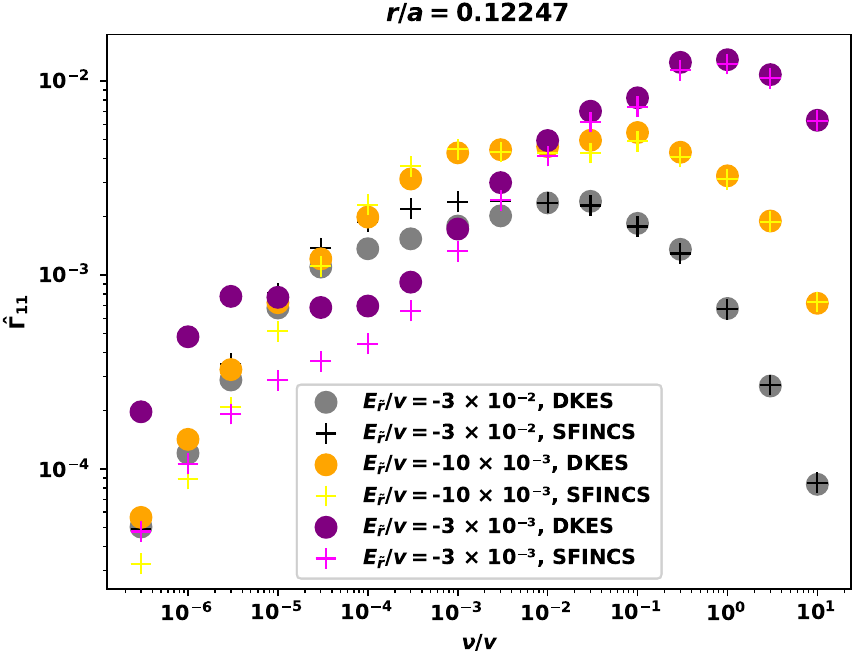}
    \includegraphics[width=.5\textwidth]{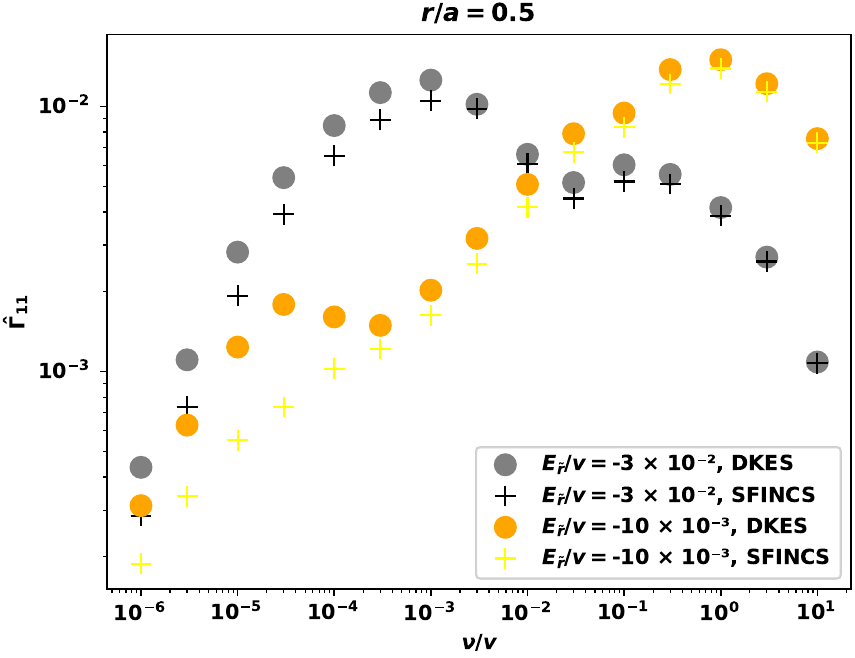}
   \caption{Comparison between SFINCS (crosses) and DKES (circles) monoenergetic radial transport coefficients at $r/a=0.12247$ and $r/a=0.5$ for the Hydra-Np04-20190108 configuration presented in Ref. \protect\citet{Beidler2024}. Comparison is shown for large electric fields near resonant values. A SFINCS resolution of $N_{\theta}=39$, $N_{\phi}=59$ and $N_{\xi}=116$ is used.}
    \label{fig:bench2}
\end{figure}

The use of SFINCS to obtain a DKES-like database requires two conversions of quantities due to the different normalisations used in the two codes. Such normalisations are detailed in \cref{appendix}. To confirm that these conversions are being performed correctly, a benchmark against the DKES results for the electron root configuration Hydra-Np04-20190108 shown in Ref. \citet{Beidler2024} is provided here. Figure \ref{fig:bench1} shows the comparison of the radial transport monoenergetic coefficients at two different radial positions between SFINCS and DKES. These results were obtained starting from the same DKES-like database set of inputs $\nu/v$ and $E_r/(vB)$. Note that due to historical reasons the $E_{r}/(vB)$ parameter in the DKES-like database files is written using the electric field parameter $E_{\tilde{r}}$ in which $\tilde{r}$ is defined in \cref{appendix}. The quantity $\hat{\Gamma}_{11}$ is the monoenergetic radial transport coefficient normalised as per the conventions of an NTSS input (see \cref{appendix}). The resolution used in these scans was obtained from the convergence test as described in the previous section and is $N_{\theta}=39$, $N_{\phi}=59$ and $N_{\xi}=116$. We can see that the overall values of $\hat{\Gamma}_{11}$ match quite well between the two codes at the two radial positions. Especially, if we account that these results are plotted on a logarithmic scale and thus span several orders of magnitude. At low collisionality, for the larger values of the normalised electric field at $r/a=0.12247$ in \cref{fig:bench1}, there is a mismatch between DKES and SFINCS. However, SFINCS produces data points closer to the expected $\sqrt{\nu^*}-\nu^*$ behaviours than DKES.  In \cref{fig:bench2} a similar comparison is performed for the larger electric field values. These values of the electric field are the values in which a resonant behaviour is observed. This can be observed, for example, by a sudden change of the radial transport coefficient when compared with the curves for the other values of the electric field (especially at middle values of collisionalities in which the electric field has, for the smaller non-resonant values, no effect on the radial coefficient). This resonance is due to the $E\times B$ poloidal drift being large enough to compete with the term which describes motion along the field line. We can see that in the case of these electric fields (see \cref{fig:bench2}), there is a large mismatch. We expect the two codes not to agree in this region. The most important physics near this resonance is described by the vanishing of the terms in the drift-kinetic equation, and thus the numerical result depends on the exact calculation of numerical values very close to zero. Since the two codes use distinct numerical methods, it is expected that they will present a different behaviour in this region.  

\begin{figure}
     \includegraphics[width=.5\textwidth]{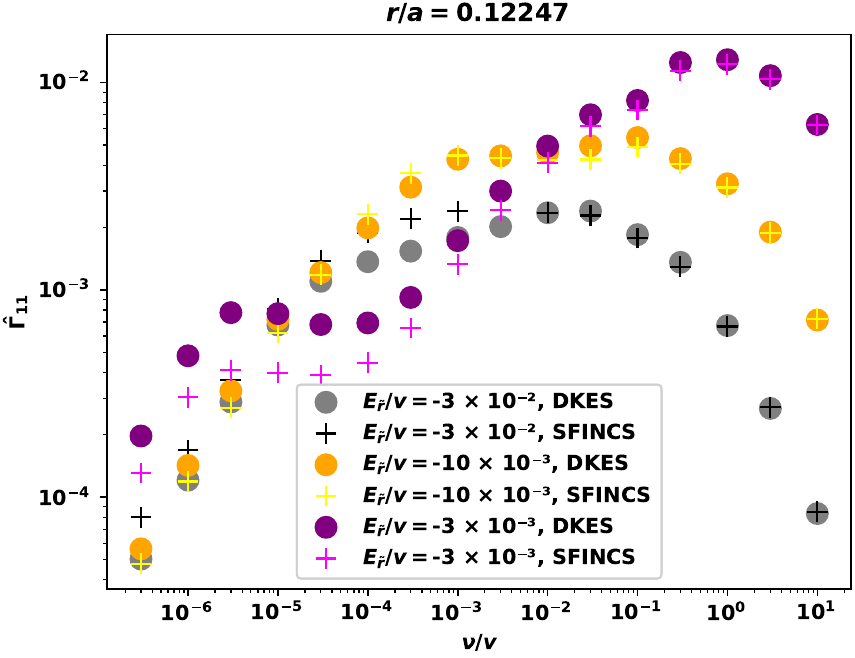}
    \includegraphics[width=.5\textwidth]{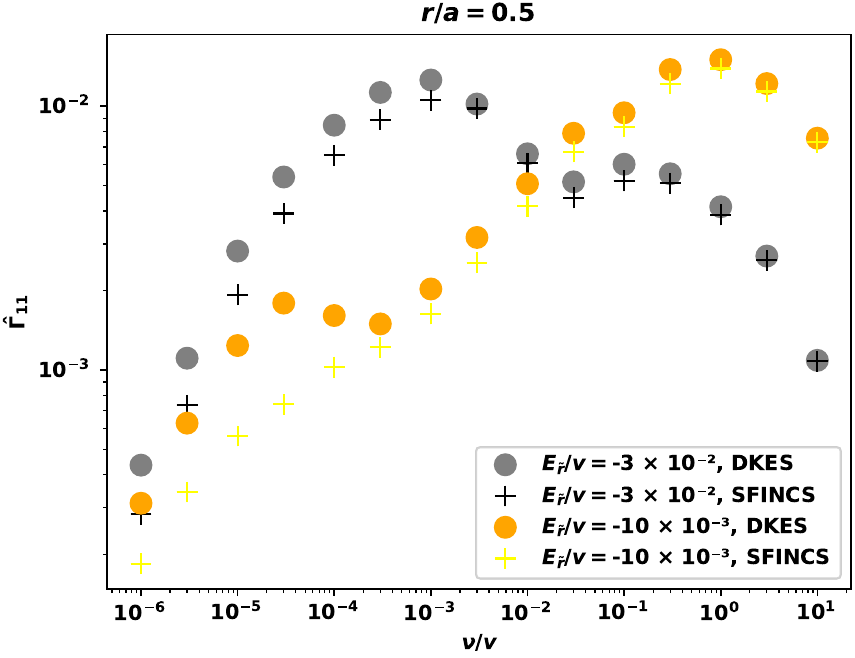}
   \caption{Comparison between SFINCS (crosses) and DKES (circles) monoenergetic radial transport coefficients at $r/a=0.12247$ and $r/a=0.5$ for the Hydra-Np04-20190108 configuration presented in Ref. \protect\citet{Beidler2024}. Comparison is shown for large electric field values near resonance. A SFINCS resolution of $N_{\theta}=39$, $N_{\phi}=431$ and $N_{\xi}=351$ is used.}
    \label{fig:bench3}
\end{figure}

Still, to better understand if this mismatch is inherent to the differences between the codes and not because we are using a resolution based on the lowest collisionality point with a zero electric field, we redo the convergence test using a point at the lowest collisionality value, but at a large normalised electric field value near the resonance for $r/a=0.12247$. Indeed, the convergence study at this point indicates a need for a larger resolution of $N_{\theta}=39$, $N_{\phi}=431$, $N_{\xi}=350$. Using this larger resolution, we redo the transport coefficients scans for these large electric fields presented in \cref{fig:bench2}. The results are again compared to the DKES results in \cref{fig:bench3}.  We can see that for $r/a=0.12247$, the increase in resolution provides a small improvement but the two codes still present a mismatch, which agrees with the expectation that the behaviour of the resonance will be captured differently by the two codes. At $r/a=0.5$, we can see from comparing \crefrange{fig:bench2}{fig:bench3} that the increase of resolution has almost no effect and the mismatch between the codes remains. Nevertheless, these large resonant electric fields are only expected to be important for large mass impurity species and thus should not affect the analysis of our optimised configurations in this work. In fact, for a deuterium ion at $T=17.8\,\text{keV}$ the electric field would need to be of the order of $E_r\sim 100\,\text{kV/m}$ to experience the resonant behaviour seen in \crefrange{fig:bench2}{fig:bench3}. 

\begin{figure}
    \centering
     \includegraphics[width=.5\textwidth]{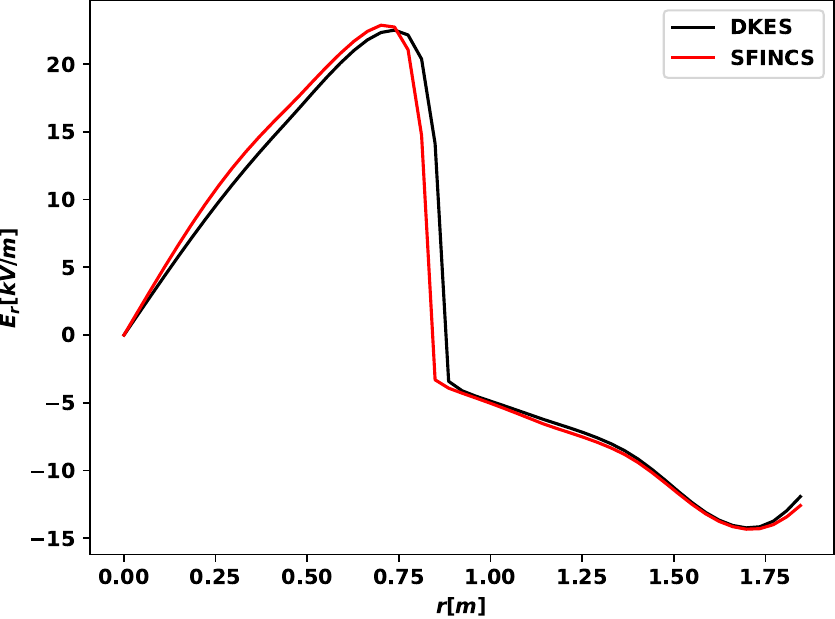}
   \caption{Comparison of the radial electric field solution obtained with NTSS from using as an input the DKES and the SFINCS generated databases of the monoenergetic radial transport coefficients for the resolution $N_{\theta}=39$, $N_{\phi}=59$ and $N_{\xi}=116$.}
    \label{fig:bench_Er}
\end{figure}

As a final check, we run NTSS with the DKES and SFINCS generated databases. We compare the two results of the electric field profile in \cref{fig:bench_Er}, in which the SFINCS database for the resolution $N_{\theta}=39$, $N_{\phi}=59$ and $N_{\xi}=116$ is used. We see that the electron root produced by the SFINCS-generated database is slightly different. This difference is expected since there are small differences between the values of the monoenergetic coefficients as was seen in \cref{fig:bench1}. However, such a difference in both the radial scans and the electron root is small. The DKES electron root has a maximum value of $E_{{r}_\text{DKES}}^{max}=22.521 \,\text{kV/m}$ and in the case of SFINCS, the maximum is $E_{{r}_\text{DKES}}^\text{max}=22.883 \,\text{kV/m}$. The electron root to ion root transition is calculated by taking the mean value of the two radial grid points in which the electric field observes a sign change. The root transition location is with DKES $r_\text{DKES}=0.867\,\text{m}$ and with SFINCS $r_\text{SFINCS}=0.830\,\text{m}$. Thus, the relative error of the maximum electric field calculation is of $1.607\%$ while the error of the root transition position is $-4.25\%$.  The largest error is found to be in the transition zone due to the sensitivity of this region to small variations of the inputs. Nevertheless, the errors are relatively small given that the calculation of such a profile involves using the results of two drift kinetic numerical solvers with different numerical methods and interpolating such results in a three-dimensional parameter space before solving the transport equations. Therefore, despite these small differences between the two codes, we expect the validation method proposed to be appropriate to test the optimisations for the presence of an electron root. 

\section{Results \textbf}
In this section, optimisation studies in order to assess the use of the electron root cost function proposed in \cref{cost_neo} are performed. All the optimisations shown in this section will be done assuming a vacuum solution of the MHD force balance equation except when stated otherwise.
\subsection{The initial configuration}
\begin{figure}
\includegraphics[width=.5\textwidth]{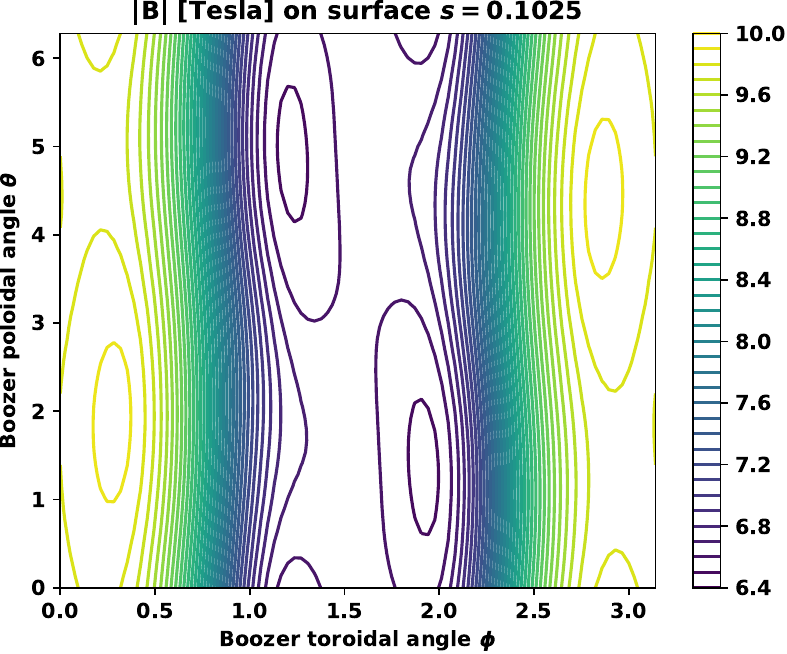}
\includegraphics[width=.5\textwidth]{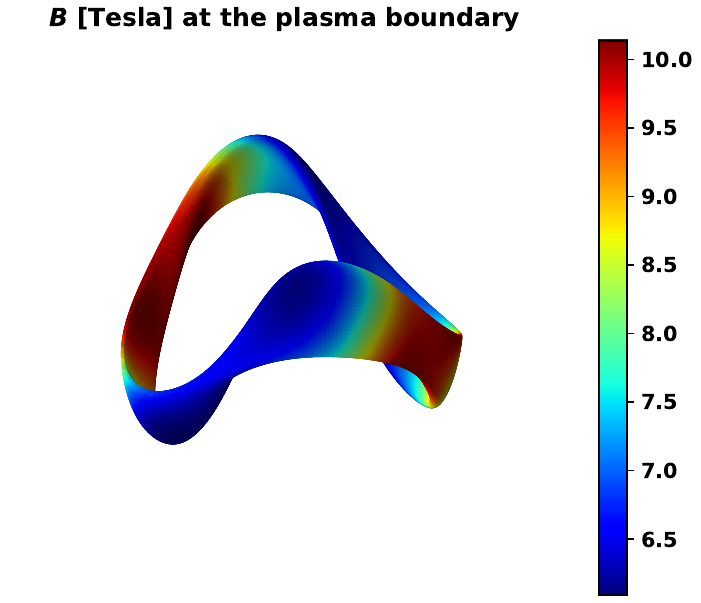}
\caption{Contours of the magnetic field strength of the initial configuration at $s=0.1025$ in Boozer coordinates and the same quantity at the plasma boundary.}
    \label{fig:B_initial}
\end{figure}

The efficiency of the QI cost function, \cref{costQI}, was shown in Ref. \citet{Goodman2023} to be highly dependent on the initial condition used. Therefore, we use one of the QI-optimised equilibria obtained in Ref. \citet{Goodman2023} as a starting point. We thus select the QI-optimised configuration obtained in that work for $n_\text{fp}=2$. From that configuration, we remove the boundary coefficients with $M>2$ and $N>2$ and optimise with maximum mode numbers $M_\text{pol}=2$ and $N_\text{tor}=2$. We also scale the minor radius to $a=1.2\,\text{m}$ and $B=8\,\text{T}$ while keeping the aspect ratio $A_\text{initial}=10$. The magnetic field strength of the resulting configuration is depicted in \cref{fig:B_initial} at different flux surfaces.
\begin{figure}
     \includegraphics[width=.5\textwidth]{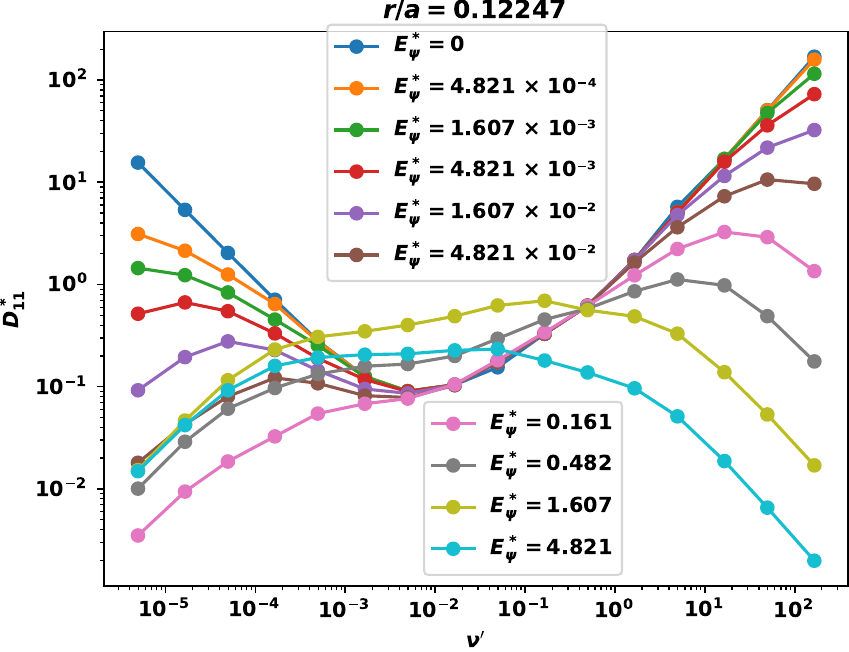}
    \includegraphics[width=.5\textwidth]{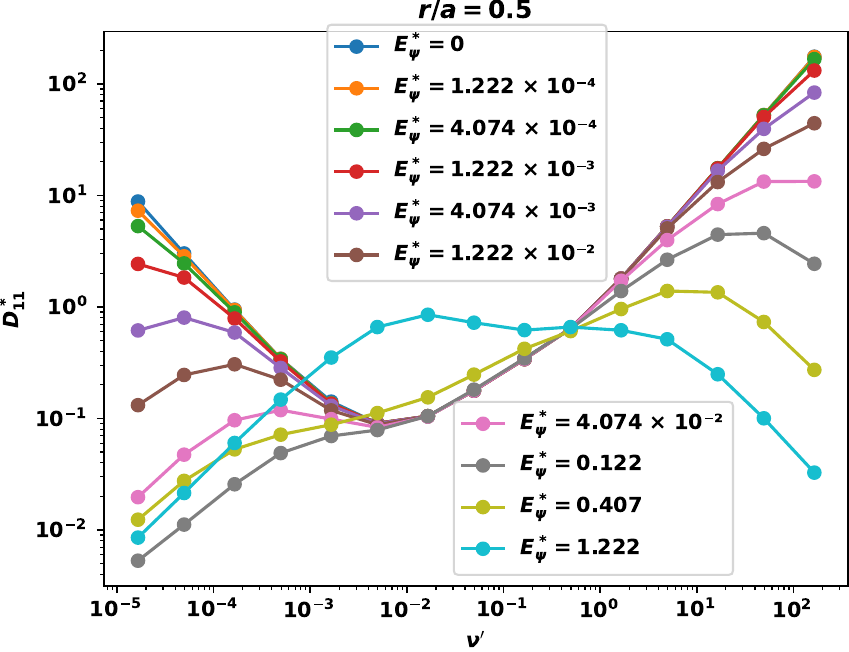}
   \caption{Scans of the monoenergetic radial transport coefficients at $r/a=0.12247$ and $r/a=0.5$ for the initial magnetic configuration.}
    \label{fig:D11_initial}
\end{figure}
The magnetic strength contours in Boozer coordinates at $s=0.1025$ indicate some deviations from precise QI. This is due to the removal of the boundary harmonics with $M>2$ and $N>2$ from the original configuration. We then optimise this initial configuration using the QI cost function alone, the electron root cost function alone, and both cost functions. This allows us to compare the different configurations that result from such optimisations. We note, however, that the mirror ratio, aspect ratio and mean iota targets will be maintained fixed in all of the different optimisations.

\begin{figure}
    \centering
     \includegraphics[width=.5\textwidth]{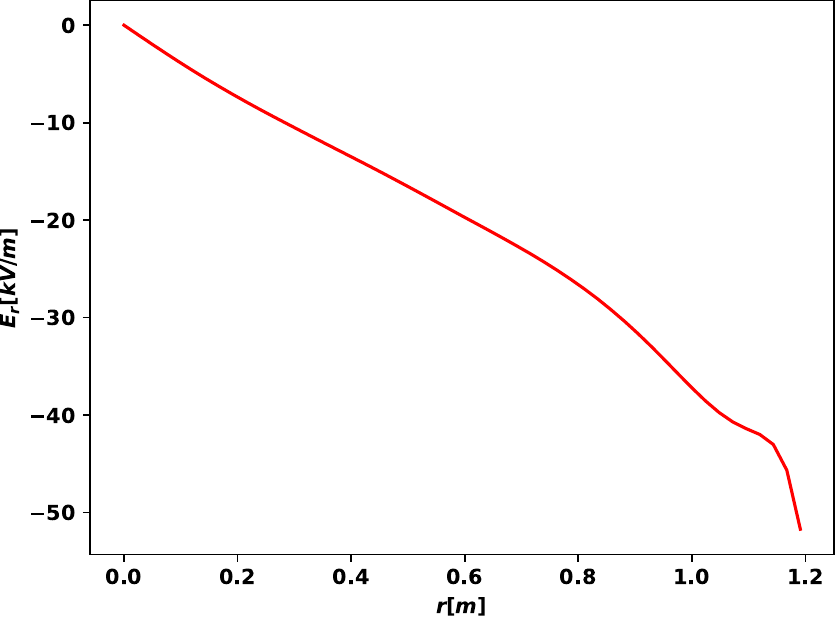}
   \caption{Radial electric field solution at the temperature and density profiles in \cref{profilesT,profilesN}, with $T_0=17.8\,\text{keV}$ and $n_0=4.21\times 10^{20}\, \text{m}^{-3}$, for the initial magnetic configuration. Solution calculated using the NTSS code by solving \cref{Er_transp}.}
    \label{fig:Er_initial}
\end{figure}

Regarding this initial condition, the monoenergetic radial transport coefficients scans at $r/a=0.12247$ and $r/a=0.5$ can be seen in \cref{fig:D11_initial}.   
With the monoenergetic radial transport coefficient database, we can solve the electric field diffusion equation, \cref{Er_transp}, with the NTSS code. We note here that the electric field diffusivity $D_{E_{\psi}}^a$ is an input of the NTSS code. In this work, we follow Ref. \citet{Beidler2024} and use $D_{E_{\psi}}^a=2\,\text{m}^2/s$ in every NTSS simulation. Maintaining the constant density and temperature profiles equal to the ones used during optimisation (see \cref{profilesT,profilesN}, with $T_0=17.8\,\text{keV}$ and $n_0=4.21\times 10^{20}\, \text{m}^{-3}$), and assuming equal concentrations of deuterium and tritium as pertain to reactor conditions, we obtain the electric field profile which can be seen in \cref{fig:Er_initial} where $a=1.191\,\text{m}$. The electric field solution is an ion root for the initial configuration at the temperature and density profiles used during optimisation. 

\subsection{QI optimisation}
\begin{figure}
     \includegraphics[width=.5\textwidth]{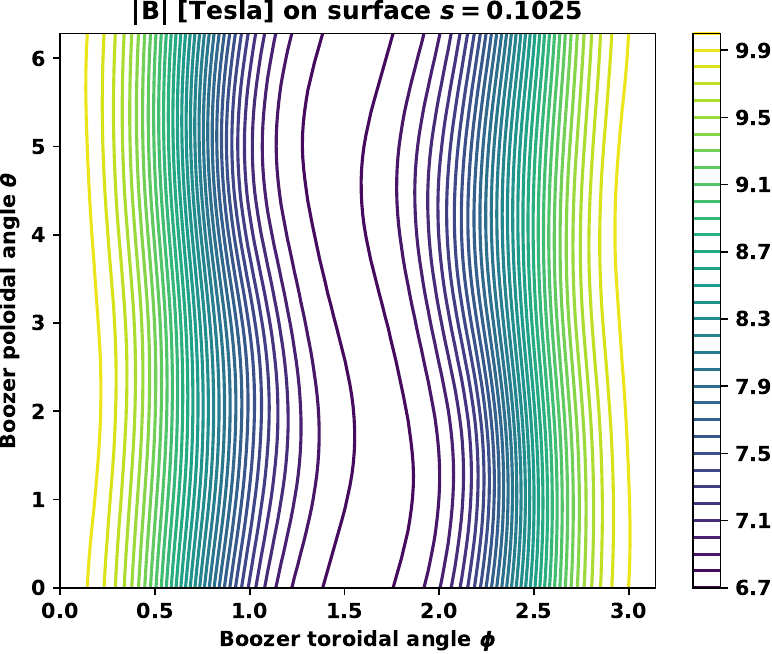}
    \includegraphics[width=.5\textwidth]{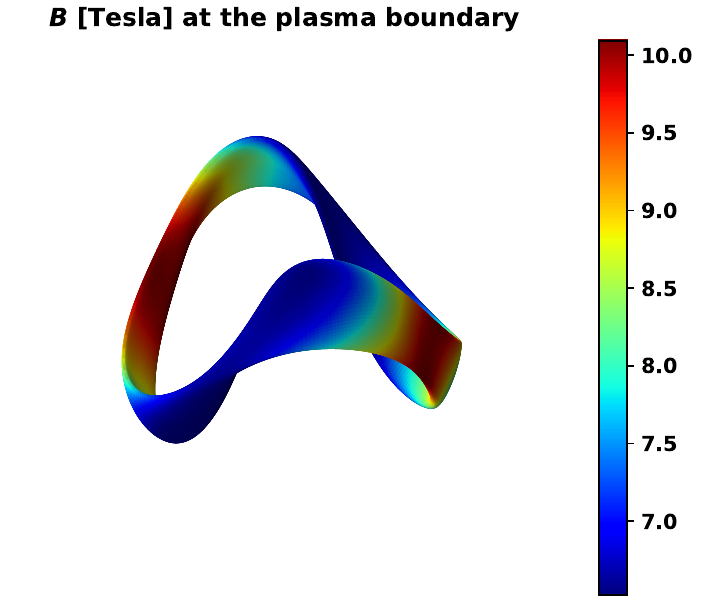}
   \caption{Contours of the magnetic field strength at $s=0.1025$ in Boozer coordinates, for the magnetic configuration optimised only for the QI cost function and the same quantity at the plasma boundary.}
    \label{fig:B_QI}
\end{figure}

\begin{figure}
    \centering
     \includegraphics[width=.5\textwidth]{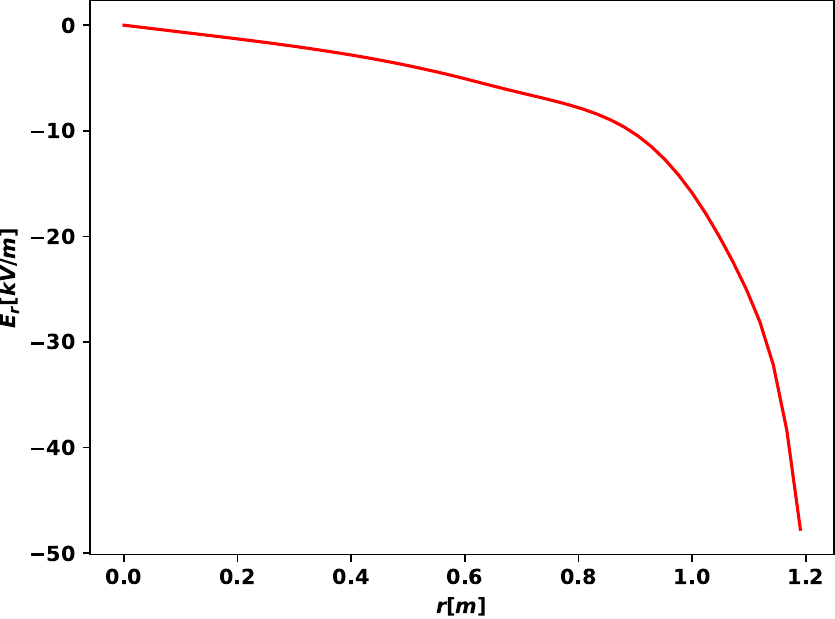}
   \caption{Radial electric field solution at the temperature and density profiles in \cref{profilesT,profilesN}, with $T_0=17.8\,\text{keV}$ and $n_0=4.21\times 10^{20}\, \text{m}^{-3}$, for the magnetic configuration optimised only for the QI cost function. Solution obtained using the NTSS code by solving \cref{Er_transp}.}
    \label{fig:Er_QI}
\end{figure}

We now optimise the initial configuration regarding a QI target only. While performing this optimisation, we still maintain the previously discussed targets of the mirror ratio, mean iota and aspect ratio in the optimisation. The optimisation resulted in a final cost function of $ 3.003\times 10^{-5}$. The final aspect ratio is $10.000$ and the volume averaged magnetic field is $7.962 \,\text{T}$. The magnetic field strength can be seen in \cref{fig:B_QI}. We can see that the magnetic field strength in Boozer coordinates at $s=0.1025$ has no local minima, contrary to what was observed in \cref{fig:B_initial} for the initial configuration. This indicates that the QI optimisation was successful in making the magnetic field closer to a QI stellarator. At other flux surfaces, the field still presents some deviations from QI. However, this is not surprising as the QI target is applied at only 4 flux surfaces $s=[1/24,5/24,9/24,13/24]$. It is thus expected that the field outside the exact location of these flux surfaces may be less close to a QI field.
\begin{figure}
     \includegraphics[width=.5\textwidth]{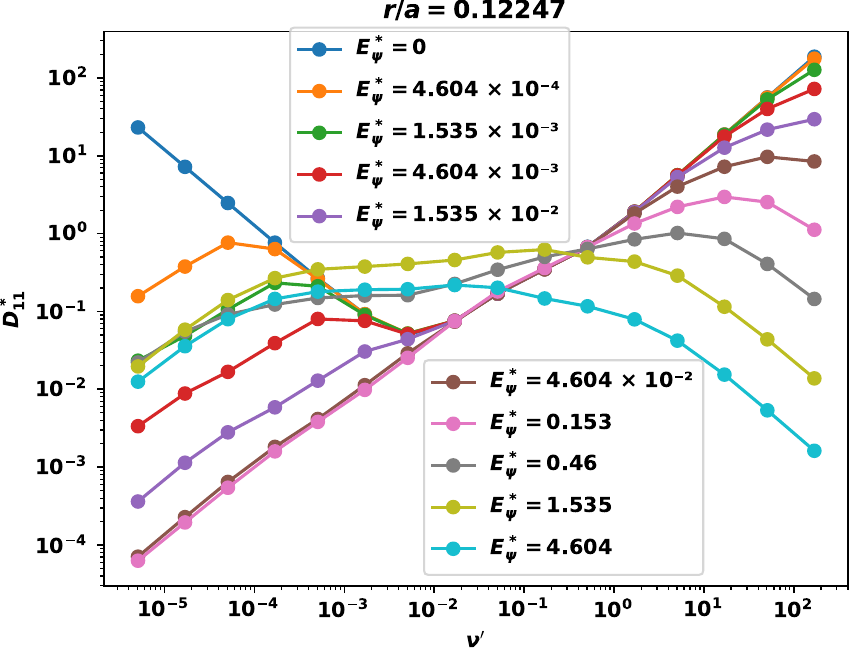}
    \includegraphics[width=.5\textwidth]{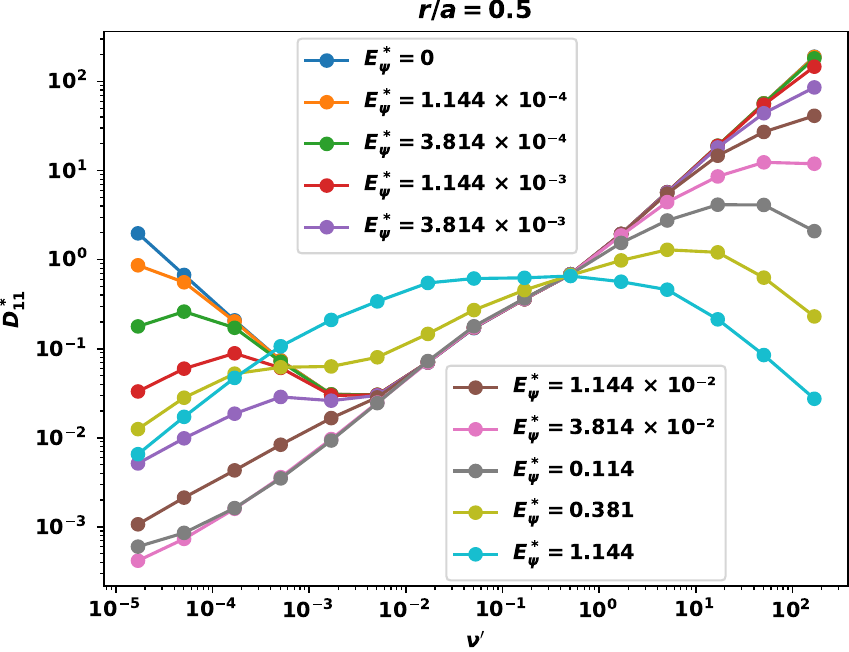}
   \caption{Scans of the monoenergetic radial transport coefficients at $r/a=0.12247$ and $r/a=0.5$ for the magnetic configuration optimised only for the QI cost function.}
    \label{fig:D11_QI}
\end{figure}
The monoenergetic radial transport coefficients scans for this optimised configuration at $r/a=0.12247$ and $r/a=0.5$ are shown in \cref{fig:D11_QI}. At both radial positions we see that the the plateau regime at intermediate collisionality values disappears after the QI optimisation when compared to the initial configuration (see \cref{fig:D11_initial}). While the $1/\nu^*$ regime is roughly the same at $r/a=0.12247$ between the two configurations, the reduction of transport for the same absolute value of the electric field $E_r$ is larger in the QI-optimised configuration due to the disappearance of the $1/\nu^*$ regimes at finite $E_r$ and low collisionalities for the optimised configuration. At $r/a=0.5$ the value of the transport coefficients in the $1/\nu^*$ regime is reduced (as expected for a magnetic field closer to perfect QI) even if at finite electric field we still see the presence of a $1/\nu^*$-like regime. By solving the electric field diffusion equation with the temperature and density profiles used during optimisation (see \cref{profilesT,profilesN} with $T_0=17.8\,\text{keV}$ and $n_0=4.21\times 10^{20}\, \text{m}^{-3}$) we obtain the electric field solution shown in \cref{fig:Er_QI} where $a=1.190\,\text{m}$. We see that an ion root is again observed for the density and temperature values chosen for the optimisations in this work. Thus, we conclude that the QI cost function alone is not enough to optimise the magnetic configuration to have an electron root solution at such plasma parameters. However, we note that the magnitude of the radial electric field solution obtained for this optimised configuration is overall smaller than the solution in \cref{fig:Er_initial} obtained for the initial configuration. This indicates that, for this configuration, ambipolarity is achieved for smaller values of the radial electric field strength. Such a result is a consequence of the reduction of the ion-relevant radial transport observed in this configuration.
\subsection{Electron root optimisation}
\begin{figure}
     \includegraphics[width=.5\textwidth]{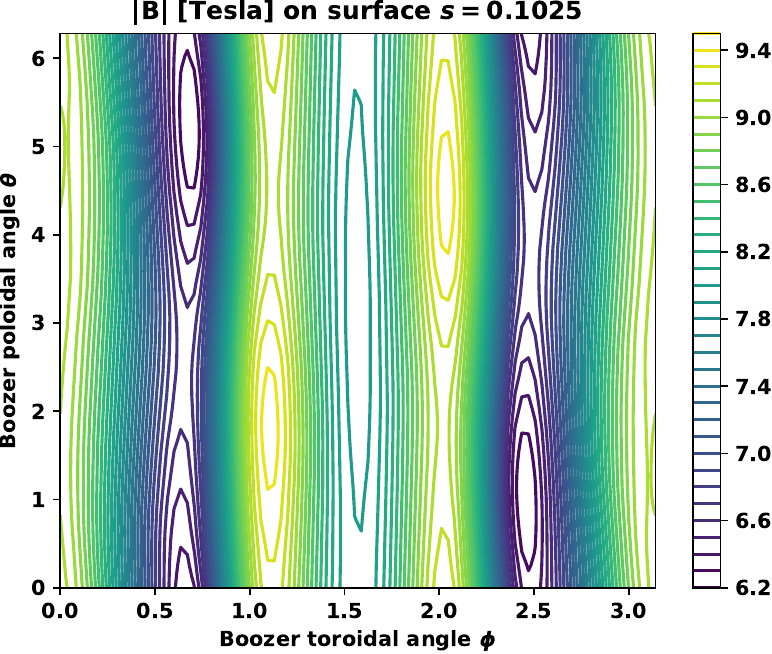}
        \includegraphics[width=.5\textwidth]{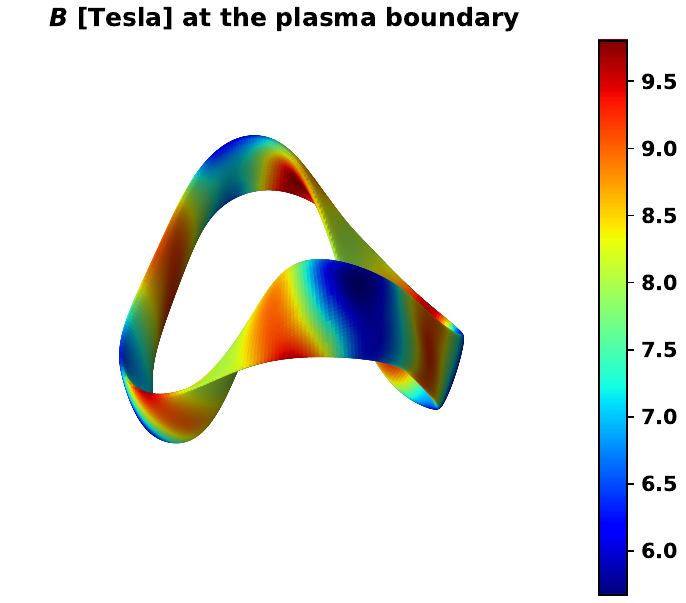}
   \caption{Contours of magnetic field strength at $s=0.1025$ in Boozer coordinates for the magnetic configuration optimised only for the electron root cost function, and the same quantity at the plasma boundary.}
    \label{fig:B_Er}
\end{figure}

We now analyse the result of an optimisation of the initial condition when we target only the electron root cost function while keeping the mirror, aspect ratio, and mean iota targets the same. The final aspect ratio of this configuration is $9.985$, the final volume average magnetic field is $7.906 \,\text{T}$  and the final total cost function is $0.079$. In particular, the final electron root cost function at $r/a=0.2$  is $0.154$, at $r/a=0.29$ is $0.117$ and at $r/a=0.35$ is $0.142$. The magnetic field strength at different flux surfaces is shown in \cref{fig:B_Er}. We can see from the magnetic field strength contours in Boozer coordinates at $s=0.1025$ that the resulting magnetic field has a preference for poloidal closed contours. However, the resulting magnetic field is quite non-optimised in regards to omnigenity. Indeed, looking more closely at the magnetic field strength at $s=0.1025$, we can observe that the magnetic spectrum has three local wells per field period and that these wells have different shapes for different field lines. 
\begin{figure}
     \includegraphics[width=.5\textwidth]{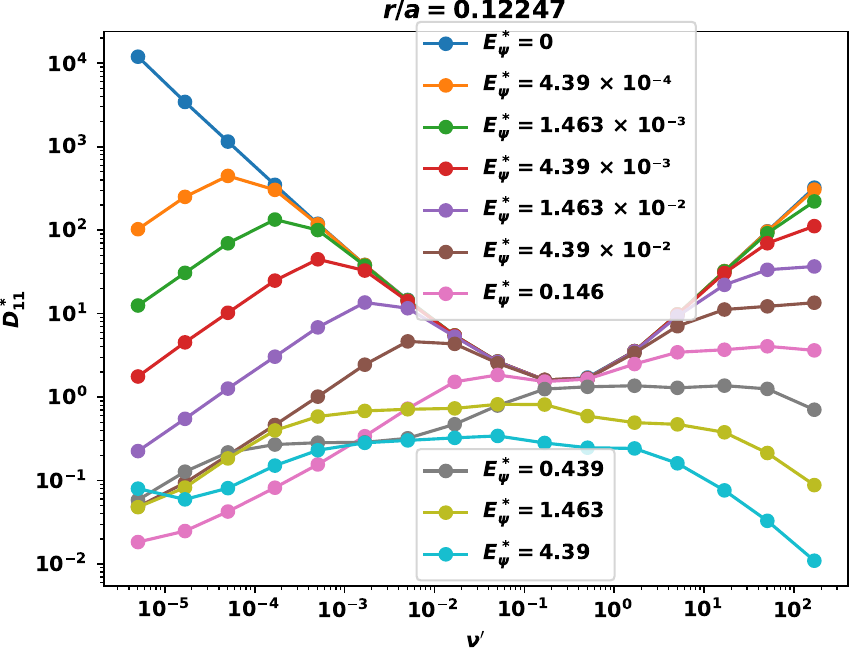}
    \includegraphics[width=.5\textwidth]{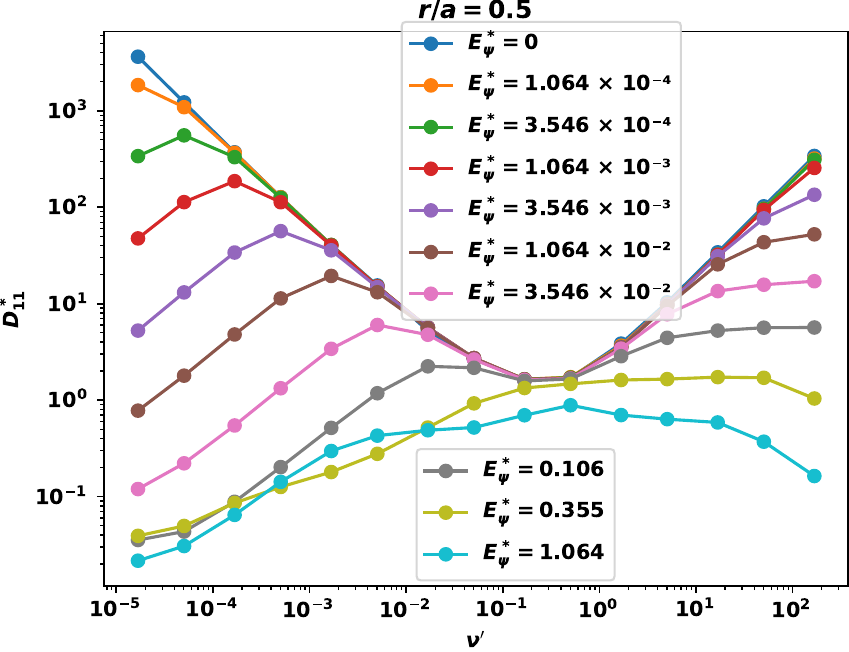}
   \caption{Scans of the monoenergetic radial transport coefficients at $r/a=0.12247$ and $r/a=0.5$ for the magnetic configuration optimised only for the electron root cost function.}
    \label{fig:D11_Er}
\end{figure}
\begin{figure}
    \centering
     \includegraphics[width=.5\textwidth]{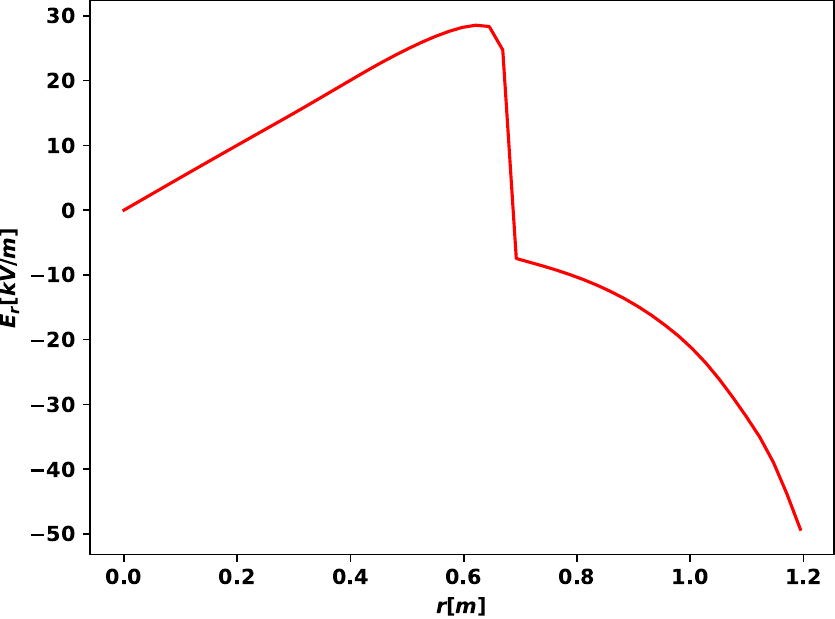}
   \caption{Radial electric field solution at the temperature and density profiles in \cref{profilesT,profilesN}, with $T_0=17.8\,\text{keV}$ and $n_0=4.21\times 10^{20}\, \text{m}^{-3}$, for the magnetic configuration optimised only for the electron cost function. Solution obtained using the NTSS code by solving \cref{Er_transp}.}
    \label{fig:Er_Er}
\end{figure}
The monoenergetic radial transport coefficients scans at different radial positions can be seen in \cref{fig:D11_Er}. We can see that there is a large increase of the $1/\nu^*$ regime when compared with the QI-optimised configuration (see \cref{fig:D11_QI}). However, at finite electric field and low collisionality, there are well-defined transitions from a $1/\nu^*$ to a linear $\nu^*$ regime, with the $\sqrt{\nu^*}$ regime disappearing almost completely. The electric field profile obtained from solving the electric field transport equation is shown in \cref{fig:Er_Er}. The temperature and density profiles used to obtain such a solution were the same as the ones used during optimisation, which we recall here, are given by \cref{profilesT,profilesN}, with $T_0=17.8\,\text{keV}$ and $n_0=4.21\times 10^{20}\, \text{m}^{-3}$. It is observed that this configuration has an electron root at these temperature and density profiles which is the original purpose of the electron root cost function. Nevertheless, as we do not target any symmetry constraints, the magnetic field can freely adapt to satisfy the electron root function and the final result is not a QI field as shown in \cref{fig:B_Er}. The electron root for this configuration has a maximum of $E_{r}^\text{max}=28.557\,\text{kV/m}$ and a root transition at $r/a=0.570$ with $a=1.195\,\text{m}$.

\subsection{QI and electron root optimisation}
\begin{figure}
     \includegraphics[width=.5\textwidth]{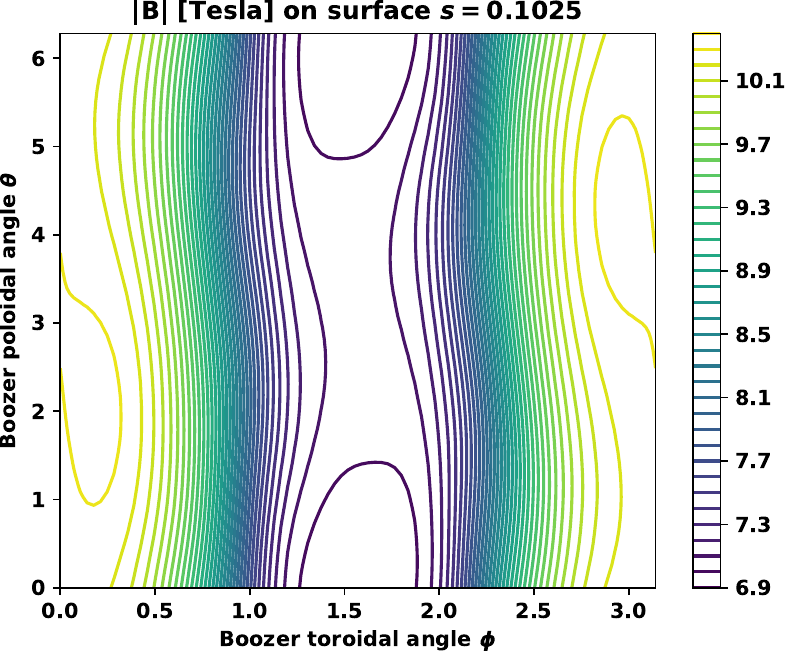}
    \includegraphics[width=.5\textwidth]{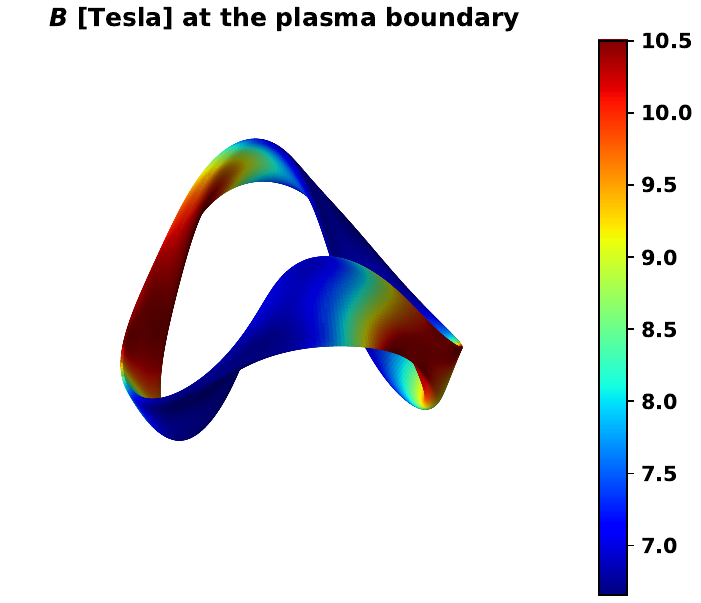}
   \caption{Contours of the magnetic field strength in Boozer coordinates at $s=0.1025$ for the magnetic configuration optimised for both QI and electron root. The same quantity at the plasma boundary is also shown.}
    \label{fig:B_QI_Er}
\end{figure}

\begin{figure}
    \includegraphics[width=.5\textwidth]{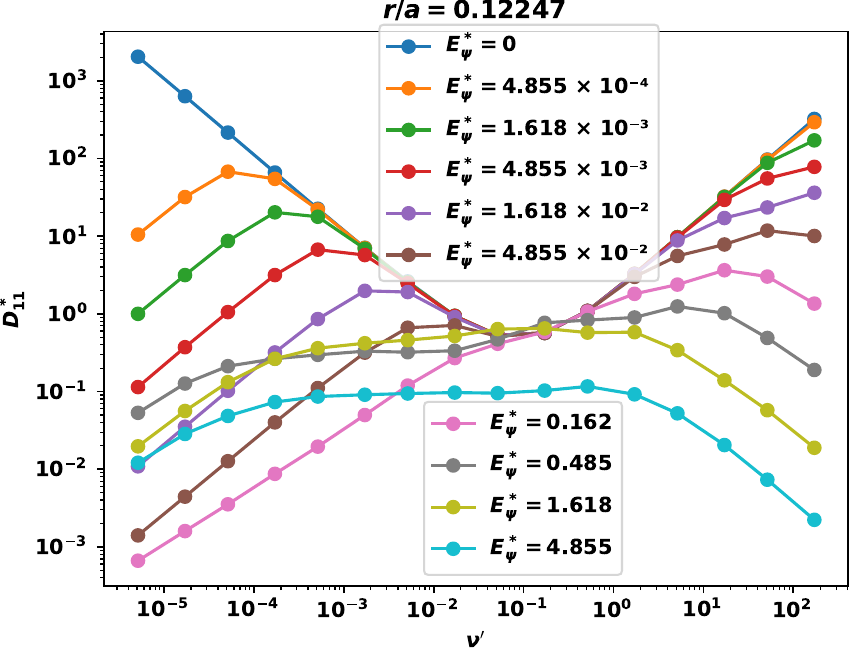}
    \includegraphics[width=.5\textwidth]{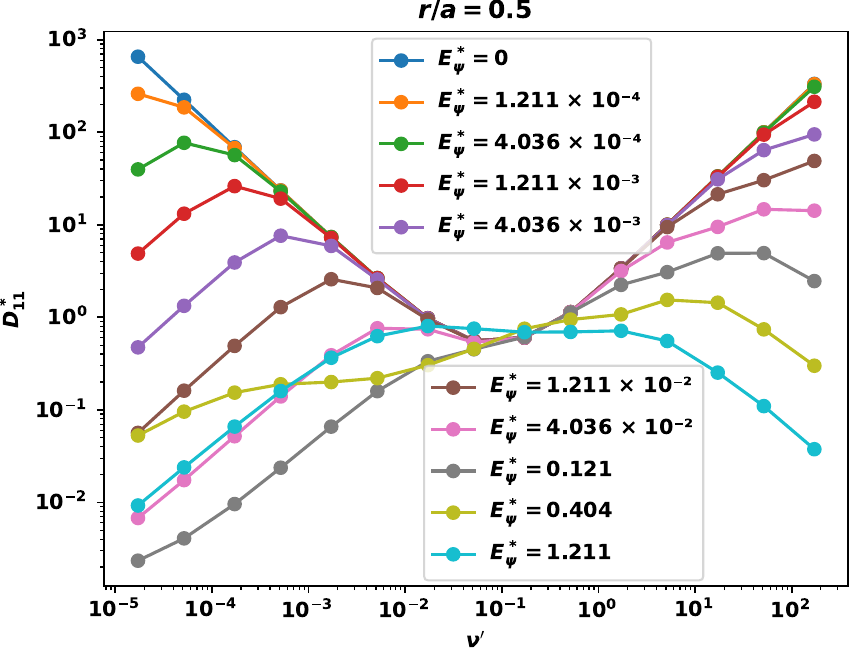}
   \caption{Scans of the monoenergetic radial transport coefficients at $r/a=0.12247$ and $r/a=0.5$ for the magnetic configuration optimised for both electron root and QI.}
    \label{fig:D11_QI_Er}
\end{figure}

\begin{figure}
    \centering
     \includegraphics[width=.5\textwidth]{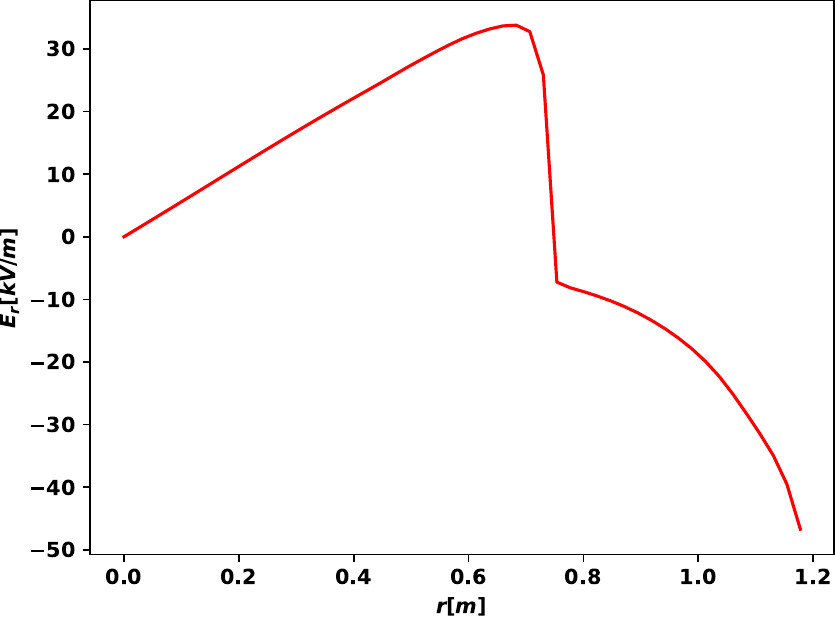}
   \caption{Radial electric field solution at the temperature and density profiles in \cref{profilesT,profilesN}, with $T_0=17.8\,\text{keV}$ and $n_0=4.21\times 10^{20}\, \text{m}^{-3}$, for the magnetic configuration optimised for both QI and electron root. Solution obtained using the NTSS code by solving \cref{Er_transp}.}
    \label{fig:Er_QI_Er}
\end{figure}

We now optimise for both QI and electron root targets keeping the same aspect ratio, mean iota and mirror ratio targets as before. The final aspect ratio of the optimised configuration is  $10.015$, the volume-averaged magnetic field strength is $8.373\, \text{T}$ and the final total cost function is $0.043$. The final electron root cost function is $0.130$ at $r/a=0.2$, $0.0917$  at $r/a=0.29$ and $0.120$ at $s=0.35$. The magnetic field strength contours in Boozer coordinates at $s=0.1025$ and the magnetic field strength at the plasma boundary are depicted in \cref{fig:B_QI_Er}. At $s=0.1025$ we can see that the magnetic field strength presents the usual contours of a QI configuration with a small local minimum. When comparing such contours with the previous ones shown for the QI-only optimised configuration in \cref{fig:B_QI}, it is clear that adding the optimisation for the electron root adds this local minimum, making the magnetic field less close to perfect QI. Nevertheless, such defects at the finer structure of the magnetic field can be tuned, by changing the weights of both QI and electron root cost functions, which were maintained at unity throughout this work. 
The monoenergetic radial transport coefficients scans at $r/a=0.12247$ and $r/a=0.5$ for this configuration are shown in \cref{fig:D11_QI_Er}. From the radial transport coefficient scans we see again the increase of the $1/\nu^*$ transport when compared against the configuration optimised only for QI (see \cref{fig:D11_QI}). However, such an increase is not as large as in the case of the optimisation performed only for electron root (see \cref{fig:D11_Er}). At the same time, the behaviour at finite electric field is again characterised by a well-defined and quick transition from the $1/\nu^*$ regime to a $\nu^*$ regime at both radial positions. As in the previous optimisation, which targeted only the electron root cost function, such behaviour is a consequence of the disappearance of the $\sqrt{\nu^*}$ transport regime at low collisionality and finite electric field. In this case, though, the $\nu^*$ regime at low collisionality for the scan at $r/a=0.5$ is better defined than in the previous optimisation done only for electron root (see \cref{fig:D11_Er}).
We solve again the electric field transport equation while keeping the same temperature and density profiles used during optimisation, which we again recall are given by \cref{profilesT,profilesN}, with $T_0=17.8\,\text{keV}$ and $n_0=4.21\times 10^{20}\, \text{m}^{-3}$. The electric field solution obtained is depicted in figure \cref{fig:Er_QI_Er}. We can see that even with a configuration with relatively good QI properties we can obtain an electron root at reactor parameters. This electron root has a maximum of $E_r^\text{max}=33.791\,\text{kV/m}$ and a root transition at $r/a=0.6300$ with $a=1.178\,\text{m}$. Notice also that the electron root obtained in this case is larger than the one obtained when optimising only for the electron root (see \cref{fig:Er_Er}). In fact, the final electron root cost function obtained was overall larger when optimising only for electron root than in this case. These results indicate that the electron root optimisation can get better results when used together with the QI optimisation. Similar observations were made in Ref. \citet{Kim2023} for the case of turbulence optimisation and in Ref. \citet{Bader2019} for the optimisation of fast particles. In the case of the electron root optimisation, such behaviour may occur because, in the optimisation using only the electron root cost function, a path that increases the $1/\nu^*$ very quickly is first chosen by the optimiser, leading to a magnetic field strength with a lack of any symmetry (see \ref{fig:B_Er}). In the case in which both QI and electron root cost functions are used, the optimiser first selects a magnetic field with good QI properties limiting the increase of the $1/\nu^*$ regime. Then, in both optimisations small adjustments are done to the magnetic field, in order to generate a well-defined $\nu^*$ regime at low collisionality. From comparing \cref{fig:D11_Er,fig:D11_QI_Er}, these small adjustments look easier to achieve by the optimisation loop when starting from a configuration closer to the QI magnetic geometry.

\begin{figure}
     \includegraphics[width=.5\textwidth]{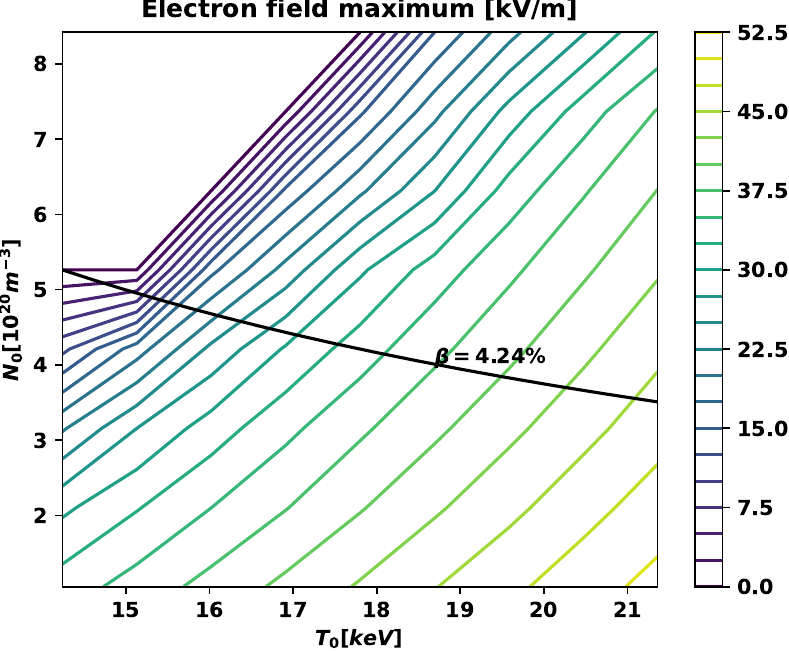}
    \includegraphics[width=.5\textwidth]{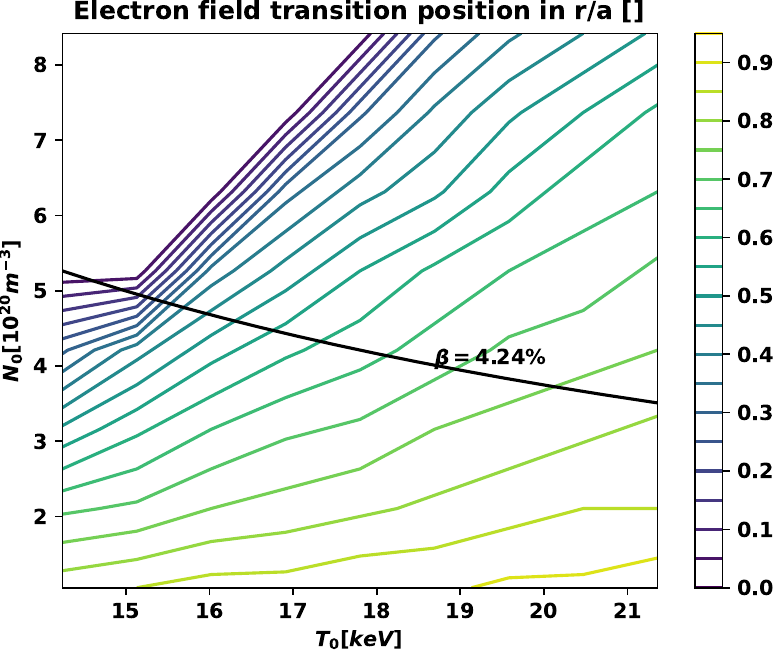}
   \caption{Scans of the maximum of the radial electric field and the location of the root transition for different temperatures and densities on-axis. The magnetic field and volume of the plasma are maintained fixed in this scan. The plotted line is a contour of constant $\beta=4.24\%$.}
    \label{fig:NT_scan}
\end{figure}

In order to check the robustness of this optimised magnetic configuration with respect to the presence of an electron root for different plasma parameters we solve the electric field diffusion equation for different temperatures and density values. The profiles used are obtained by scaling the profiles in \cref{profilesT,profilesN} used during the optimisation. We plot the results of this scan for the maximum of the electric field and the location of the root transition (normalised to the minor radius). The results can be seen in \cref{fig:NT_scan}. We note that when the electric field has a maximum of zero, no electron root is present. At the same time, a zero value is attributed to the relative position of the transition to indicate the cases in which no electron root is present. As expected the region of plasma parameters in which the electron root ceases to exist lies at low temperatures and high densities (delimited by the darkest blue contour), which represents the higher collisionality region. Nevertheless, the presence of the electron root for this optimised configuration is seen to be robust as it is present for a large number of temperatures and densities relevant to reactor scenarios. In particular, for a density on-axis of $n=4.21  \times 10^{20} \,\text{m}^{-3}$ we can see that for the entire range of temperatures scanned ($14.24-21.36 \,\text{keV}$) the electron root is present. The trend in the scan also indicates that larger temperatures than $21.36\,\text{keV}$ (which were not simulated here but can also be relevant for reactor scenarios) should have an electron root. The maximum of the radial electric field strength increases with increasing temperature and decreasing density, with the maximum values of $E_{r}^\text{max}=50.789 \,\text{kV/m}$ being achieved at the largest temperature of $21.36 \,\text{keV}$ and lower density of $1.05\times 10^{20} \,\text{m}^{-3}$ in the scan. More surprising perhaps is the position of the root transition, which from observing the scan in \cref{fig:NT_scan}, increases more with decreasing density than with increasing temperature. At the lowest densities in the scan, the electron-ion root transition occurs quite near the plasma boundary, with the electron root spanning through the majority of the minor radius extension. 

\begin{figure}
     \includegraphics[width=.5\textwidth]{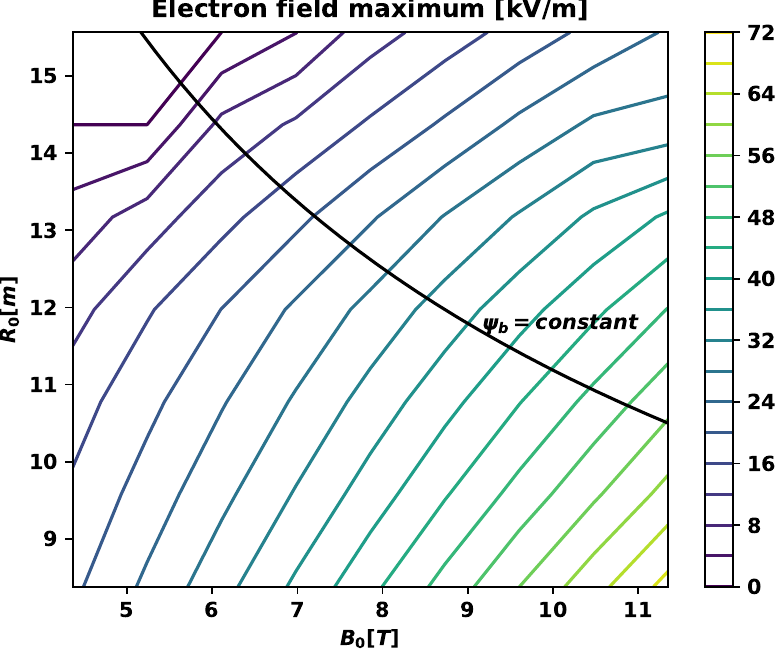}
    \includegraphics[width=.5\textwidth]{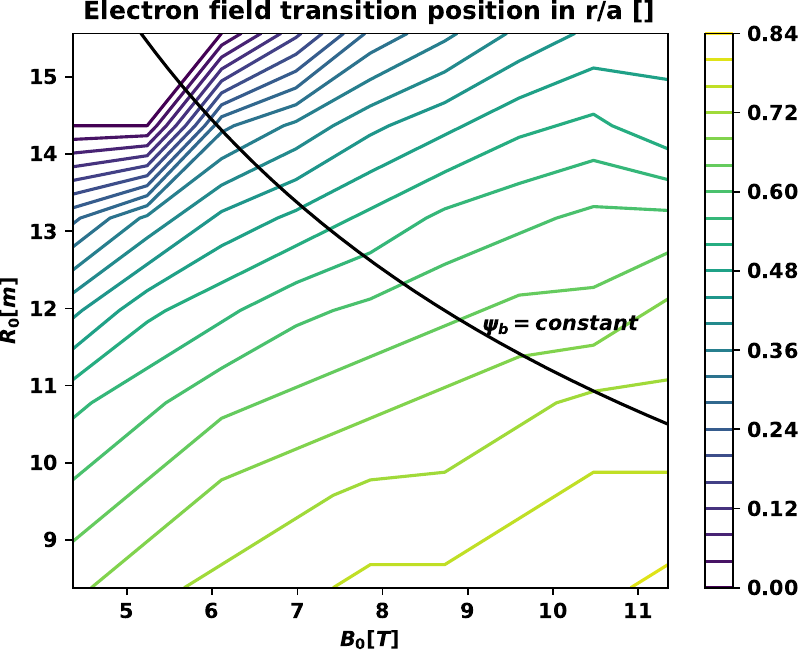}
   \caption{Scans of the maximum of the radial electric field and the location of the root transition for different scaling of the major radius and magnetic field strength on-axis. The plasma $\beta=4.24\%$ is constant for the entire scan. The plotted line is a contour of constant toroidal flux at the boundary.}
    \label{fig:BA_scan}
\end{figure}

\begin{figure}
     \includegraphics[width=.5\textwidth]{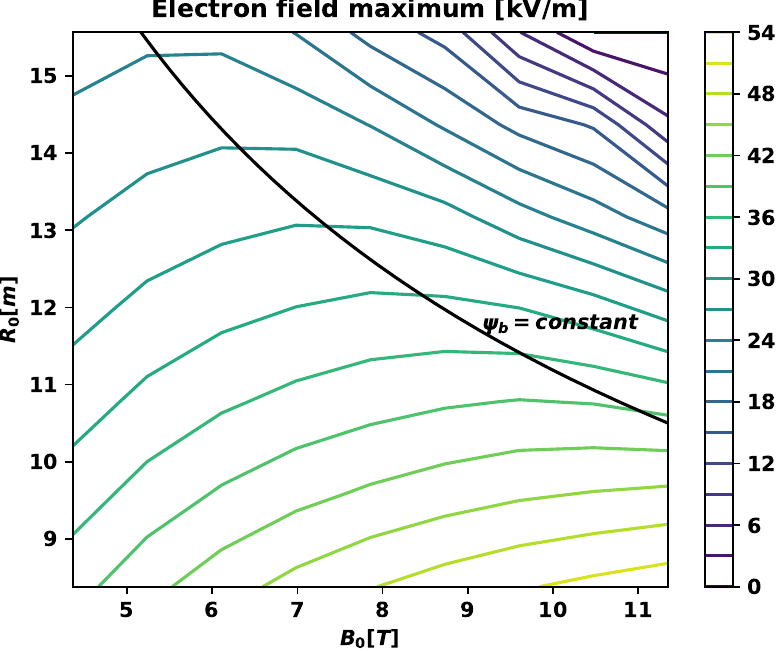}
    \includegraphics[width=.5\textwidth]{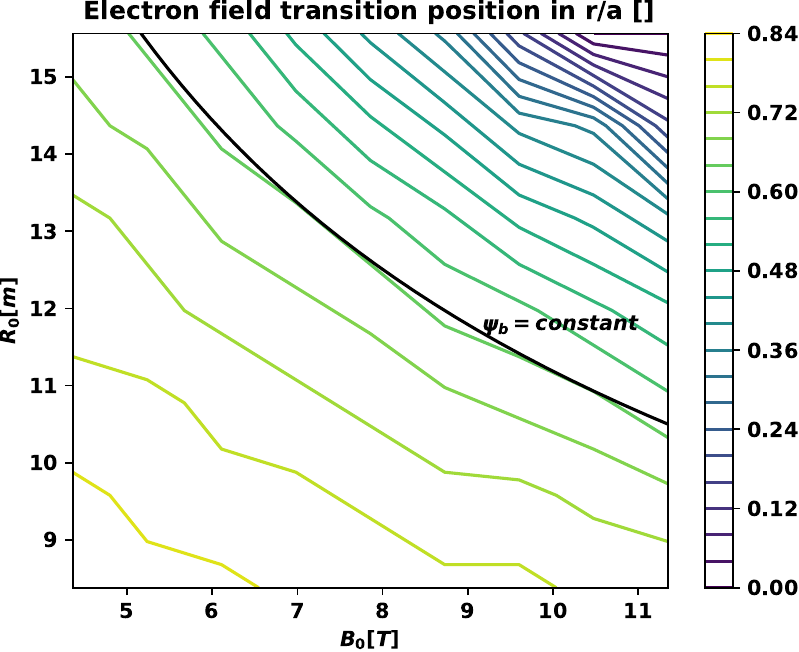}
   \caption{Scans of the maximum of the radial electric field and the location of the root transition for different scaling of the major radius and magnetic field strength on-axis. Contrary to the scans in \cref{fig:BA_scan}, the plasma $\beta$ varies in these scans. The plotted line is a contour of fixed toroidal flux at the boundary.}
    \label{fig:BA_scan_nobeta}
\end{figure}

A similar scan was done by scaling the major radius and the magnetic field. In this scan, the aspect ratio is maintained constant by scaling the minor radius with the major radius, and $\beta=4.24\%$ is also kept constant by scaling the density and temperature profiles by the same scaling factor applied to the magnetic field. The resulting scans are shown in \cref{fig:BA_scan}. We see that the area in which the electron root ceases to exist is located in the region of large major radius and low magnetic field (delimited by the darkest blue contour). It is observed that in the range of scanned major radius and magnetic field values, the electron root exists at almost all parameters. It is interesting to see that the maximum of the electron root has its larger value at a high magnetic field and small major radius values, with a maximum electric field of $E_{r}^\text{max}=69.307 \, \text{kV/m}$ being obtained. The position of the root transition increases more with decreasing major radius than with the increasing magnetic field, and a value of up to $r/a=0.810$ can be obtained showing that a large region of the plasma can achieve electron root for a small major radius plasma at a wide range of magnetic field values. 
Both the density and temperature scans in \cref{fig:NT_scan}, and the radius and magnetic field scans in \cref{fig:BA_scan} show the possibility of an electron root solution at a wide range of both plasma and configuration parameters, which indicates the robustness of the optimised magnetic configuration to the presence of an electron root. It is also observed that the electron root maximum and the position of the root transition increase with the magnetic field. However, one would expect the radial transport coefficient in the $1/\nu^*$ regime to decrease with the square of the magnetic field strength and the $\nu^*$ regime radial transport coefficient to be independent of $B$. These scalings are expected to make the electron root disappear as the magnetic field increases. To better investigate the presence of such behaviour, a scan in the values of the magnetic field and major radius allowing $\beta$ to vary is performed and can be seen in \cref{fig:BA_scan_nobeta}. This is achieved by maintaining the on-axis density and temperature fixed at the same values used in the optimisations, i.e., $T_0=17.8\,\text{keV}$ and $n_0=4.21\times 10^{20}\, \text{m}^{-3}$. We point out here that in these two last scans in \cref{fig:BA_scan_nobeta}, $\beta$ varies between very small and quite large values. The larger beta values obtained for the lower magnetic fields are outside of the expected values for a stellarator fusion reactor and thus we do not expect these to be achievable in reality. Nevertheless, we add this scan for completeness and to see the behaviour of the electron root when the magnetic field is increased without adjusting the density and temperature profiles.

In the scan shown in \cref{fig:BA_scan_nobeta}, we see that the electron root ceases to exist in the region of large magnetic field and large major radius. Accordingly, it is observed that the increase of the magnetic field at a large major radius decreases the values of the maximum of the electron root. However, at small major radius values an increase in the magnetic field still increases the electron root maximum for the values scanned here. We note that for the smaller major radius, the normalised position of the root transition decreases with the magnetic field. Such behaviour indicates that at smaller sizes of the machine, the region of the plasma in which the electron root exists will get smaller with increasing magnetic field even if the electron root maximum may get large. Thus, at some high magnetic field, larger than the ones scanned here, the electron root transition will eventually disappear. The electron roots with a larger position of the transition are located at a low major radius and low magnetic field. These scans retrieve the expected behaviour of the electron root being more difficult to achieve with an increasing magnetic field. Nevertheless, for this electron root optimised configuration, an electron root is still achievable at a value of magnetic field of $B=11\,\text{T}$ for a wide range of major radii. We point out that such magnetic field value is already quite large for state-of-the-art technologies. Outside of these scans, for the lowest value of the major radius shown here, it was found that the electron root ceases to exist between $B=37\,\text{T}$ and $B=38\,\text{T}$ which are unrealistic values for the currently available technologies. 

\subsection{Finite-$\beta$ optimisation}
\begin{figure}
     \includegraphics[width=.5\textwidth]{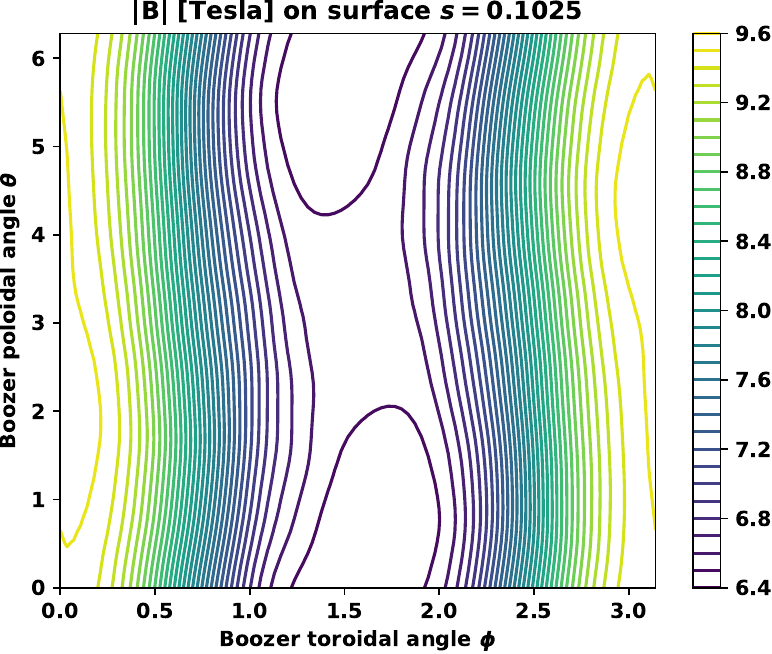}
        \includegraphics[width=.5\textwidth]{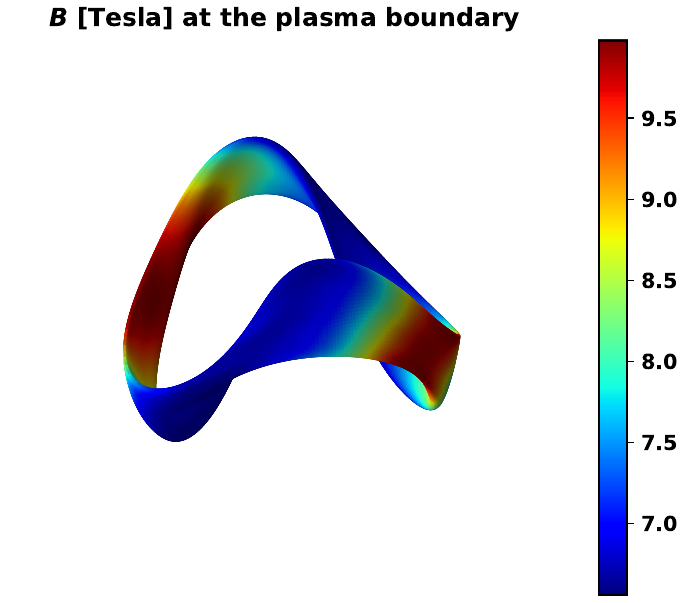}
   \caption{Contours of the magnetic field strength in Boozer coordinates at $s=0.1025$ for the magnetic configuration optimised for both QI and electron root with finite-$\beta$ corrections. The same quantity at the plasma boundary is also shown.}
    \label{fig:B_finiteb}
\end{figure}

\begin{figure}
    \includegraphics[width=.5\textwidth]{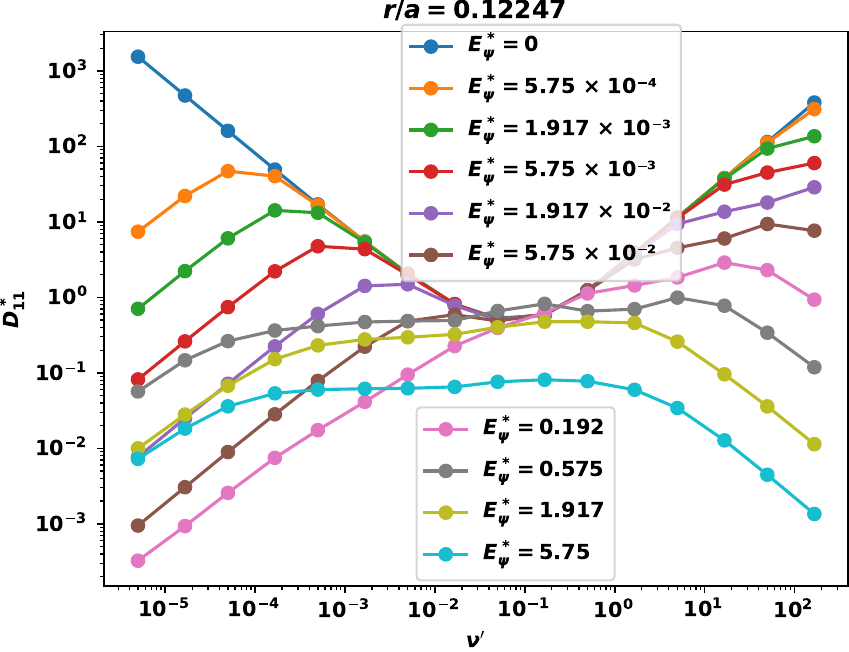}
    \includegraphics[width=.5\textwidth]{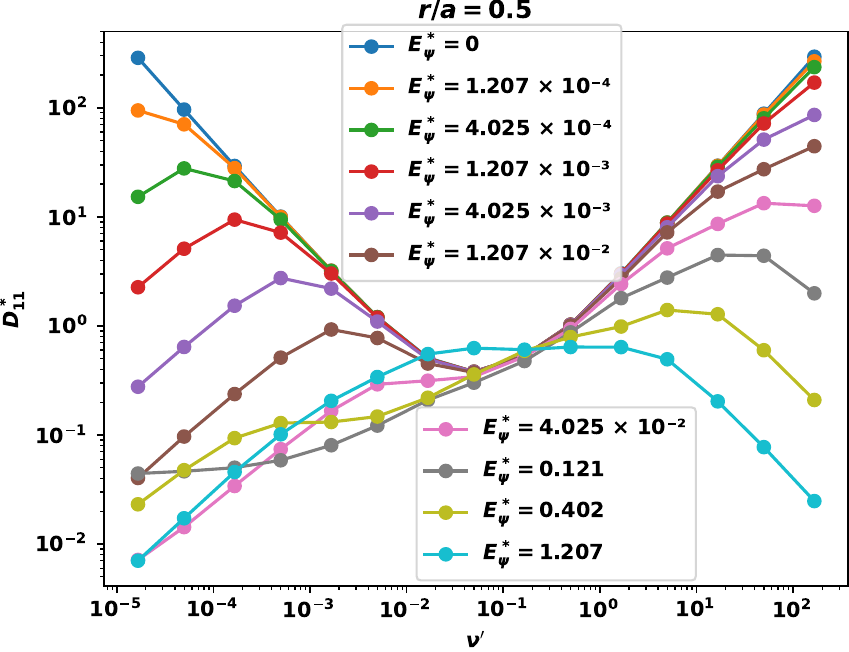}
   \caption{Scans of the monoenergetic radial transport coefficients at $r/a=0.12247$ and $r/a=0.5$ for the magnetic configuration optimised for both electron root and QI with finite-$\beta$ effects.}
    \label{fig:D11_finiteb}
\end{figure}

\begin{figure}
    \centering
     \includegraphics[width=.5\textwidth]{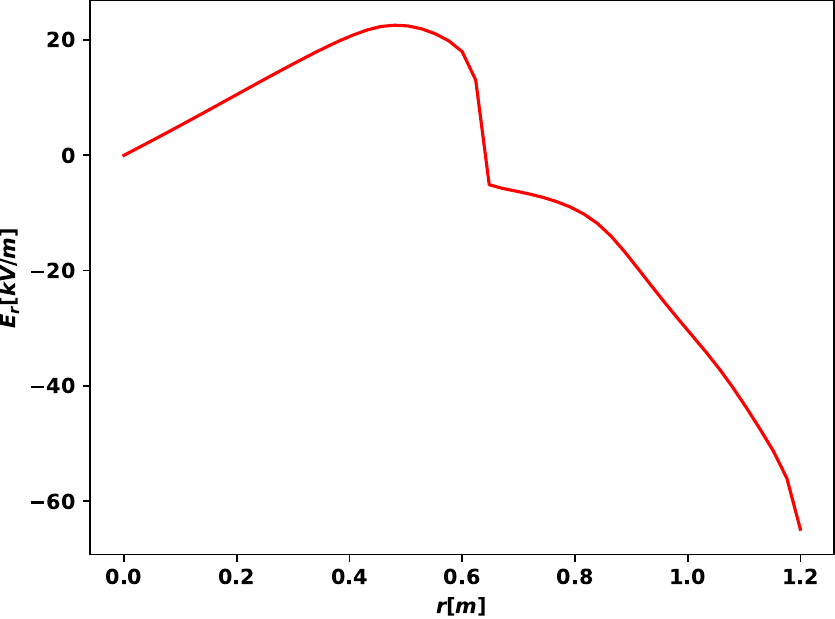}
   \caption{Radial electric field solution at the temperature and density profiles in \cref{profilesT,profilesN} with $T_0=17.8\,\text{keV}$ and $n_0=4.21\times 10^{20}\, \text{m}^{-3}$ for the magnetic configuration optimised for both QI and electron root with finite-$\beta$ effects. Solution obtained with the NTSS code by solving \cref{Er_transp}.}
    \label{fig:Er_finiteb}
\end{figure}

Up to now, the optimisations shown have considered vacuum VMEC solutions only. This does not account for the finite-$\beta$ corrections from the plasma pressure in the MHD equation. Such considerations are worth investigating because the electron root optimisation depends on the solution of the drift kinetic equation, which is obtained by considering the existence of finite prescribed density and temperature profiles. A pressure gradient is added to the MHD force balance, and we optimise for both QI and electron root targets with the pressure profile given by $P=2nT$, where the density and temperature profiles are again the ones in \cref{profilesT,profilesN}, with $T_0=17.8\,\text{keV}$ and $n_0=4.21\times 10^{20}\, \text{m}^{-3}$. We note that the optimisation for this case is done using $51$ flux surfaces in VMEC instead of the standard $24$ flux surfaces because it was observed that a smaller resolution would lead to an optimised configuration which would present significant variations in the $\iota$ profile when the final optimised configuration radial resolution was increased from $24$ to $201$ flux surfaces. The resulting configuration has an aspect ratio of $10.004$ and a volume-averaged magnetic field of $7.760\, \text{T}$. The final total cost function is $0.053$ and the final electron root cost function is $0.113$ at $r/a=0.2$, $0.116$  at $r/a=0.29$ and $0.1452$ at $s=0.35$. The magnetic field strength in Boozer coordinates at $s=0.1025$ and magnetic field strength at the plasma boundary for this configuration can be seen in \cref{fig:B_finiteb}. By comparing the contours of the magnetic field strength at $s=0.1025$ with the one obtained with the optimisation at vacuum (see \cref{fig:B_QI_Er}), we can see that both have similar behaviours, with the finite beta contours also showing a local minimum. 
The monoenergetic radial transport coefficient scans at $r/a=0.12247$ and $r/a=0.5$ can be seen in \cref{fig:D11_finiteb}. From solving the electric field diffusion equation with the same profiles as in \cref{profilesT,profilesN} used at optimisation, with $T_0=17.8\,\text{keV}$ and $n_0=4.21\times 10^{20}\, \text{m}^{-3}$, we obtain the electric field profile which is shown in \cref{fig:Er_finiteb}. We can see that an electron root is achievable when considering a consistent finite-$\beta$ correction in the MHD force balance during optimisation. This electron root is smaller than the one obtained at the same temperature and densities with the vacuum optimisation (see \cref{fig:Er_QI_Er}). The maximum of the electron root is in this case $E_{r}^\text{max}=22.569\,\text{kV/m}$ and the root transition occurs at $r/a=0.530$ with $a=1.200\,\text{m}$. This result shows that an electron root optimisation is possible with finite-$\beta$ corrections, especially when we consider that this optimised configuration has a volume averaged $\beta \approx 5\%$.

\subsection{Confinement properties}
\begin{figure}
     \includegraphics[width=.5\textwidth]{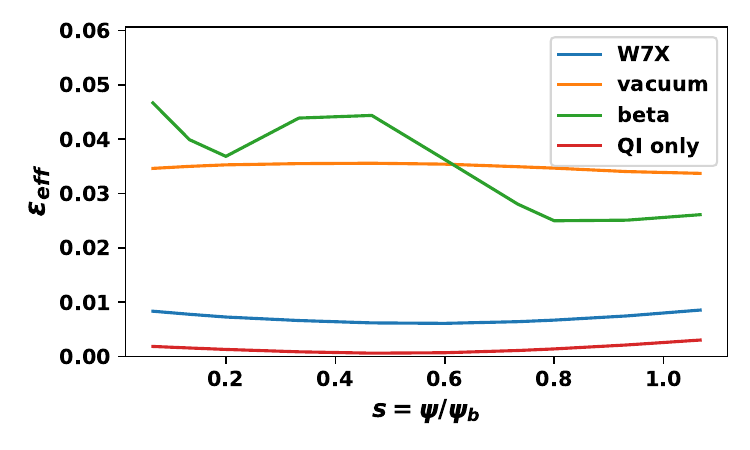}
    \includegraphics[width=.5\textwidth]{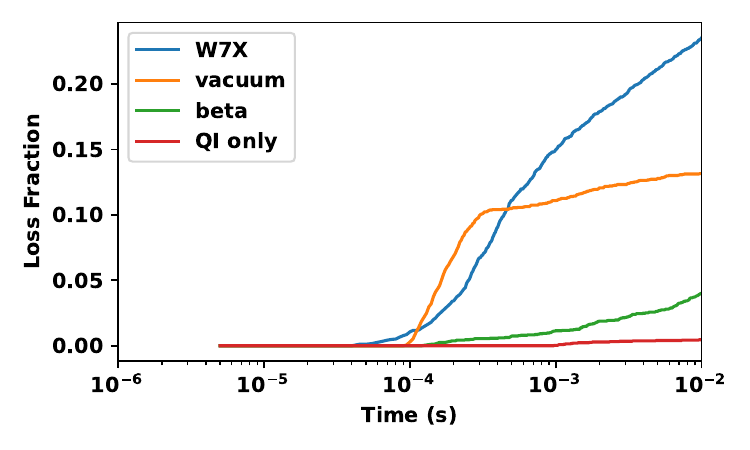}
   \caption{Epsilon effective and loss fraction of fusion alpha particles for the vacuum standard W7-X configuration, for the configuration optimised only for QI and for the vacuum and finite-$\beta$ configurations optimised for both QI and electron root. For the loss fraction of alpha particles, the volume-averaged magnetic field and minor radius of the configurations were scaled to $5.7\, \text{T}$ and $1.7\,\text{m}$, respectively.}
    \label{fig:fast_par}
\end{figure}

We now take a look at the two common figures of merit used to evaluate the confinement of fast particles, the effective ripple (epsilon effective), $\varepsilon_\text{eff}$, and the fraction of lost alpha particles from fusion reactions. The epsilon effective can be inferred from the slope of the $1/\nu^*$ previous monoenergetic radial scans in figures \cref{fig:D11_initial,fig:D11_QI,fig:D11_Er,fig:D11_QI_Er,fig:D11_finiteb}, which was discussed in the previous sections. Nevertheless, here we show the value of $\varepsilon_\text{eff}$ obtained with the code NEO \citep{Nemov1999}. The loss fraction of alpha particles is obtained with the code SIMPLE \citep{Albert2020}. This code does not include the effects of a strong electric field, but such an effect is expected to be small for the fast alpha particles due to their large velocity.  We note that we did not optimise directly for either of these metrics, only indirectly using the QI cost function, as the main objective was to isolate and understand the effects and effectiveness of the electron root cost function. Indeed, as we observe in \cref{fig:fast_par}, we can see that for both vacuum and finite-$\beta$ configurations optimised for both QI and electron root, the epsilon effective is less than $0.05$ for all the flux surfaces, but is around 3 times larger than the same quantity for W7-X. However, the loss fraction of fast particles saturates at a lower level for both vacuum and finite-$\beta$ configurations when compared to the vacuum standard W7-X configuration. The loss fraction saturates around $12\%$ for the vacuum configuration and around $3\%$ for the finite-$\beta$ configuration. The decrease in the loss of fast particles of the finite-$\beta$ configuration when compared to the vacuum configuration is probably related to the effect of the tangent magnetic fields which originate from the current due to the finite pressure gradient.  We conclude that even if we do not achieve zero losses for precise QI-only optimised configurations, we still obtain a relatively small loss of fast particles compared to W7-X. \cref{fig:fast_par}. 
We also note that the configuration optimised only for the QI cost function shows values of epsilon effective smaller than the ones observed for W7-X and the loss of fast particles is also the smallest of all the configurations.

\subsection{Power balance considerations}
So far we have calculated the electric field solution by solving only the electric field diffusion equation, \cref{Er_transp}, which means that we are solving NTSS transport equations with density and temperature profiles kept constant. It is also important to understand if an electric field can be realised while achieving the required target of fusion power for a fusion reactor. While an extensive study of such conditions is not the main topic of this work, we would like to understand if such solutions of the transport equations can be achieved with the optimised configurations for the QI and electron root optimised configurations shown in the previous sections. This is of particular importance if we take into account that we did not optimise for a small $\varepsilon_\text{eff}$. This means that the $1/\nu^*$ regime, in which the heat flux scales with $T^{9/2}$, is not directly minimised in these optimisations. While we expect an electron root in the core region to reduce the ion transport due to the effect of the electric field in bounding the radial drift, it is still true that the $1/\nu^*$ regime is important in the electron to ion root transition zone and possibly in the ion root region if the electric field there is small. Therefore, it is important to verify if, at the reactor temperatures we are aiming for, we can still contend with the heat flux generated by the possible $1/\nu^*$ regime in the root transition region. 

\begin{figure}
     \includegraphics[width=.5\textwidth]{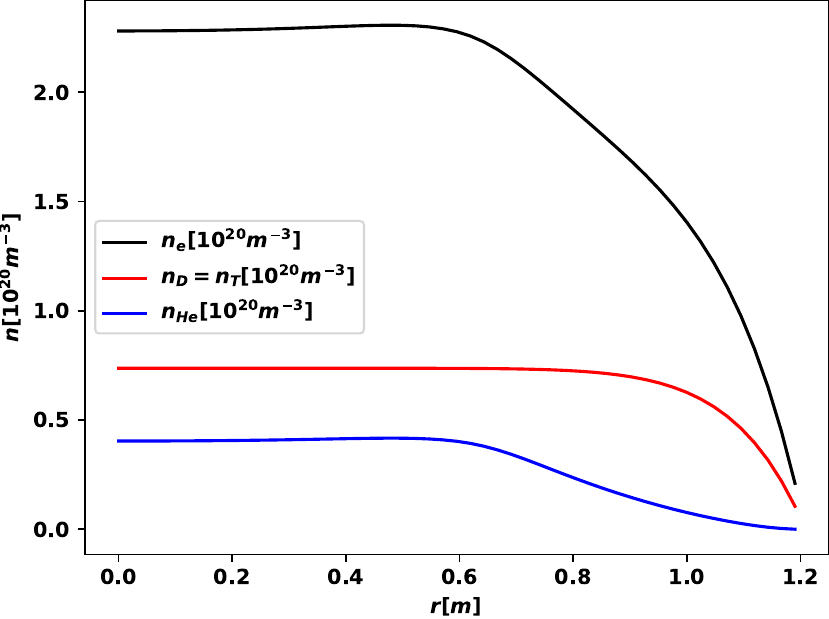}
    \includegraphics[width=.5\textwidth]{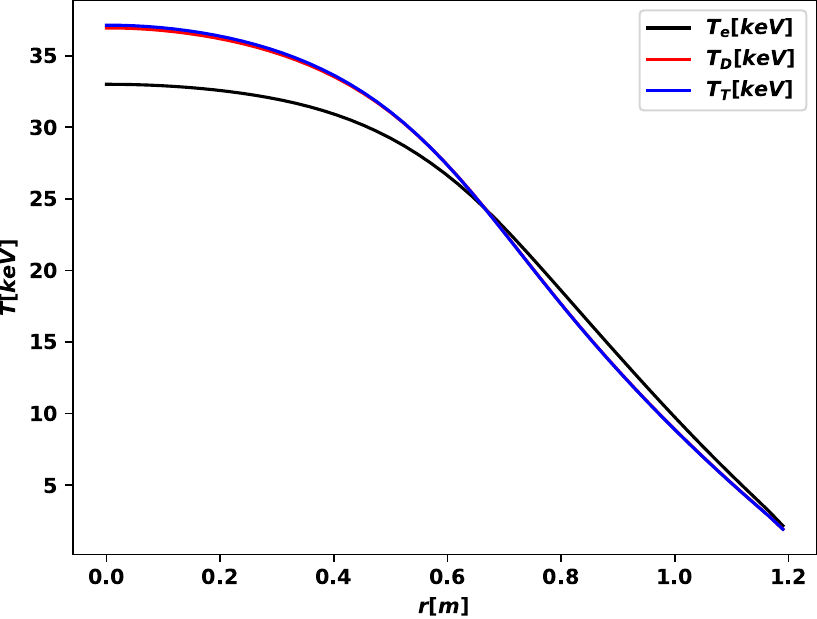}
   \caption{Density and temperature profiles obtained with the NTSS code by performing a power balance simulation for the configuration optimised only for QI.}
    \label{fig:NT_QI}
\end{figure}

\begin{figure}
    \centering
     \includegraphics[width=.5\textwidth]{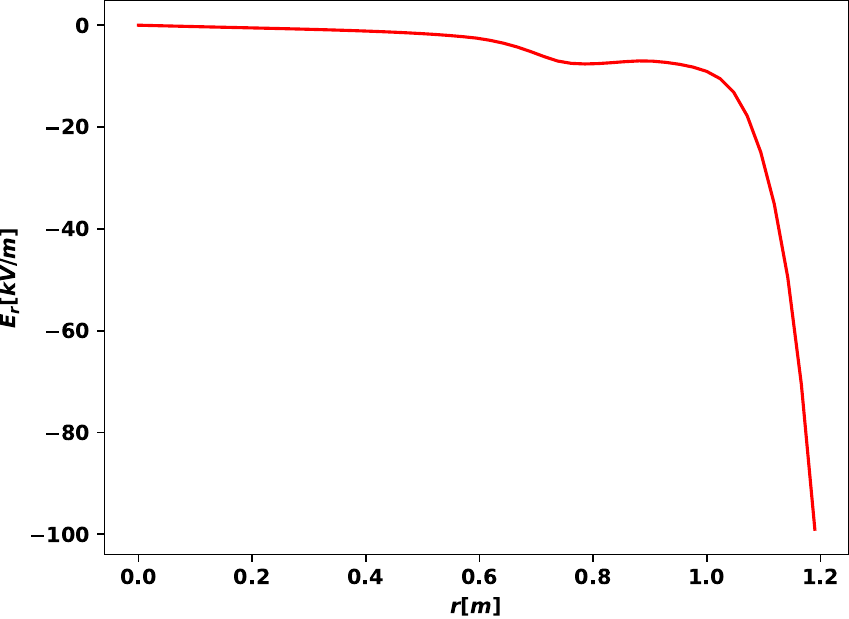}
   \caption{Radial electric field profile obtained from the NTSS code by performing a power balance simulation for the configuration optimised only for QI.}
    \label{fig:Er_HeatQI}
\end{figure}

\begin{figure}
     \includegraphics[width=.5\textwidth]{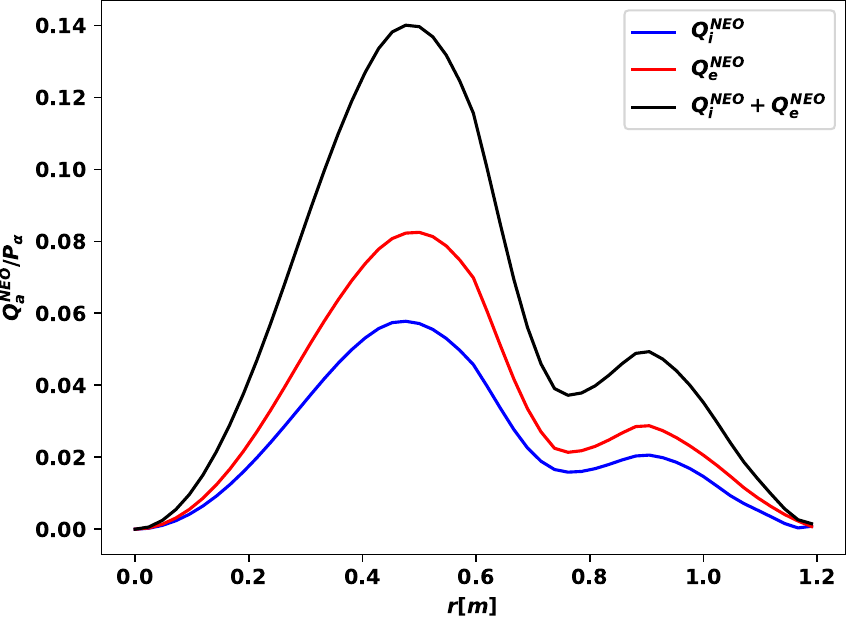}
    \includegraphics[width=.5\textwidth]{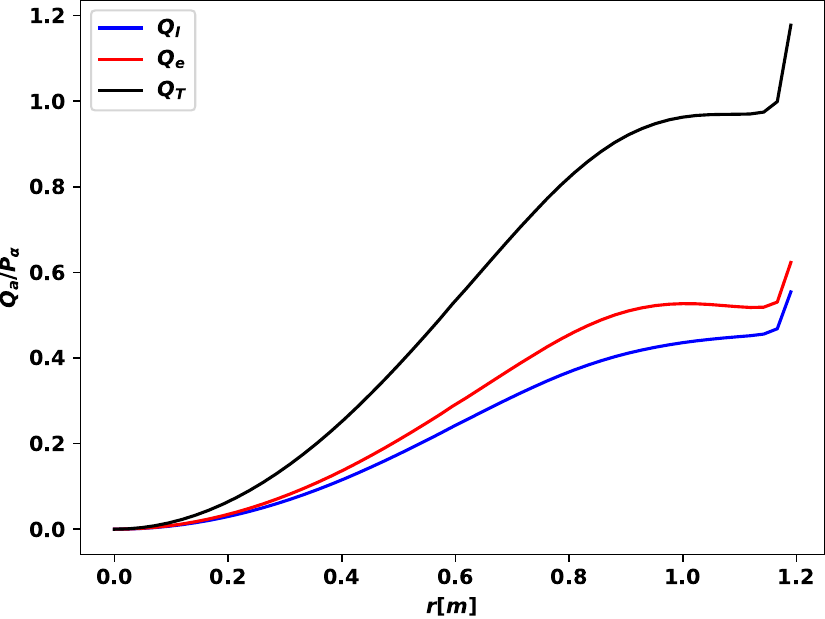}
   \caption{Neoclassical (left) and total (right) energy fluxes obtained for the power balance simulation performed for the magnetic configuration optimised only for QI. The simulation was performed by the NTSS code and the energy fluxes are normalised to the volume-averaged alpha power of $P_{\alpha}=302\,\text{MW}$.}
    \label{fig:Heat_QI}
\end{figure}

With this goal in mind, we perform simulations with the NTSS code for the configuration optimised only for QI and for the vacuum and finite-$\beta$ configurations optimised for both QI and electron root, to target an alpha power close to $300\,\text{MW}$. We perform these simulations by fixing only the density profiles of deuterium and tritium while allowing the temperatures of all species and the electric field to evolve. We also allow the helium density to evolve as in the simulations performed in Ref. \citet{Beidler2024}, with a relative concentration of deuterium and tritium of $0.4$ and a helium relative concentration of $0.1$. Following Ref. \citet{Beidler2024} we also add anomalous transport via a model in which $\chi_a=\mathcal{C}(P \text{[MW]})^{3/4}/(n [10^{20} \text{m}^{-3}])$, with the choice of $\mathcal{C}=6.5\times 10^{-3}\text{m}^2/s$ and $\mathcal{D}_a=0.01\chi_a$. We note that this is a very simple model but it is added here to capture the same conditions used for the simulation shown in Ref. \citet{Beidler2024} in which an electron root was observed, thus allowing for a better comparison. For the configuration optimised only for QI we use \cref{profilesT,profilesN} scaled to $T_0=7.12\,\text{keV}$ and $n_0=1.47\times 10^{20}\,\text{m}^{-3}$ as the initial profiles. We obtain a plasma with $302\,\text{MW}$ of volume-averaged alpha power after subtracting the Bremsstrahlung losses and an energy confinement time of $\tau_E=1.68\,\text{s}$ is obtained which is a factor of $2.32$ above the International Stellarator Scaling (ISS04) scaling \citep{Yamada2005}. The temperature and density profiles for the different species can be observed in \cref{fig:NT_QI}. The radial electric field solution obtained is shown in \cref{fig:Er_HeatQI} and is an ion root despite the low density and high temperatures of the ion species. Such a result is expected as this configuration was not optimised for electron root. The neoclassical and total (neoclassical and anomalous) energy fluxes, normalised to the volume-averaged alpha power $P_{\alpha}=302\,\text{MW}$, are depicted in \cref{fig:Heat_QI}. The neoclassical energy fluxes are seen to be small for both ion and electron species as expected from a QI-optimised configuration, with the sum of ion and electron neoclassical energy fluxes staying below $0.15P_{\alpha}$. The sum of ion and electron total energy fluxes (neoclassical and anomalous) stays below the value of the volume-averaged alpha power, except at the boundary, at which, the large energy flux variation is likely due to the artificial choice of the temperature and density values at the boundary. 

\begin{figure}
     \includegraphics[width=.5\textwidth]{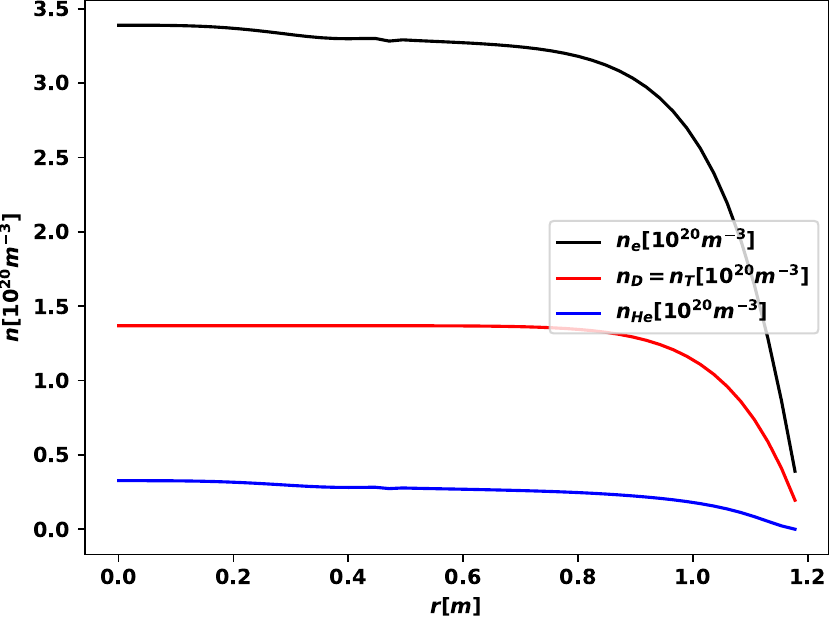}
    \includegraphics[width=.5\textwidth]{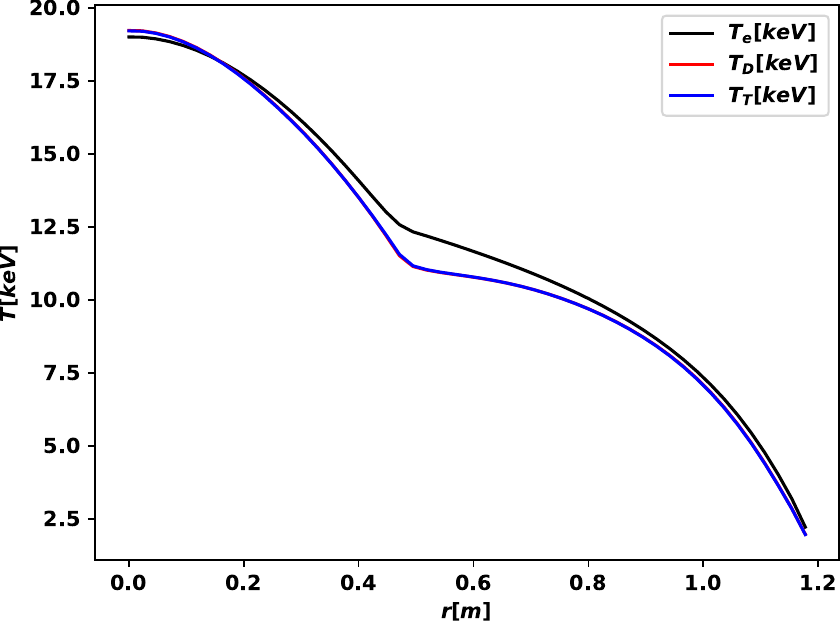}
   \caption{Density and temperature profiles obtained from a power balance simulation, performed by the NTSS code for the vacuum configuration optimised for both QI and electron root.}
    \label{fig:NT_HeNeo}
\end{figure}

\begin{figure}
\centering
     \includegraphics[width=.5\textwidth]{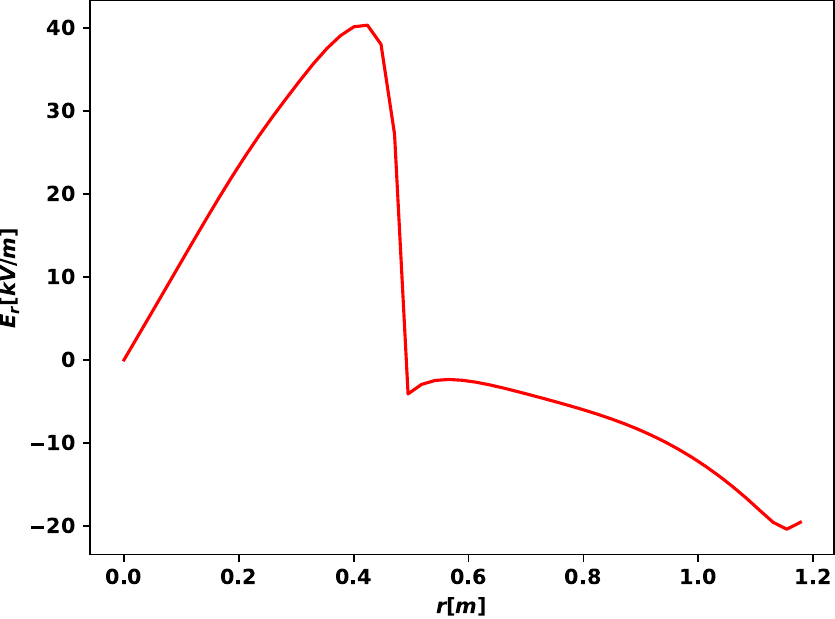}
   \caption{Radial electric field profile obtained from a power balance simulation, performed with the NTSS code for the vacuum configuration optimised for both QI and electron root.}
    \label{fig:Er_HeNeo}
\end{figure}

\begin{figure}
     \includegraphics[width=.5\textwidth]{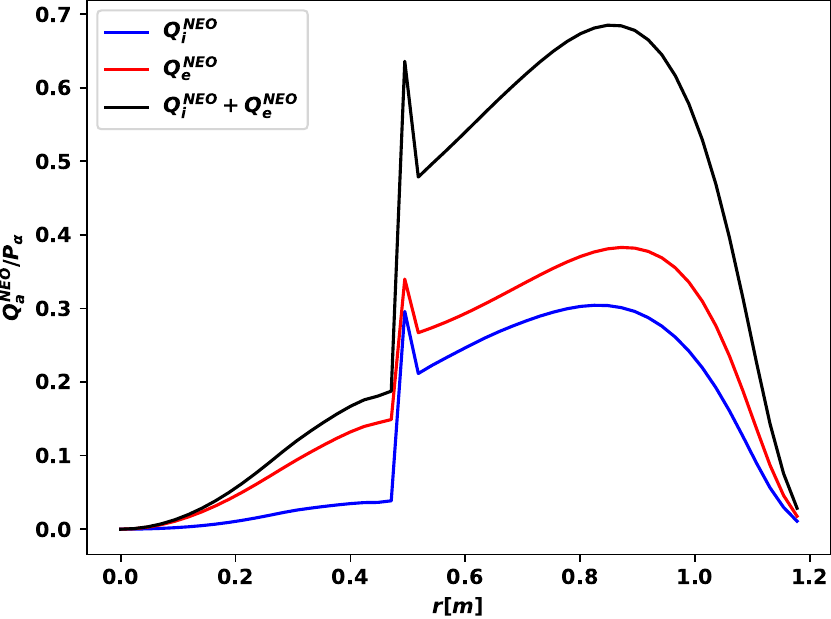}   
     \includegraphics[width=.5\textwidth]{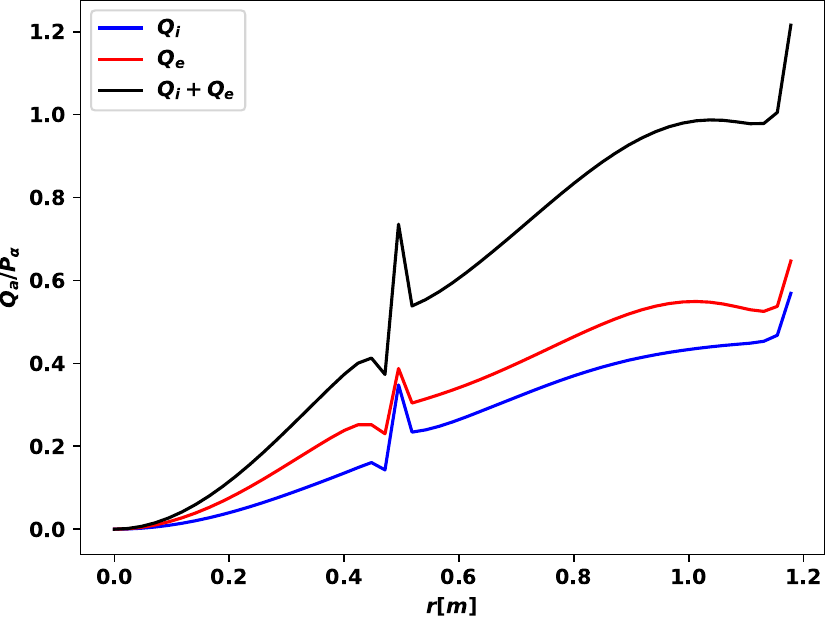}
   \caption{Neoclassical (left) and total (right) energy fluxes obtained for the power balance simulation for the vacuum magnetic configuration optimised for both QI and electron root. The simulation was performed by the NTSS code and the energy fluxes are normalised to the volume-averaged alpha power of $P_{\alpha}=305\,\text{MW}$}
    \label{fig:Heat_HeNeo}
\end{figure}

We now perform a similar simulation for the vacuum configuration optimised for both QI and electron root. We use again the profiles shape given in \cref{profilesT,profilesN} for the initial profiles. In this case, such profiles are scaled so that the initial density on-axis is $n_0=2.74\times 10^{20} \,\text{m}^{-3}$ and the initial temperature on-axis for all species is $T_0=12.46\,\text{keV}$. We obtain an equilibrium plasma with $305\,\text{MW}$ of volume-averaged alpha power after subtracting the Bremsstrahlung losses. An energy confinement time of $\tau_E=1.38\text{s}$ is obtained which is a factor of $1.50$ above the ISS04 scaling. The final density and temperature profiles can be seen in \cref{fig:NT_HeNeo}. The electric field profile obtained is depicted in \cref{fig:Er_HeNeo}, where we observe that an electron root is obtained while achieving a target alpha power close to $300\,\text{MW}$. This electron root has a maximum of $E_r^\text{max}=40.297\,\text{kV/m}$ and the root transition is located at $r/a=0.410$ with $a=1.178\,\text{m}$. We note that such an electron root is obtained for lower temperature and higher density values (and thus higher collisionalities) than the ones obtained in \cref{fig:NT_QI} for the configuration optimised only for QI. This shows the effect of the electron root cost function in helping to achieve an electron root, even when the power balance is considered. The neoclassical and total (neoclassical and anomalous) energy fluxes normalised to the volume-averaged alpha power $P_{\alpha}=305\,\text{MW}$ are shown in \cref{fig:Heat_HeNeo}. We observe that inside the electron root region, there is a large reduction of the neoclassical energy fluxes, as expected. This reduction leads to the sum of ion and electron neoclassical energy fluxes being below $0.2P_{\alpha}$ in the electron root region. We point out that such values of neoclassical energy fluxes are very close to the maximum values of energy fluxes shown in \cref{fig:Heat_QI} for the configuration optimised only for QI. Such observation is quite interesting as it shows that such reduced energy fluxes in the electron root zone are obtainable for this configuration, despite the fact that it has an overall much larger $\varepsilon_\text{eff}$ than the configuration optimised only for QI (see \cref{fig:fast_par}). Additionally, such values of energy flux obtained in the electron root are also below the maximum values observed for W7-X standard and high mirror configurations in Ref. \citet{Beidler2021}. Such a result indicates that if we are able to obtain a configuration with an electric field solution such that an electron root exists and spans over a large region of the plasma minor radius, we may be able to obtain reduced levels of neoclassical energy transport without the need of targeting the $\epsilon_\text{eff}$. The total energy fluxes for this configuration are again below $P_{\alpha}$, except at the boundary. This happens despite the neoclassical energy flux being larger in the ion root region than for the configuration optimised only for QI. This is just a consequence of the $300\,\text{MW}$ solution for the QI-only optimised configuration (see \cref{fig:NT_QI}) being obtained at a smaller density, which makes the anomalous transport larger for the simple anomalous transport model considered. In the root transition, the energy fluxes have a small discontinuity due to the abrupt variation of the electric field. The maximum value of energy flux at the discontinuity is around $0.6P_{\alpha}$ which is still at tolerable levels despite the large value of the epsilon effective seen in \cref{fig:fast_par}. 

\begin{figure}
     \includegraphics[width=.5\textwidth]{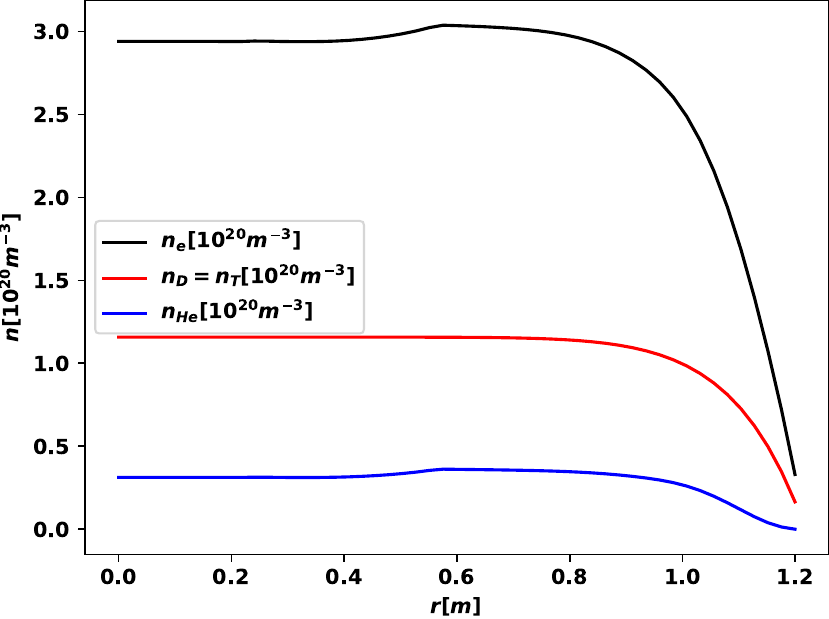}
    \includegraphics[width=.5\textwidth]{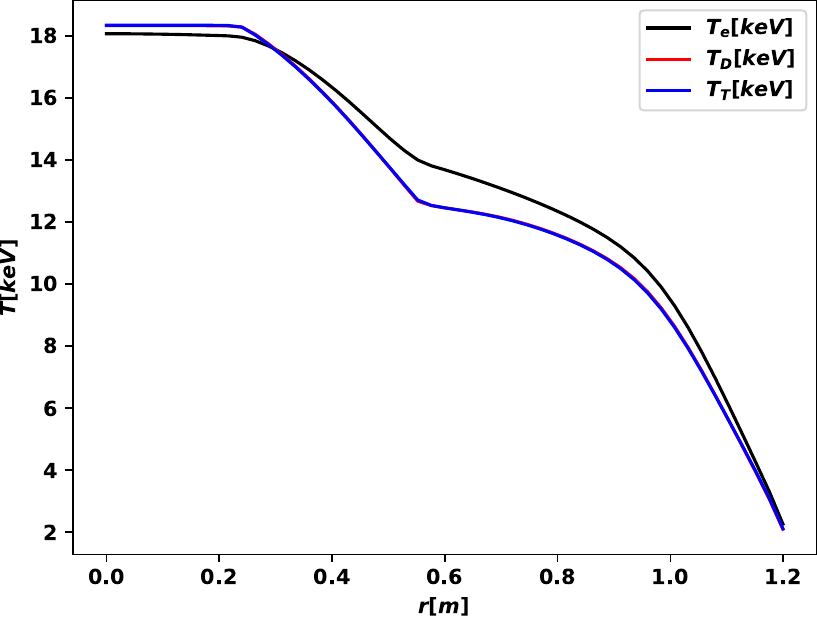}
   \caption{Density and temperature profiles obtained with the NTSS code by performing a power balance simulation for the finite-$\beta$ configuration optimised for both QI and electron root.}
    \label{fig:NT_He_finiteb_300}
\end{figure}

\begin{figure}
    \centering
     \includegraphics[width=.5\textwidth]{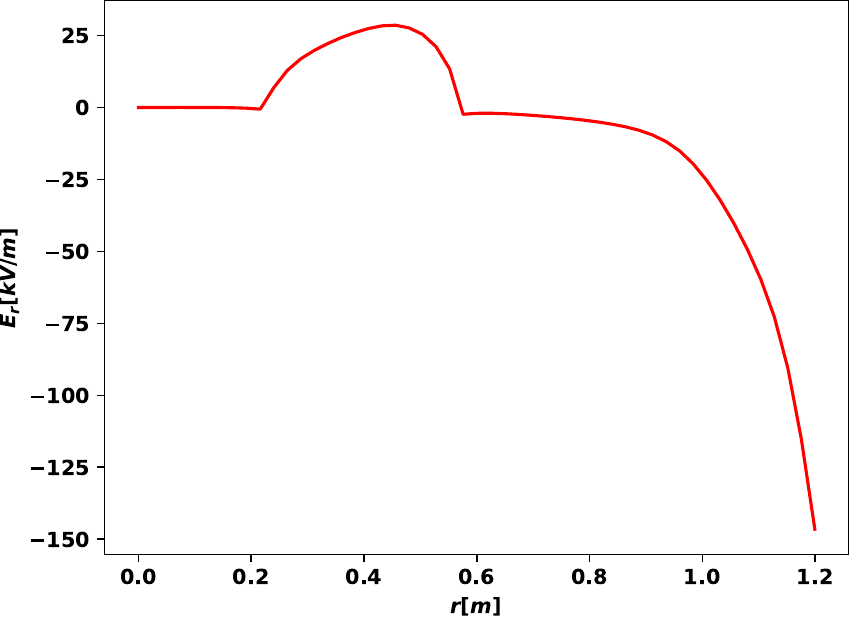}
   \caption{Radial electric field profile obtained with the NTSS code by performing a power balance simulation for the finite-$\beta$ configuration optimised for both QI and electron root.}
    \label{fig:Er_He_finiteb_300}
\end{figure}

\begin{figure}
     \includegraphics[width=.5\textwidth]{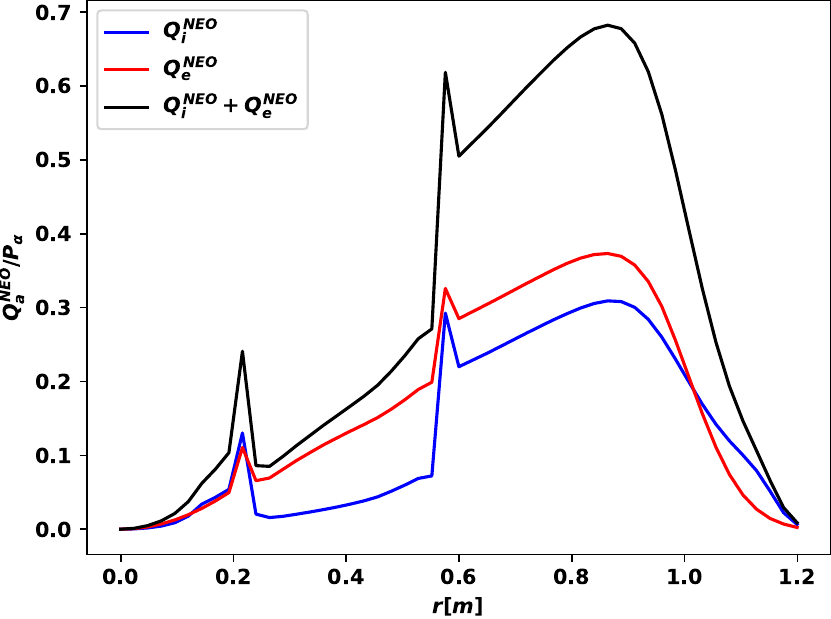}   
     \includegraphics[width=.5\textwidth]{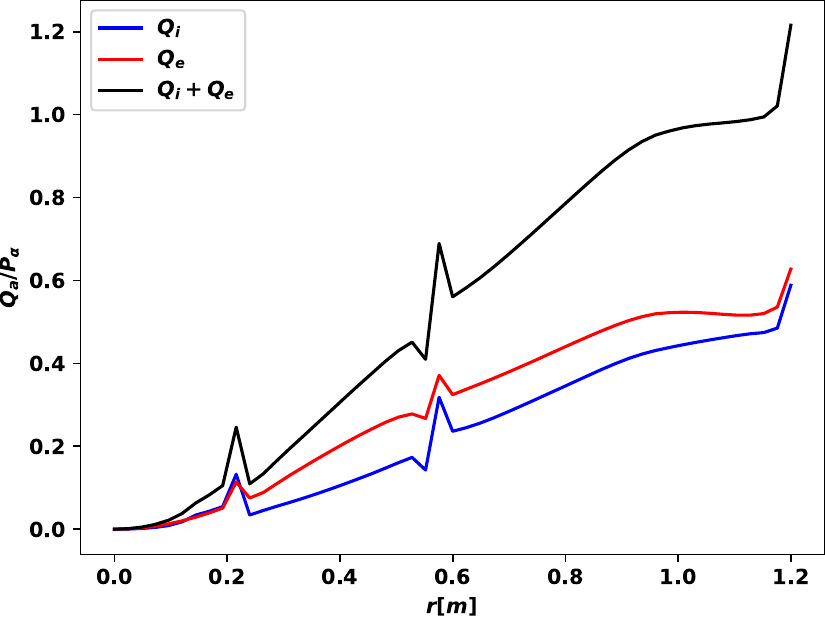}
   \caption{Neoclassical (left) and total (right) energy fluxes obtained for the power balance simulation performed for the finite-$\beta$ magnetic configuration optimised for electron root and QI. The simulation was performed with the NTSS code and the energy fluxes are normalised to the volume-averaged alpha power of $P_{\alpha}\approx 314\,\text{MW}$.}
    \label{fig:Heat_finiteb_300}
\end{figure}

A similar simulation is now performed using the finite-$\beta$ configuration optimised for both QI and electron root. In this case, because we optimised for a specific pressure profile, using density and temperature profiles that do not match such pressure is not fully consistent. However, due to the aspect ratio and volume of the initial configuration, using the density and temperature profiles consistent with optimisation will lead to a very large alpha power that will be unrealistic from the engineering point of view. Moreover, even if NTSS is initialised with consistent temperature and density profiles, it will not lead to final profiles that are consistent with the pressure profile used in the MHD force balance. The most correct approach would be to obtain several optimised configurations for increasing $\beta$ and use them in an equal number of NTSS simulations to mimic a startup phase of the plasma, but such a task is out of the scope of this paper. We thus resort to the same strategy as in the previous vacuum configurations. We use again as initial profiles the density and temperature profiles of \cref{profilesT,profilesN} scaled to an on-axis temperature of $T_0=12.55 \,\text{keV}$ and density of $n_0=2.31\times 10^{20} \,\text{m}^{-3}$. We obtain a volume-averaged alpha power of $P_{\alpha}=314\,\text{MW}$ after subtracting the Bremsstrahlung losses and an energy confinement time of $\tau_E=1.45\,\text{s}$ which is a factor of $1.79$ above the ISS04 scaling. The final temperature and density profiles obtained in this simulation are presented in \cref{fig:NT_He_finiteb_300} and the electric field solution can be seen in \cref{fig:Er_He_finiteb_300}. 
In this case, we see that an electron root region is again present in the core region, but two ion root regions are present. One ion root region is the usual region spanning from the boundary of the plasma until the root transition occurs at $r=0.564 \,\text{m}$. Another one is an ion root living between the magnetic axis and the root transition at $r=0.228 \,\text{m}$. The electron root has a maximum of $E_r^\text{max}=28.577\,\text{kV/m}$. The electron root and ion root regions situated between $r=0.228 \,\text{m}$ and the plasma boundary are similar to the ones observed in \cref{fig:Er_HeNeo}. The ion root zone situated between the magnetic axis and $r=0.228 \,\text{m}$ has very small (almost residual) electric field values and is related to a flattening of the temperature and electron density profiles in this region (see \cref{fig:NT_He_finiteb_300}). In particular, due to the flattening of such profiles in the first ion root region, there is a hollowing of the electron density profile (see \cref{fig:NT_He_finiteb_300}) with a positive electron density gradient occurring in the region of the electron root situated between $r=0.228 \,\text{m}$  and $r=0.564 \,\text{m}$. The final neoclassical and total (neoclassical and anomalous) energy fluxes normalised to the volume-averaged alpha power $P_{\alpha}=314\,\text{MW}$ can be seen in  \cref{fig:Heat_finiteb_300}. Two discontinuity regions can be seen in the energy flux solutions at the location of the two root transitions situated at $r=0.228 \,\text{m}$  and $r=0.564 \,\text{m}$ as expected. 
We observe again that in the electron root regions, the neoclassical energy fluxes are reduced with this case showing overall values of the sum of electron and ion neoclassical energy fluxes below $0.25P_{\alpha}$ in the electron root regions. In the root transition situated at $r=0.564 \,\text{m}$, the sum of ion and electron neoclassical energy fluxes reaches a value of around $0.6P_{\alpha}$ as seen for the previous case of the vacuum configuration optimised for QI and electron root (see \cref{fig:Heat_HeNeo}). The total energy fluxes have a similar behaviour as in the previous simulation staying below $P_{\alpha}$ except at the boundary. 

\subsection{Impurity transport properties}
\begin{figure}
     \includegraphics[width=.5\textwidth]{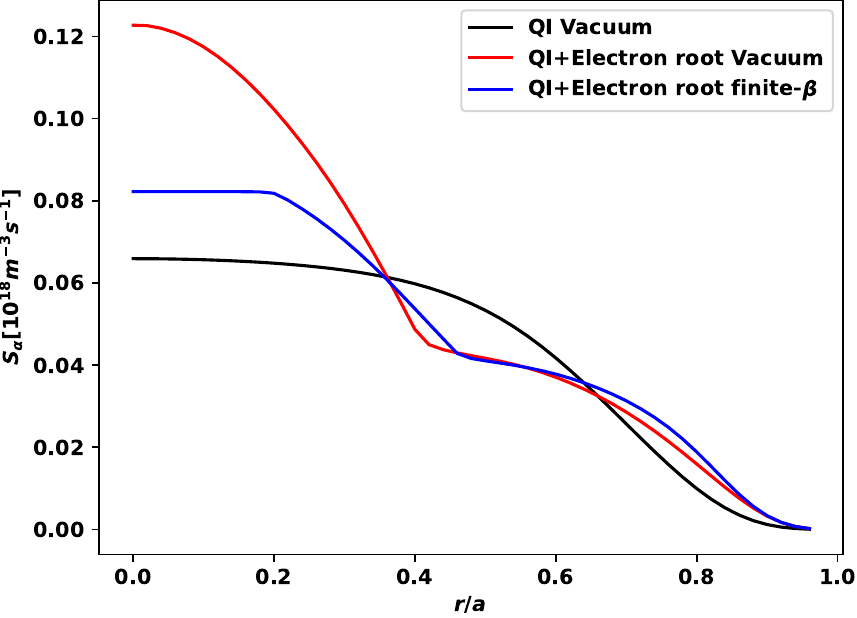}
    \includegraphics[width=.5\textwidth]{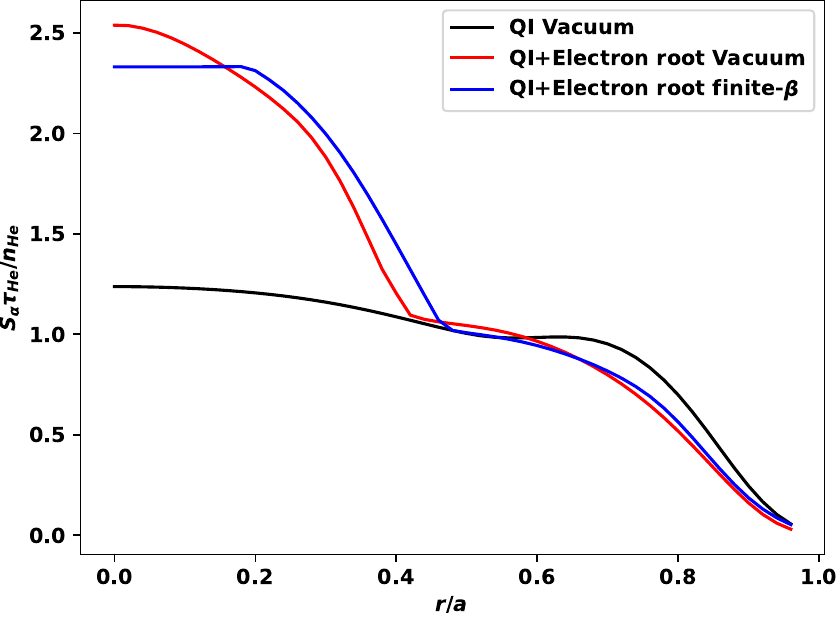}
   \caption{Source of fusion $\alpha$ particles (left) and the ratio $S_{\alpha}\tau_{He}/n_{He}$ (right) for the configuration optimised only for QI, and the vacuum and finite-$\beta$ configurations optimised for both QI and electron root. Such quantities are obtained from the power balance simulations performed with NTSS.}
    \label{fig:Helium}
\end{figure}

In this section, we discuss briefly the properties of impurities. Two types of impurities are of concern for future stellarator reactors, helium ash and heavy tungsten impurities. We compare the properties of these two species for the configuration optimised only for QI and the vacuum and finite-$\beta$ configurations optimised for both QI and electron root. To analyse the helium ash, we follow Ref. \citet{Beidler2024} and use the ratio $S_{\alpha}\tau_{He}/n_{He}$ to quantify if the helium ash is transported effectively outwards. Here the source of helium ash $S_{\alpha}$ is taken to be the source of fusion alpha particles and the helium ash confinement time is calculated as $\tau_{He}=\int_0^1 d\rho n_{He}\rho / \int_0^1 d\rho S_{\alpha} \rho$. Such quantities can be obtained from the power balance simulations performed in the previous section. As mentioned in Ref. \citet{Beidler2024}, we expect that, if the quantity $S_{\alpha}\tau_{He}/n_{He}$ is well above one, the helium ash will be transported more effectively outwards. The quantities $S_{\alpha}$ and $S_{\alpha}\tau_{He}/n_{He}$ are shown in \cref{fig:Helium} for the three configurations of interest. We can see that in the electron root region, $S_{\alpha}\tau_{He}/n_{He}$ is considerably larger for both the vacuum and finite-$\beta$ configurations optimised for both QI and electron root when compared to the configuration optimised only for QI. We can see in \cref{fig:Helium}, that in the configurations optimised for both QI and electron root the quantity, $S_{\alpha}\tau_{He}/n_{He}$ can reach values greater than two. Such an increase of $S_{\alpha}\tau_{He}/n_{He}$ in those configurations, happens both in zones, which show a value $S_{\alpha}$ greater and lower than in the configuration optimised only for QI. Such observations indicate that the electron root region obtained for the QI and electron root optimised configurations is beneficial with respect to the exhaust of helium ash. 

\begin{figure}
    \centering
     \includegraphics[width=.5\textwidth]{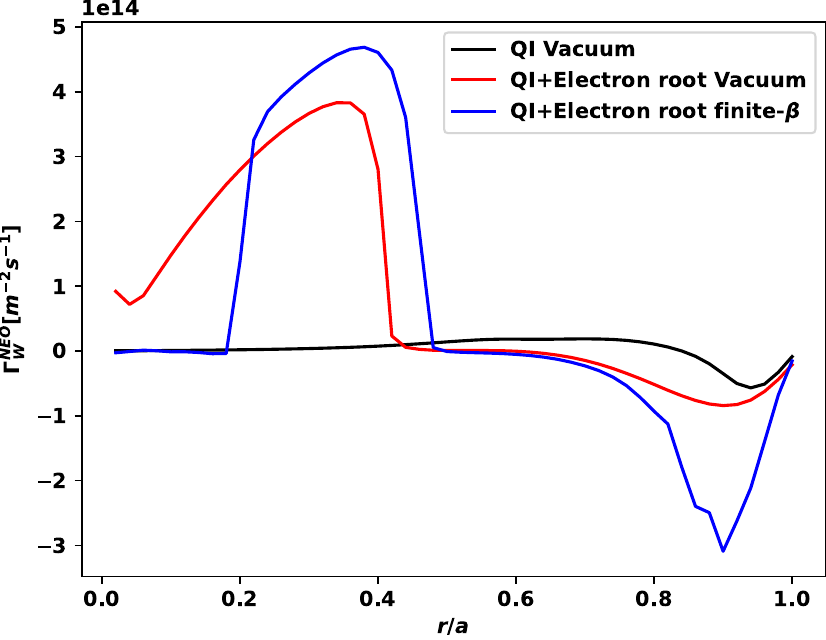}
   \caption{Neoclassical particle-flux-density of tungsten obtained from a SFINCS simulation with tungsten, deuterium, tritium and electrons. The particle-flux-density is shown for the configuration optimised only for QI and both the vacuum and finite-$\beta$ configurations are optimised for QI and electron root. A full Fokker-Planck collision operator is used for the four species and the profiles obtained in the power balance simulations are used.}
    \label{fig:tungsten_ii}
\end{figure}

We now look into the properties of heavy tungsten particles, which we take here to be in a single ionisation state with charge $Z_W=40$ and a mass number of $184.840$. We also consider $T_W=T_i=(T_D+T_T)/2$ so that the tungsten temperature is such that $T_W\sim T_i$. To keep this treatment simple we assume that the tungsten is a trace species and its density is such that $n_W=10^{-5}n_i$. This assumption allows $Z_W^2n_W/n_i\sim 10^{-2}\ll 1$ so that tungsten will not have an influence in the background species either through modification of the ambipolarity constraint or by collisions. This consideration is in line with the "safe" concentration limit of tungsten seen in tokamak experiments \citep{Putterich2013} and it should serve here to elucidate about tungsten physics before any accumulation or peaked density profiles are observed. Such an assumption also allows the usage of the profiles obtained from the previous power balance simulations. We use such profiles in SFINCS to solve the drift-kinetic equation in the form \cref{DK}, considering four species, i.e., tungsten, deuterium, tritium and electrons, with a full linearised Fokker-Planck collision operator. Such an approach to solve the drift kinetic equations is necessary here because tungsten is expected to be highly collisional, due to its high charge, and its particle-flux-density is expected to be affected by collisions with the main thermal ions. The tungsten particle-flux-density for the QI-only optimised configuration and the vacuum and finite-$\beta$ configurations optimised for both QI and electron root can be seen in \cref{fig:tungsten_ii}. We observe that the tungsten particle-flux-density for the configuration optimised only for QI has a region of positive values despite having an ion root (see \cref{fig:Er_HeatQI}). This is due to a combination of the screening provided by the temperature gradient and the fact that the values of the electric field, in such region of the ion root solution of this configuration, are small. However, the tungsten particle-flux-density for this configuration is overall around one order of magnitude smaller than the particle-flux-densities for the other configurations. We see that for both the vacuum and finite-$\beta$ configurations optimised for QI and electron root, there is a large value of the tungsten particle-flux-density in the electron root region (see \cref{fig:Er_HeNeo} and \cref{fig:Er_He_finiteb_300}). Therefore, as expected such configurations show stronger beneficial effects on the removal of tungsten particles from the plasma core. We end this section by mentioning that though small the positive values of the QI configuration could provide for this plasma equilibrium an outward flux of tungsten. Nevertheless, such an effect is weaker than for the other configurations in the electron root region and is dependent on the temperature profiles achieved. The large tungsten particle-flux-densities observed in the electron root region for the configurations optimised for electron root indicate that in such conditions the beneficial outward impurity flux should be present for most temperature profiles. 

\subsection{Coil configurations}
\begin{figure}
    \centering
     \includegraphics[width=.5\textwidth,trim={1cm 3.5cm 0cm 0cm},clip]{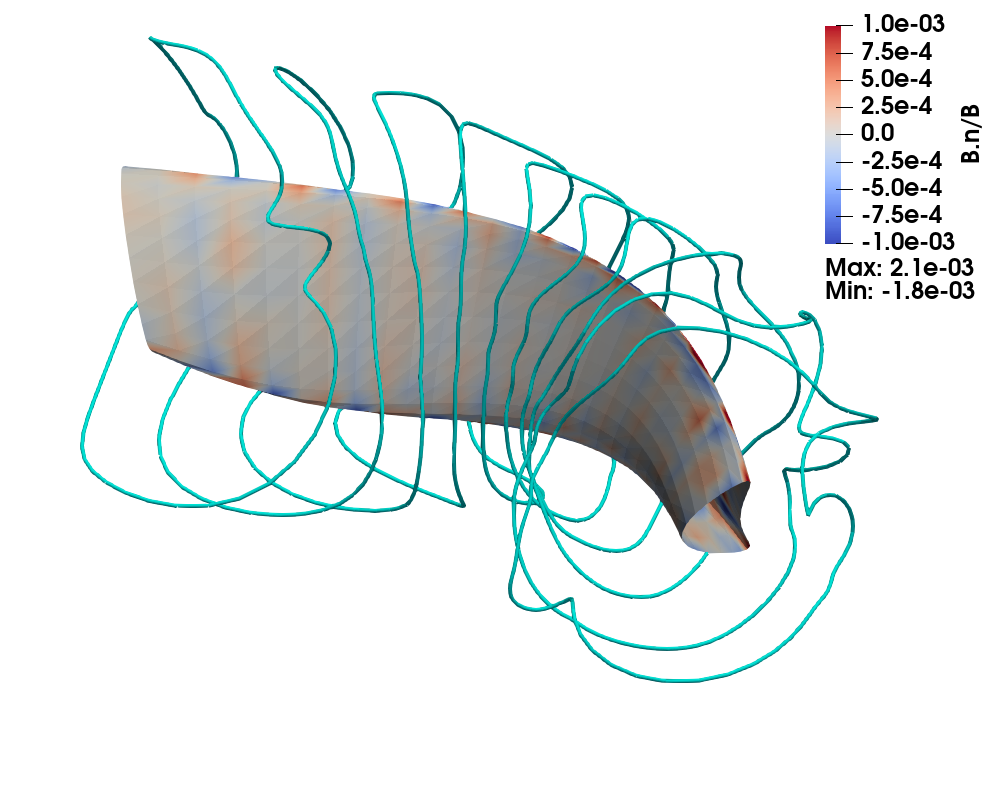}
   \caption{Optimised coils obtained for half of a field period for the vacuum configuration optimised for both QI and electron root. A surface plot of the plasma boundary representing the normal magnetic field at the boundary normalised to the magnetic field is also shown.}
    \label{fig:coils_vacuum}
\end{figure}

For completeness, we show two sets of coils obtained for the QI and electron root optimised configurations. One set of coils is obtained for the vacuum configuration and another for the finite-$\beta$ configuration. Both of these were obtained with the SIMSOPT code using a second-stage optimisation approach. In the finite-$\beta$ configuration, a virtual casing principle is used to separate the magnetic field due to the coils from the plasma contribution. The coils obtained for the vacuum configuration case together with the plasma boundary are presented in \cref{fig:coils_vacuum} for half of a field period. The surface contour plot of the boundary represents the residual normal magnetic field at the boundary of the plasma (normalised to the magnetic field) which should vanish when the coil field matches the target field from the optimisation. The set of coils is composed of four independent sets of ten coils, each one representing half of a field period, due to the constraints given by assuming stellarator symmetry and $n_\text{fp}=2$. 
\begin{figure}
    \centering
    \includegraphics[width=.25\textwidth,trim={5.1cm 6.8cm 12.5cm 0.4cm},clip]{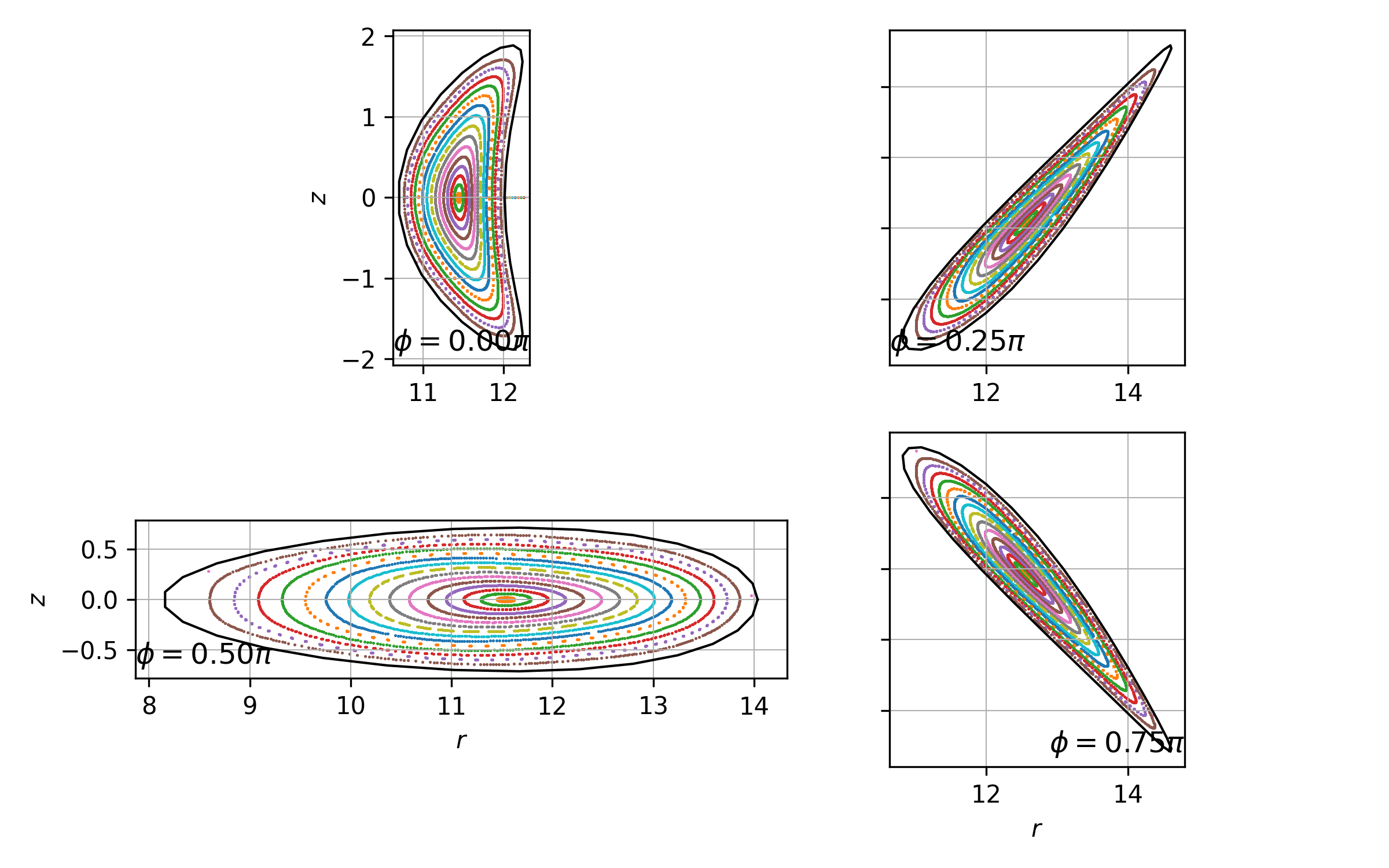}
\caption{Poloidal cross-section at $\phi=0$ of the 3D Poincaré plots for the optimised set of coils obtained for the vacuum configuration optimised for both QI and electron root.}    
    \label{fig:poincare_vacuum}
\end{figure}
They have a minimum separation between them of 0.1 m, and each has a total length of 38.5, 36.9, 28.2, 26.7, 28.2, 31.4, 34.3, 38.2, 37.2 and 40.2 m, respectively.
They have a maximum curvature of 1.9 m$^{-1}$ and a maximum mean squared curvature of 0.6 m$^{-1}$.
One of the coils has a fixed current of $10^6$A, and the remaining coils have a maximum current of $1.31 \times 10^6$ A and a minimum current of $5.6 \times 10^5$ A.
The poloidal cross-section, at toroidal angle $\phi=0$, of the respective Poincaré plots is shown in \cref{fig:poincare_vacuum} that demonstrates that such coils can accurately reproduce the optimised fixed boundary equilibrium.

\begin{figure}
    \centering
     \includegraphics[width=.5\textwidth,trim={1cm 6.1cm 0cm 0cm},clip]{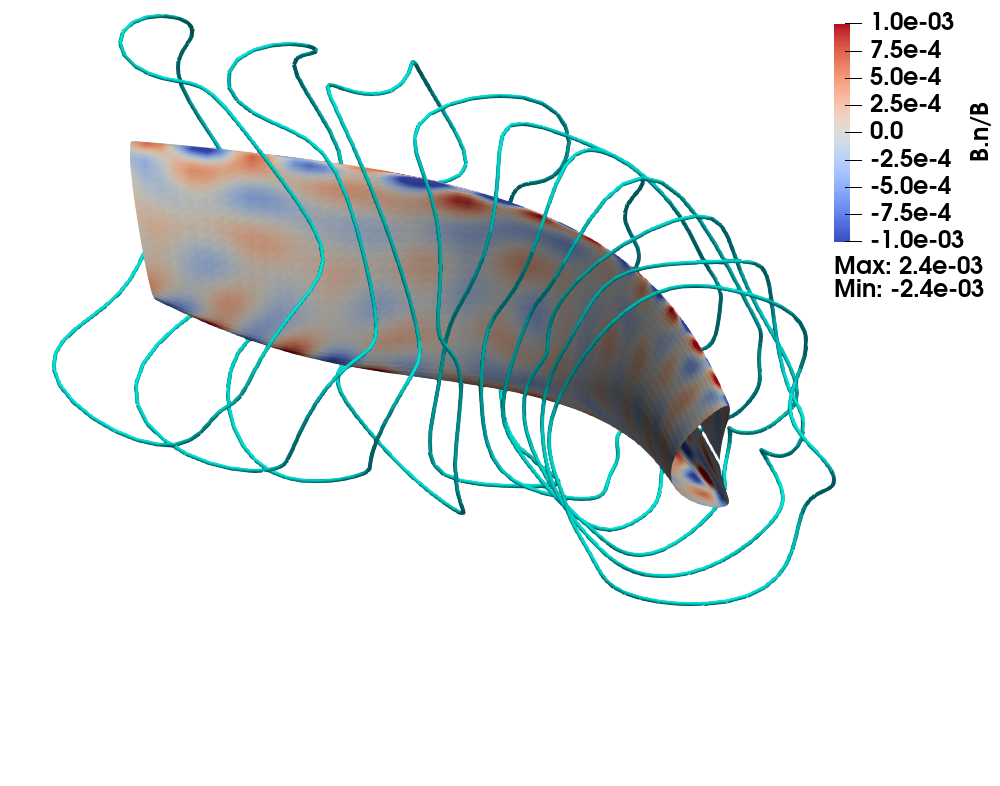}
   \caption{Optimised coils obtained for the finite-$\beta$ for half of the field period of the configuration optimised for both QI and electron root. A surface plot of the plasma boundary representing the normal magnetic field at the boundary normalised to the magnetic field is also shown.}
    \label{fig:coils_beta}
\end{figure}

The set of optimised coils for half of a field period obtained for the finite-$\beta$ configuration optimised for QI and electron root can be seen in \cref{fig:coils_beta}. This set of coils is again one of the four equal sets of ten different coils that compose the total array of coils for the entire plasma.
They have a minimum separation between them of 0.26 m, and each has a total length of 25.6, 26.7, 27.2, 28.1, 32.2, 37.3, 39.1, 41.5, 39.5 and 38.2 m, respectively.
They have a maximum curvature of 1.7 m$^{-1}$ and a maximum mean squared curvature of 0.5 m$^{-1}$.
The coils have a fixed total sum of currents of $1.42 \times 10^8$A with each current of approximately 10\% of the total current.
The surface contour plot of the magnetic field normal to the plasma boundary normalised to the magnetic field is also shown in \cref{fig:coils_beta} for half of a field period.
The normal magnetic field shown is comprised of the coil magnetic field computed using the Biot-Savart law minus the target magnetic field obtained using the virtual casing principle.
No Poincaré plots are presented for the finite-$\beta$ configuration because they would only represent the vacuum contribution of the magnetic field. To consider both the vacuum and the plasma contributions a finite-$\beta$ optimisation considering the free boundary solution of the MHD force balance is required. However, such a simulation is out of the scope of this paper. 

\subsection{Intuitive picture of electron root optimisation}
To obtain an intuitive picture of what is being achieved by the electron root cost function during optimisation, we may look into the behaviour of the monoenergetic radial transport scans of the configurations which are optimised with the electron root cost function and into the cost function itself. There are two ways the ratio $L_{11}^i/L_{11}^e$ can be minimised. One is by increasing the radial transport coefficient of the electrons and another one is by decreasing the ion radial transport coefficient at the values of collisionality of interest. Ideally, we expect the optimisation to achieve an electron root by targeting both ways of minimising the cost function. From the monoenergetic radial transport coefficient scans in \cref{fig:D11_Er,fig:D11_QI_Er,fig:D11_finiteb}, we find that two trends appear. The first is the increasing of the $1/\nu^*$ regime and the second is the existence of two well-defined regimes at low collisionality and finite electric field. These are the $1/\nu^*$ and $\nu^*$ regimes, as a consequence of the disappearance of the $\sqrt{\nu^*}$ regime. The electron transport increase is related to the the increase in the $1/\nu^*$ regime, in which this species usually lives due to its large thermal velocity. Therefore, the disappearance of the $\sqrt{\nu^*}$ must be related to the ion species' role in minimising the electron root cost function. To better understand these trends we can look at how the distribution function solution is obtained for these regimes. 
Following Ref. \citet{Shaing2015}, the contribution to the non-ambipolar radial transport is completely described by the solution of the bounce-averaged drift kinetic equation. If we write the drift equation in terms of the variables $(\psi,\alpha,\phi)$, and neglect the mirror term (which is small at large aspect ratios) the local bounce-averaged drift kinetic equation can be written as 
\begin{equation} \label{DK_alpha}
\langle \bm{v_d}\cdot \nabla \alpha\rangle_B\frac{\partial f_{1}}{\partial \alpha}+\langle \bm{v_d}\cdot \nabla \psi\rangle_B\frac{\partial f_{M}}{\partial \psi}=\langle C(f_{1})\rangle_B,
\end{equation} 
where the bounce average of a quantity $X$ is defined as $\langle X\rangle_B =\int^{\phi_B}_{-\phi_B}\frac{B X d \phi}{v_{\parallel}}/\int^{\phi_B}_{-\phi_B}Bd \phi/v_{\parallel}$. For large aspect ratio stellarators, the most important drift across field lines is the poloidal $E\times B$ drift. In the $1/\nu^*$ regime, the poloidal $E\times B$ drift frequency is smaller than the collision frequency and the bounce-averaged drift kinetic equation to be solved is
\begin{equation} \label{DK_1nu}
\langle \bm{v_d}\cdot \nabla \psi\rangle_B\frac{\partial f_{M}}{\partial \psi}=\langle C(f_{1})\rangle_B,
\end{equation} 
which represents a balance between the bounce-averaged drift and the collisions. The collisions limit the radial excursions of the helical trapped particles. Thus, if the collision frequency increases, the radial excursion decreases and we have a $1/\nu^*$ scaling for the radial transport coefficient. The electrons with their large thermal velocity are usually well represented by this regime.  
At lower collisionalities, the collisions get weaker and the $E\times B$ drift becomes important in reducing the radial excursion resulting from the radial drift. The equation to solve is then 
\begin{equation} \label{DK_nu}
\langle \bm{v_d}\cdot \nabla \alpha\rangle_B\frac{\partial f_{1}}{\partial \alpha}+\langle \bm{v_d}\cdot \nabla \psi\rangle_B\frac{\partial f_{M}}{\partial \psi}=0,
\end{equation} 
which is the balance between the poloidal and radial drifts. Ions feel this effect before the electrons because their thermal velocity is smaller. Therefore, their radial excursion is smaller and easier to limit by the poloidal drift. As seen in Ref. \citet{Rodriguez2023}, the bounce-averaged drift can be weighted by the magnetic quantity  
\begin{equation} \label{bouncedrift}
\Delta_{Y}=Y(\phi_B)-Y(-\phi_B)=0, \hspace{5mm} Y=\frac{\nabla \psi \times B \cdot \nabla B}{B\cdot \nabla B}.
\end{equation} 
We see that this quantity can diverge at the maxima and minima of $B$. This is explained by the fact that, at the maxima, barely trapped particles can reach their bounce tips at $B_\text{max}$, but do so, by taking a long time to get from the bottom of the well to the top. Such time approaches infinity as they approach the orbit tip, which leads to the divergence in \cref{bouncedrift}. The solution of \cref{DK_nu} which is given  by 
\begin{equation} \label{f_nu}
\frac{\partial f_{1}}{\partial \alpha}=-\frac{\langle \bm{v_d}\cdot \nabla \psi\rangle_B\frac{\partial f_{M}}{\partial \psi}}{\langle \bm{v_d}\cdot \nabla \alpha\rangle_B},
\end{equation} 
is thus only well-defined for particles which are not close to the trapping/passing boundary. For these particles, the radial transport is related to the different total drift they feel when moving in the magnetic wells. These different values of the drifts allow trapped particles to be de-trapped and then trapped again without collisions. Such a regime leads to a radial transport that scales linearly with $\nu^*$ and decreases with the electric field. This regime is not to be confused with the banana-like regime in which the radial transport also scales with $\nu^*$, but is independent of the electric field. The banana regime also exists in stellarators but does not contribute to the solution of the bounce-averaged drift kinetic equation, because it generates an intrinsically ambipolar radial transport. 
The resonance of \cref{DK_nu} for the barely trapped particles close to the passing/trapped boundary is resolved by adding the collision term to \cref{DK_nu}. This means that these particles are so barely trapped that even a very weak collision frequency is enough to de-trap them. This resonance layer physics is well known and leads to the $\sqrt{\nu^*}$ regime. Because it has a resonant behaviour, the part of the distribution function that represents the solution of the barely trapped particles is usually the most important contribution to the total solution of the distribution, unless the magnetic configuration is such that the bounce-averaged drift is kept finite and small near $B_\text{max}$. The vanishing of this $\sqrt{\nu^*}$ regime observed in \cref{fig:D11_Er,fig:D11_QI_Er,fig:D11_finiteb} indicates that the electron root cost function is minimised by keeping the bounce-averaged drift contribution for the barely trapped ions small. 

Therefore, the cost function for the electron root is a balance of minimising the radial transport of barely trapped ions and maximising the radial transport of the deeply trapped electrons. We can obtain, for the different magnetic configurations presented in this work, approximated measures of such quantities and see if a relation is seen between them and the electron root optimisation. To do so, we measure the bounce-averaged radial drift using the magnetic quantity in \cref{bouncedrift}. We calculate such quantity for the consecutive maxima, $\Delta_{Y_\text{max}}$ or minima, $\Delta_{Y_\text{min}}$, present in the magnetic field spectrum of the configuration between $0$ and $10\pi$. We then perform an average over the absolute value of such quantity for all the pairs of maxima or minima, respectively. Because different field lines can have distinct magnetic wells, we also perform an average over different field lines, in which we consider $51$ values of the field line label $\alpha$ uniformly distributed between $0$ and $2\pi$. The final result is shown for the different configurations in \cref{fig:intuitive}. The ratio $\Delta_{Y_\text{max}}/\Delta_{Y_\text{min}}$, which represents the relative strength of the bounce averaged drift of barely trapped particles to the bounce averaged radial drift of the deeply trapped particles, is also shown in \cref{fig:intuitive}. We see that configurations that do not have an electron root for the density and temperature profiles used during optimisation, i.e.\, the initial configuration and the configuration only optimised for QI (see \cref{fig:Er_initial,fig:Er_QI}), have a ratio $\Delta_{Y_\text{max}}/\Delta_{Y_\text{min}}$ close or larger than $1$. On the other hand, configurations that are optimised for an electron root, have a value of the ratio $\Delta_{Y_\text{max}}/\Delta_{Y_\text{min}}$ smaller than $0.5$. In fact, the vacuum configuration optimised for both QI and electron root has the lower value of $\Delta_{Y_\text{max}}/\Delta_{Y_\text{min}}$ and also shows the largest electron root for the density and temperature profiles used during optimisation. Therefore, it is seen that the ratio $\Delta_{Y_\text{max}}/\Delta_{Y_\text{min}}$ has similar behaviour to the electron root cost function, which then justifies the intuitive picture of the electron root optimisation presented.
\begin{table}
\centering
\def\arraystretch{1.5}
\begin{tabular}{ |l||c|c|c|  }
  \hline
  \noalign{\vskip -0.085in} 
 Configuration & $\Delta_{Y_\text{max}}$ & $\Delta_{Y_\text{min}}$ & $\Delta_{Y_\text{max}}/\Delta_{Y_\text{min}}$ \\[-1.5ex]
\hline
\noalign{\vskip -0.085in} 
 Initial& $1683$ & $1765$ & $0.954$ \\
 Only QI&   $845$ & $559$   & $1.512$\\
 Only Er &$303$ & 656&  $0.462$\\
 QI+Er (vacuum)& $2467$   & $10899$ & $0.226$\\
 QI+Er (finite-$\beta$)&  $1021$  & $2328$& $0.439$\\
 [-1.5ex]\hline
\end{tabular}
\caption{Average of the quantity in \cref{bouncedrift} between consecutive maxima ($\Delta_{Y_\text{max}}$) and minima ($\Delta_{Y_\text{min}}$) over $51$ field lines. The ratio $\Delta_{Y_\text{max}}/\Delta_{Y_\text{min}}$ is also shown. Such quantities are presented for the different optimised magnetic configurations obtained in this work.}
\label{fig:intuitive}
\end{table}

\section{Conclusion}
In this work, we propose a new optimisation method to obtain magnetic configurations that favour an electron root solution for the radial electric field. We motivate the cost function and the optimisation method using neoclassical transport theory and describe a method of testing the efficacy of the cost function. Using both methods, we show that the electron root optimisation leads to magnetic configurations which favour an electron root in the plasma. We obtain a vacuum magnetic configuration optimised for both QI and electron root. This magnetic configuration has an electron root solution at a wide range of reactor-relevant density and temperature values. We show that it is possible to achieve an electron root in such configurations for strong values of magnetic field strength on-axis, such as $B_0=8 \,\text{T}$ and above. In fact, it is seen that such a configuration is robust to the presence of an electron root solution of the electric field for several values of $n_0$, $T_0$, $R_0$ and $B_0$, without the need to rely on tweaking the shape of the profiles. Such electric field solutions are obtained for the reactor-relevant case of $T_e=T_i$, contrary to the case of current experimental devices, like W7-X, where such electron roots are obtained by decoupling electron from ion temperatures and reaching very peaked electron temperatures with the use of ECRH. A finite-$\beta$ configuration optimised for both QI and electron root was also obtained. It was shown that such a configuration has an electron root at the temperature and density profiles consistent with the ones targeted during optimisation, showing that such solutions can also be pursued when considering finite-$\beta$ effects. 

By performing power balance simulations we conclude that the vacuum and finite-$\beta$ configurations optimised for both QI and electron root, show electron root solutions for a volume-averaged alpha power target of $300\text{MW}$. Such solutions indicate that the maximum value of neoclassical energy fluxes in the electron root region can be similar to the maximum values of neoclassical energy fluxes observed in the configuration optimised only for QI and even lower than the maximum of the neoclassical energy fluxes observed in Ref. \citet{Beidler2021} for W7-X configurations. Such a result is observed, despite the fact that both W7-X and the configuration optimised only for QI have much lower epsilon effective values than the ones optimised for both QI and electron root. Such observation indicates that if the configurations are optimised for an electron root to be present in a large extension of the plasma radius, the $\varepsilon_\text{eff}$ may not be necessary as an optimisation metric, which may allow for a less restrictive parameter space during multi-objective optimisations. 

Concerning impurities, evidence of beneficial effects for the exhaust of helium ash and the reduction of core tungsten accumulation is observed for the solutions of these power balance simulations performed for the vacuum and finite-$\beta$ configurations optimised for both QI and electron root. Additionally, the large electric field gradients observed in the root transitions of these solutions are expected to generate large $E\times B$ shear flows that may play a role in reducing turbulence. On a side note, we remark here that if such a turbulence reduction is verified in large root transitions, it would mean that an ion and electron temperature decoupling would cease to exist, and thus, an electron root obtained via this decoupling would also cease to exist. Such argument further motivates the optimisations done in this work, in which configurations with robust electron root solutions at $T_e=T_i$ are achieved. 
We point out that varying the parameters $n_0$, $T_0$, $a_0$ and $B_0$ was seen to have an impact on the electron root maximum and root transition position. Thus, further tuning of such parameter space can be used to enhance the beneficial effects of the electron root on impurities and (possibly) turbulence. Both the impurity behaviour and the possible turbulence reduction can greatly affect plasma performance, thus making electron root solutions worth studying as an operation scenario for future stellarator fusion reactors.  

Future paths for this work include doing multi-objective optimisations, by adding other cost functions to \cref{cost1,cost_neo}. Targets such as MHD stability, fast particle confinement, bootstrap current or turbulence could be considered. Such work would ideally be performed together with the analysis of the different trade-offs between the electron root cost function and the other cost functions. Another important subject is the study of the impact of these new optimised electron root configurations on heavy impurities such as tungsten. Though here we have shown the beneficial effects of the electron root solutions obtained on the improvement of helium ash exhaust and the reduction of tungsten accumulation, further impurity simulations should be performed considering the effect of tungsten density asymmetries. Such simulations can be used to analyse the effect of different electron root parameters, such as the maximum and root transition location,  on heavy impurity transport. Ideally, this would result in optimal points in plasma and configuration parameters space for helium ash exhaust and tungsten reduction. Finally, a study on how the electron root solutions obtained with these optimisations can impact turbulence should be performed. A study of the impact of different electron root parameters, such as root transition steepness, location and width should be studied to find optimal values of the plasma and configuration space for the possible reduction of turbulence. As a final remark, an important question concerns whether we can optimise directly for such parameters and if we can define the limits to obtain a reduction of turbulence. Regarding such a topic, we point out that the gradient length of the root transition should ideally stay within an ion gyro-radius to have a strong effect in reducing turbulence.

\section*{Acknowledgments}
We thank M. Landreman, H. M. Smith, A. G. Goodman and B. F. Lee for helpful discussions and the SIMSOPT, DKES, SFINCS and NTSS teams for their valuable contributions.
This work has been carried out by funding from Proxima Fusion.
Views and opinions expressed are however those of the author(s) only and do not necessarily reflect those of the funding entity.
R. J. is supported by the Portuguese FCT-Fundação para a Ciência e Tecnologia, under Grant 2021.02213.CEECIND and DOI  \href{https://doi.org/10.54499/2021.02213.CEECIND/CP1651/CT0004}{10.54499/2021.02213.CEECIND/CP1651/CT0004}.
IPFN activities were supported by FCT - Fundação para a Ciência e Tecnologia, I.P. by project reference UIDB/50010/2020 and DOI  \href{https://doi.org/10.54499/UIDB/50010/2020}{10.54499/UIDB/50010/2020}, by project reference UIDP/50010/2020 and DOI \href{https://doi.org/10.54499/UIDP/50010/2020}{10.54499/UIDP/50010/2020} and by project reference LA/P/0061/202 and  DOI \href{https://doi.org/10.54499/LA/P/0061/2020}{10.54499/LA/P/0061/2020}.

\appendix
\section{DKES-SFINCS normalisations} \label{appendix}
The DKES-like database used as an input in the NTSS code, assumes the DKES code to use as input parameters, the inverse mean free path $\nu/v$ and the normalised electric field $E_{\tilde{r}}/(vB_0)$ where typically one sets $B_0=1$. The coordinate $\tilde{r}$ in DKES is defined so that 
\begin{equation} \label{Psi_r}
\frac{d \psi}{d \tilde{r}}=rB_0, \hspace{10mm} \frac{d \tilde{r}}{d r}=\frac{2 \psi_B}{a^2B_0},
\end{equation}
with the minor radius $a=\sqrt{V(\psi_B)/(2\pi^2R_0(\psi_B))}$. Using these definitions the SFINCS monoenergetic input variables $\nu^{\prime}$ and $E_{\psi}^*$ can be written in terms of the DKES inputs as 
\begin{align*} \label{nu_prime}
\nu^{\prime}=\frac{G+\iota I}{B_0}\left[\frac{3\sqrt{\pi}}{4}\left(\left.\text{erf}(x=1)-\frac{\text{erf}(x=1)}{2x^2}+\frac{2}{2x}\frac{d [\text{erf}(x)]}{dx}\right|_{x=1}\right)\right]^{-1}\frac{\nu}{v} \numberthis
\end{align*}
and
\begin{equation} \label{Estar}
E_{\psi}^*=-\frac{G}{\iota}\left(\frac{d\psi}{d\tilde{r}}\right)^{-1}\frac{E_{\tilde{r}}}{v},
\end{equation}
respectively. In \cref{nu_prime}, $\text{erf}(x)=2/\sqrt{\pi}\int_0^x \exp (-y^2) dy$ is the error function, and the terms containing such function come from the definition of the deflection frequency in Ref. \citet{HelanderSigmar}. The radial transport coefficient as per the DKES-like database files is written in terms of the SFINCS monoenergetic radial coefficient $D_{11}^S$ as 
\begin{equation} \label{GammaHat}
\hat{\Gamma}_{11}=\frac{\sqrt{\pi}G^2 B_0}{8(G+\iota I)\Big{(}\frac{d \psi}{d \tilde{r}}\Big{)}^2}D_{11}^S.
\end{equation}
In Ref. \citet{Beidler2011} the normalised monoenergetic radial coefficient is defined as $D_{11}^*=D_{11}/D_{{11}_p}$, with $D_{{11}_p}$ the value of $D_{11}$ in the plateau regime of an equivalent tokamak. This quantity can be defined in terms of $D_{11}^S$ as
\begin{equation} \label{D11star}
D_{11}^*=-\frac{\iota R_0 G^2 B_0^3\Big{(}\frac{d \tilde{r}}{d r}\Big{)}^2}{\sqrt{\pi} (G+\iota I)\Big{(}\frac{d \psi}{d \tilde{r}}\Big{)}^2}D_{11}^S.
\end{equation}

\bibliographystyle{jpp}
\bibliography{references}
\end{document}

%% file: preamble.tex
\usepackage{graphicx}
\usepackage{epstopdf, epsfig}
\usepackage[utf8]{inputenc}
\usepackage[T1]{fontenc}
\usepackage{amsmath}
\usepackage[usenames,dvipsnames]{color}
\usepackage{float}
\usepackage{mathtools}
\usepackage{amsmath,amssymb,amsbsy}
\usepackage{bbm,bm,mathrsfs,yfonts}
\usepackage{hyperref}
\usepackage[capitalize]{cleveref}
\usepackage[toc,page]{appendix}
\usepackage[export]{adjustbox}
\usepackage[normalem]{ulem} 

\newcommand\redsout{\bgroup\markoverwith{\textcolor{red}{\rule[0.5ex]{2pt}{0.4pt}}}\ULon}
\newcommand{\be}{\begin{equation}}
\newcommand{\ee}{\end{equation}}

\newcommand\numberthis{\addtocounter{equation}{1}\tag{\theequation}}